\documentclass[12pt
]{article}

\usepackage{graphicx}
\usepackage{amsfonts}
\usepackage{amsmath}
\usepackage[usenames]{color}
\usepackage{amssymb}
\usepackage[mathscr]{eucal} 

\usepackage[all]{xy}

\def\Z{\mathbb{Z}}

\def\R{\mathbb{R}}
\def\C{\mathbb{C}}

\begin{document}
\baselineskip 0.6cm
\newcommand{\vev}[1]{ \left\langle {#1} \right\rangle }
\newcommand{\bra}[1]{ \langle {#1} | }
\newcommand{\ket}[1]{ | {#1} \rangle }
\newcommand{\Dsl}{\mbox{\ooalign{\hfil/\hfil\crcr$D$}}}
\newcommand{\nequiv}{\mbox{\ooalign{\hfil/\hfil\crcr$\equiv$}}}
\newcommand{\nsupset}{\mbox{\ooalign{\hfil/\hfil\crcr$\supset$}}}
\newcommand{\nni}{\mbox{\ooalign{\hfil/\hfil\crcr$\ni$}}}
\newcommand{\EV}{ {\rm eV} }
\newcommand{\KEV}{ {\rm keV} }
\newcommand{\MEV}{ {\rm MeV} }
\newcommand{\GEV}{ {\rm GeV} }
\newcommand{\TEV}{ {\rm TeV} }

\def\diag{\mathop{\rm diag}\nolimits}
\def\tr{\mathop{\rm tr}}

\def\Spin{\mathop{\rm Spin}}
\def\SO{\mathop{\rm SO}}
\def\O{\mathop{\rm O}}
\def\SU{\mathop{\rm SU}}
\def\U{\mathop{\rm U}}
\def\Sp{\mathop{\rm Sp}}
\def\SL{\mathop{\rm SL}}

\def\change#1#2{{\color{blue}#1}{\color{red} [#2]}\color{black}\hbox{}}


\begin{titlepage}
 
\begin{flushright}
UT-09-02\\
LTH 820 \\
IPMU09-0014
\end{flushright}
 
\vskip 1cm
\begin{center}
 
{\large \bf Codimension-3 Singularities and Yukawa Couplings in F-theory}

\vskip 1.2cm
 
Hirotaka Hayashi$^1$, Teruhiko Kawano$^1$, Radu Tatar$^2$ and 
Taizan Watari$^{3}$
  
\vskip 0.4cm
{\it $^1$Department of Physics, University of Tokyo, Tokyo 113-0033, Japan  
\\[2mm]
 
$^2$Division of Theoretical Physics, Department of Mathematical
Sciences, The University of Liverpool, Liverpool, L69 3BX, England,
U.K. \\[2mm]
 
$^3$Institute for the Physics and Mathematics of the Universe, 
University of Tokyo, Kashiwano-ha 5-1-5, 277-8592, Japan}
\vskip 1.5cm

\abstract{F-theory is one of the frameworks where all the Yukawa couplings
of grand unified theories are generated and their computation is possible.
The Yukawa couplings of charged matter
multiplets are supposed to be generated around codimension-3 singularity
points of a base complex 3-fold, and that has been confirmed for the
simplest type of codimension-3 singularities in recent studies. However, 
the geometry of F-theory compactifications is much more complicated. 
For a generic F-theory compactification, such issues as flux configuration 
around the codimension-3 singularities, field-theory formulation of 
the local geometry and behavior of zero-mode wavefunctions have virtually 
never been addressed before. We address all these issues in this article, and
further discuss nature of Yukawa couplings generated at such singularities.
In order to calculate the Yukawa couplings of {\it low-energy effective theory},
however, the {\it local} descriptions of wavefunctions on
complex {\it surfaces} and a {\it global} characterization of zero-modes
over a complex {\it curve} have to be combined together. We found 
the relation between them by re-examining 
how chiral charged matters are characterized in F-theory compactification.
An intrinsic definition of spectral surfaces in F-theory turns out
to be the key concept. As a biproduct, we found a new way to understand
the Heterotic--F theory duality, which improves the precision of existing
duality map associated with codimension-3 singularities.
 } 

\end{center}
\end{titlepage}


 \tableofcontents

\section{Introduction and Summary}

The Standard Model of particle physics with neutrino masses 
contains 20--22 flavor parameters, depending on whether the 
neutrino mass is Dirac or Majorana. The Yukawa couplings are 
the coefficients of scalar--fermion--fermion couplings in 
low-energy effective Lagrangian, and they can take 
arbitrary values. This is in sharp contrast to the 
interactions involving gauge bosons, where gauge 
symmetry controls everything and very few free 
parameters are left. The effective-field-theory model 
building succeeded in translating observed 
patterns of flavor parameters into the language 
of flavor symmetries and their breaking patterns, 
but it is hard to understand where the flavor 
symmetries come from. 

In a wide class of string theory compactifications, however, 
quarks, leptons and Higgs boson arise from components of super 
higher dimensional Yang--Mills multiplets  and Yukawa couplings 
descend from super Yang--Mills interactions. 
This fact provides hope to get more theoretical control 
over flavor structure of quarks and leptons. Even though
string theory might eventually provide only a  
translation of known flavor structure in language of 
geometry, yet string theory compactifications with flavor 
structure under control would serve as a reliable framework 
in discussing issues as gravity mediated supersymmetry 
breaking, where both Planck suppressed interactions and 
flavor violating interactions should be brought under control. 

The SU(5) model of grand unification with low-energy supersymmetry 
is perfectly 
consistent with the measured value of gauge coupling constants, 
and major aspects of flavor structure are understood in terms of 
approximate Abelian flavor symmetries (also known as Froggatt--Nielsen
mechanism) consistent with Georgi--Glashow SU(5) unification.
Yukawa couplings of up-type and down-type quarks are written as 
\begin{equation}
 \Delta {\cal L} = 
\lambda_{ij} {\bf 10}^{ab}_i \; {\bf 10}^{cd}_j \; h({\bf 5})^{e} 
 \epsilon_{abcde}+ 
\lambda'_{kj} \bar{\bf 5}_{a; k} \cdot {\bf 10}^{ab}_{j} 
  \cdot \bar{h}(\bar{\bf 5})_b
\label{eq:Yukawa}
\end{equation}
in SU(5) GUT's, where $a,b,\cdots e$ are $\SU(5)$ indices and 
$i, j, k$ generation (flavor) indices. If all the matter fields---Higgs 
boson and matter fermions---arise from super Yang--Mills multiplet 
of higher dimensions with gauge group $G$ containing $\SU(5)$, and 
if all the Yukawa couplings originate from super Yang--Mills
interactions of gauge group $G$, then $G$ has to 
contain $E_6$ \cite{TW1}. This means that the candidates for  such 
framework are either one of Calabi--Yau 3-fold compactification 
of Heterotic $E_8 \times E_8$ string theory, M-theory compactification 
on $G_2$-holonomy manifolds or Calabi--Yau 4-fold compactification 
of F-theory. Those three frameworks are mutually related by string 
duality, but not all the moduli space of the three frameworks 
completely overlap.

Moduli stabilization is well understood in F-theory Calabi--Yau 4-fold 
compactification.\footnote{Reference \cite{TW1} discusses structure of 
Yukawa matrices in $G_2$-holonomy compactification of 11-dimensional
supergravity with SU(5) unification. The Yukawa matrix $\lambda_{ij}$ 
of first term of (\ref{eq:Yukawa}) tends to have small values 
in diagonal entries relatively to off-diagonal entries. To our knowledge, 
this is not a phenomenologically successful pattern of Yukawa matrix. 
In order to avoid this problem in $G_2$-holonomy compactification 
of M-theory, some of the assumptions in \cite{TW1} have to be relaxed.} 
4-form fluxes stabilize complex structure moduli (including moduli 
of 7-branes), which correspond in Calabi--Yau 3-fold compactification 
of Heterotic string theory to vector-bundle moduli as well as 
complex structure moduli. Thus, (F-term part of) Yukawa couplings 
can be discussed in F-theory with everything under control, nothing 
in a black box.

It is now known how to determine the matter spectrum of effective theory 
below the Kaluza--Klein scale in generic supersymmetric compactification 
of F-theory \cite{Curio, DI, DW-1, BHV-1, HayashiEtAl}. Charged matter 
chiral multiplets are identified with global holomorphic sections of 
certain line bundles on complex curves. 
Complex structure of elliptic fibered
Calabi--Yau 4-fold determines the divisors of those line bundles. 
Thus, one can even determine wavefunctions of the zero-mode chiral 
multiplets along the complex curves for a given complex
structure of a Calabi--Yau 4-fold, using rather standard techniques
in algebraic geometry \cite{HayashiEtAl}. Flux compactification can 
be used to determine the complex structure of a Calabi--Yau 4-fold.

A technical problem still lies in a process of calculating Yukawa 
couplings using the wavefunctions of zero modes. 
There is no question that Yukawa couplings are generated; 
massless charged matter multiplets in F-theory are identified with 
M2-branes wrapped on vanishing 2-cycles, and algebra of those 2-cycles 
determines what kind of Yukawa interactions are generated \cite{TW1}.
The question is how to calculate them. Yukawa couplings among charged 
matter multiplets are now believed to be attributed to codimension-3 
singularity points in a base 3-fold \cite{BHV-1, DW-1, HayashiEtAl}.
Unlike Type IIB string theory, however, F-theory (or M-theory) lacks
microscopic descriptions that are valid to any short 
distance scale, 
and an idea is necessary in how to deal with physics localized 
at/around the codimension-3 singularity points.

The references \cite{DW-1, BHV-1} proposed a field theory 
formulation of super Yang--Mills sector associated with 
intersecting 7-branes, which we find is useful particularly 
for this purpose. 
Instead of developing microscopic quantum theory of F-theory, field 
theory on 8-dimensional spacetime is used, and unknown 
microscopic effects can be dealt with as higher order operators 
in $\alpha'$ (or $\kappa_{11D}$) expansion with unknown coefficients. 
Such a framework should be sufficient at least in determining certain 
terms in low-energy effective theory below the Kaluza--Klein scale 
at a controlled level of precision.

In this article, we construct gauge-theory local models\footnote{
We mean by a ``local model'' a theoretical tool to analyze physics 
associated with a local geometry of a Calabi--Yau. We do not use the word 
``model'' in the way it is used in the Standard ``Model'', Georgi--Glashow 
``model'', gauge mediation ``models'' etc. See section \ref{sec:local-model} 
for more explanations on what we mean by ``local models'', as well as 
what one can do with them.} of 
various types of codimension-3 singularities of F-theory
compactification, and analyze the behavior of zero-mode wavefunctions 
around the codimension-3 singularity. Although the field theory
framework of \cite{DW-1, BHV-1} is used for this analysis, we found 
that one has to introduce branch cuts into the field theory and 
fields have to be twisted by Weyl reflection at the branch cuts, 
in order to deal with 
F-theory compactification on a Calabi--Yau with fully generic 
complex structure. This observation has never been made in the 
recent articles on F-theory, and hence the analysis of such local 
models has never been done before.\footnote{
Analysis of \cite{BHV-1, BHV-2} on codimension-3 singularities
covers only non-generic choice of complex structure.} 

It has been known that the zero-mode wavefunctions decay 
in a Gaussian profile $e^{-|u|^2}$ off the matter curves \cite{KV}.
Here, $u$ is a normal coordinate to the matter curve. This result 
however, was obtained only for generic points on matter curves. 
Around codimension-3 singularity points on matter curves, we found 
that this statement is not always true. Depending upon the types of 
codimension-3 singularity and representations of zero-modes, 
the wavefunctions decay off the matter curves either in the Gaussian 
profile $e^{-|u|^2}$ or in a slightly moderate form $e^{-|u|^{3/2}}$. 
All these results are obtained in sections~\ref{sec:warm-up} 
and \ref{sec:GUT}; the latter section deals with codimension-3 
singularities that appear in GUT model building. 

The field-theory local models, however, can be used only to 
study behavior of zero-mode wavefunctions, and interactions 
of the zero modes only in a {\it small region} in a complex 
{\it surface}. Matter zero-modes of effective theory in 4-dimensions, 
on the other hand, are identified with {\it global} holomorphic sections 
of line bundles on complex {\it curves}. Without looking at this 
global issue, we do not even know how many matter multiplets are 
in the low-energy spectrum. We thus need both descriptions, and hence 
need to figure out how these two different descriptions of chiral matter 
are related. This is one of the goals of section \ref{sec:Higgs}. 
In other words, this is to clarify the relation 
between \cite{DW-1, BHV-1} and \cite{HayashiEtAl}.

Section \ref{sec:Higgs} also provides a number of observations 
that are also interesting purely from theoretical perspectives. 
We found that the field expectation values satisfying supersymmetry 
conditions of the field theory formulation \cite{DW-1, BHV-1} 
define objects that are called canonical-bundle valued Higgs bundles
in Mathematical literature. A spectral cover is defined for a 
canonical bundle valued Higgs bundle, which is a divisor of 
a total space of a canonical bundle. Although discriminant locus of 
elliptic fibered Calabi--Yau manifold in F-theory corresponds to 
a $(p,q)$ 7-brane, and is usually regarded as generalization of 
D7-branes in Type IIB string theory, yet we find that the spectral 
cover is more natural generalization of the notion of 
a system of intersecting D7-branes. 
The notion of Higgs bundle and its spectral cover allows us 
i) to naturally extend the characterization of chiral matter multiplets 
in the Type IIB string theory \cite{KS, DKS} into F-theory,\footnote{
Although charged matter multiplets in a given representation have 
their own matter curve, there is not always an irreducible piece 
of the discriminant locus specifically for the representation.
This subtlety becomes clear only when geometry around codimension-3 
singularity is studied carefully.} and 
ii) to provide a totally new way to understand the duality between the 
Heterotic string theory and F-theory, and a new way to determine 
the duality map.

With this better understanding in the Heterotic--F theory
duality, we can study what is really going on in F-theory in geometry 
around codimension-3 singularity points (section \ref{ssec:rami_flux}). 
The duality tells that ramification of spectral 
cover and 3-form potential field (in M-theory language) are the only 
ingredients to think about. Twist of a line bundle on a spectral surface
has long been predicted in Mathematical literature, and there is nothing
unexpected from ramification. We found that 3-form field background is 
single valued, even when 2-cycles are twisted around a point of 
codimension-3 singularity.

In section \ref{sec:Yukawa}, we return to physics application. 
Using the field theory formulation of \cite{DW-1, BHV-1} and 
the zero-mode wavefunctions of the local model in section \ref{sec:GUT},
we find that the up-type Yukawa matrix of Georgi--Glashow SU(5) models 
generated from a codimension-3 singularity with $E_6$ enhanced singularity 
is rank-1 at the leading order for generic choice of complex structure 
of Calabi--Yau 4-fold. This solves the problem raised by \cite{BHV-2}.
We also argue that the overlap of Gaussian tails of zero-mode wavefunctions 
may have significant contributions to the Yukawa eigenvalues of 
lighter generations, a possibility that was not pointed out 
in \cite{HV-2}.

There were recent references discussing subjects that have overlap 
with our work \cite{Ibanez, HV-2}. 

An incomplete list of other articles that discussed physics of 
F-theory in the last several months includes \cite{Conlon, Heckman:2008es, 
Marsano:2008jq, Marsano:2008py, Heckman:2008qt, HV-1, Wijnholt:2008db, 
Jiang:2008yf, Blumenhagen:2008zz, Blumenhagen:2008aw, Heckman:2008jy, 
Bourjaily} and other articles cited elsewhere in this article.

{\bf Reading Guide}: 
Local behavior of zero-mode wavefunctions are studied in 
sections~\ref{sec:warm-up} and \ref{sec:GUT}, and are used in 
the discussion of Yukawa couplings in section \ref{sec:Yukawa}. 
Section~\ref{sec:Higgs} provides theoretical foundation to 
an idea that bridges a small gap between 
sections \ref{sec:warm-up}--\ref{sec:GUT} and section \ref{sec:Yukawa}, 
but the idea itself may look reasonable on its own. 
Thus, section \ref{sec:Higgs} can be skipped for those who are 
interested in Yukawa couplings and think that the idea is acceptable.

Section \ref{sec:Higgs}, on the other hand, deals with theoretical 
aspects of F-theory, and sheds a new light on Heterotic--F theory 
duality. Sections~\ref{sec:warm-up} and \ref{sec:GUT} provide 
``experimental data'' to the theory in section~\ref{sec:Higgs}, 
but we prepared this article so that section \ref{sec:Higgs} can 
be read separately from the other parts of this paper. 
It is an option for those not interested in Yukawa couplings 
so much to read only section \ref{sec:Higgs}.

This section is meant also to serve as a brief summary of this article. 

\section{Field-Theory Local Model of F-theory Geometry}
\label{sec:local-model}

An effective theory with ${\cal N} = 1$ supersymmetry is 
obtained when F-theory is compactified on an elliptic fibered 
Calabi--Yau 4-fold 
\begin{equation}
 \pi: X \rightarrow B_3.
\end{equation}
We assume that $X$ is given by a Weierstrass 
equation
\begin{equation}
 y^2 = x^3 + x f + g, 
\label{eq:Weierstrass-F}
\end{equation}
and $f$ and $g$ are global holomorphic sections of 
${\cal O}(-4 K_{B_3})$ and ${\cal O}(-6 K_{B_3})$, 
respectively.
The discriminant of this elliptic fibration is given by 
\begin{equation}
 \Delta = 4 f^3 + 27 g^2 = 0, 
\end{equation}
which is a complex codimension-1 subvariety (divisor) of $B_3$.

The discriminant locus $\Delta$ may have several 
irreducible components, 
\begin{equation}
 \Delta = \sum_i n_i S_i, 
\end{equation}
where $S_i$ are divisors of $B_3$, and $n_i$ their 
multiplicities. Calabi--Yau 4-fold $X$ develops 
singularity along the discriminant loci $S_i$ 
with multiplicity $n_i > 1$. When $X$ has $A_{N-1}$ 
singularity along an irreducible component $S$, 
then its multiplicity is $n = N$. The multiplicity 
is $n=N+2$ for $D_N$ singularity, and $n=N+2$ 
for $E_N$ ($N = 6,7,8$) \cite{MV1, MV2} (and references therein).

When a complex structure moduli of $X$ is chosen 
allowing such a locus $S$ of $A$-$D$-$E$ singularity
in $\Delta$, its low-energy effective 
theory has a gauge field with the corresponding 
gauge group. Such a compactification becomes a candidate 
for the description of supersymmetric unified theories, 
supersymmetric Standard Model or hidden sector triggering 
dynamical supersymmetry breaking. When the discriminant 
locus is written as 
\begin{equation}
 \Delta = n S + D', 
\label{eq:SandOthers}
\end{equation}
then matter multiplets charged under the gauge group on $S$ arise 
at the intersection $S \cdot D'$.

Some aspects of low-energy effective theory do depend on the 
full geometry of $X$. One example is the Planck scale (Newton constant) of the 
3+1-dimensional Minkowski space. 
Many aspects of gauge theory associated with the discriminant 
locus $S$, however, depend only on the geometry of $X$ around $S$. 
In order to study the profile of wavefunctions of zero-mode 
matter multiplets, one needs to study only the geometry 
along the  $S \cdot D'$ codimension-2 loci of $B_3$. There have 
also been indications that Yukawa couplings essentially 
originate from codimension-3 loci of $B_3$. Thus, 
one can go a long way in phenomenology by studying only the 
local geometry of F-theory compactification.

References \cite{DW-1, BHV-1} formulated 
a field theory on ``$S$'' that can be used to derive 
low-energy effective theory of zero modes along $S$ 
(including those along $S \cdot D'$).
Field contents on $S$ are summarized in Table~\ref{tab:BHV-fields}.
\begin{table}[t]
\begin{center}
\begin{tabular}{l|c|cc|cc}
& vector & chiral & multiplet & anti-chiral & multiplet \\
\hline
Bosonic fields & $A_\mu$ & $\varphi_{mn} du_m \wedge du_n$ & 
 $A_{\bar{m}} d\bar{u}_{\bar{m}}$ & 
 $\overline{\varphi}_{\bar{m}\bar{n}} 
  d \bar{u}_{\bar{m}} \wedge d\bar{u}_{\bar{n}}$ 
& $A_{m} du_m$ \\
Fermionic fields & $\eta$ & 
 $\chi_{mn} du_m \wedge d u_n$ & 
 $\psi_{\bar{m}} d \bar{u}_{\bar{m}}$ & 
 $\bar{\chi}_{\bar{m}\bar{n}} d \bar{u}_{\bar{m}} \wedge d
 \bar{u}_{\bar{n}}$ & $\bar{\psi}_{m} d u_m$ 
\end{tabular} 
\caption{\label{tab:BHV-fields} Field contents on $S$. All fields 
have their values in the Lie algebra $\mathfrak{g}$ that is specified by 
the $A$-$D$-$E$ singularity along $S$. If one of the fields on $S$ 
has a zero mode, then it corresponds in the effective theory in 
4-dimensions to a multiplet of ${\cal N} = 1$ supersymmetry specified 
in the first row of the corresponding column. $u_{m}$
 ($m=1,2$) are holomorphic local coordinates on $S$. 
Fermionic fields also have spinor indices of $\SO(3,1)$ Lorentz group, 
but they are suppressed in this Table. The (2,0)-form field $\varphi$ 
and (0,2)-form field $\overline{\varphi}$ are replaced by 
(1,0)-form $\varphi$ and (0,1)-form $\overline{\varphi}$, when one
 considers compactification to 5+1 dimensions.}
\end{center}
\end{table}
The field
$\varphi_{mn}(u_1,u_2) d u_m \wedge d u_n$ 
on $S$ corresponds to transverse fluctuation $\zeta^l(u_1,u_2)$ 
of D7-branes in Type IIB orientifold compactification 
on a Calabi--Yau 3-fold $X'$.
If $X$ is a  Type IIB Calabi--Yau orientifold of $X'$, $X'$ has a global
holomorphic 3-form $\Omega = \Omega_{mnl}
d u_m \wedge d u_n \wedge d u_l$, where now a set 
of local coordinates $u_m$ ($m = 1,2,3$) is chosen on $X'$.
The field $\zeta^l$ in Type IIB Calabi--Yau orientifold corresponds to 
$\varphi_{mn}$ in formulation of \cite{DW-1, BHV-1} through 
$\Omega_{mnl}\zeta^l$. Since the base 3-fold 
$B_3$ is not necessarily a $\Z_2$ quotient of a Calabi--Yau 
3-fold in general F-theory compactification, the global holomorphic 
3-form $\Omega$ does not necessarily exist. References
\cite{DW-1, BHV-1} found that using $\varphi$ instead of $\zeta$ is 
the right way to formulate the field theory.

\begin{figure}[t]
\begin{center}
\begin{tabular}{ccc}
  \includegraphics[width=.3\linewidth]{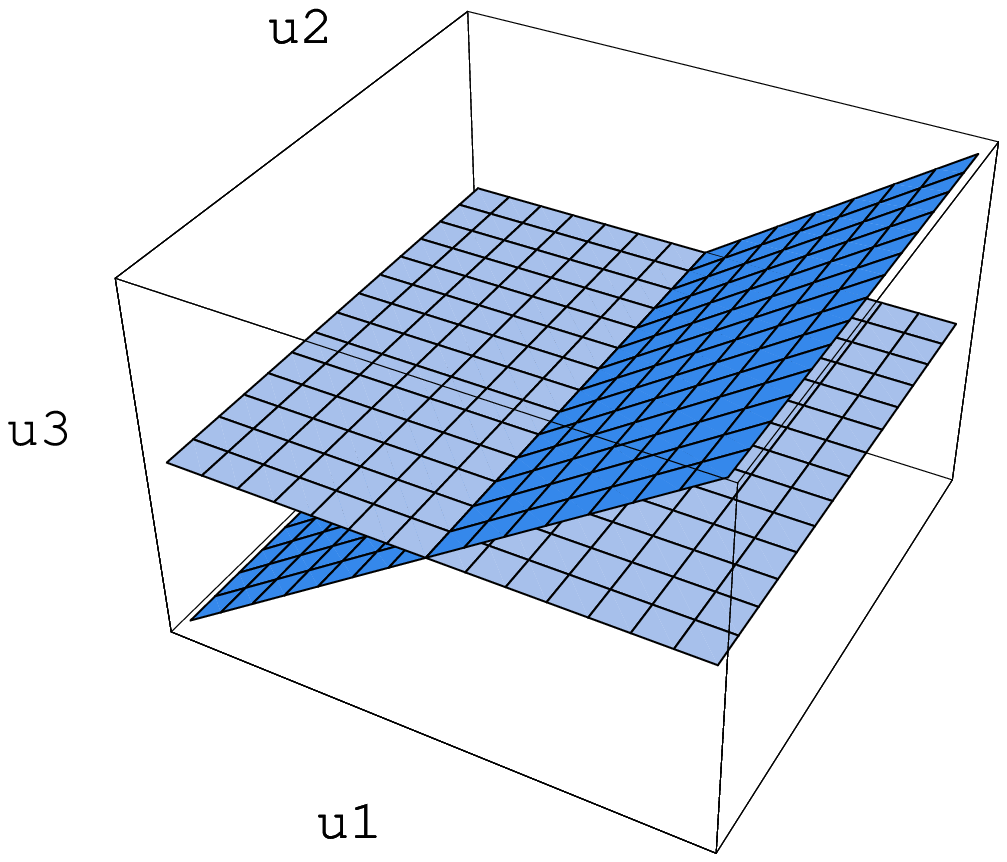} 
   & $\qquad$ &
  \includegraphics[width=.3\linewidth]{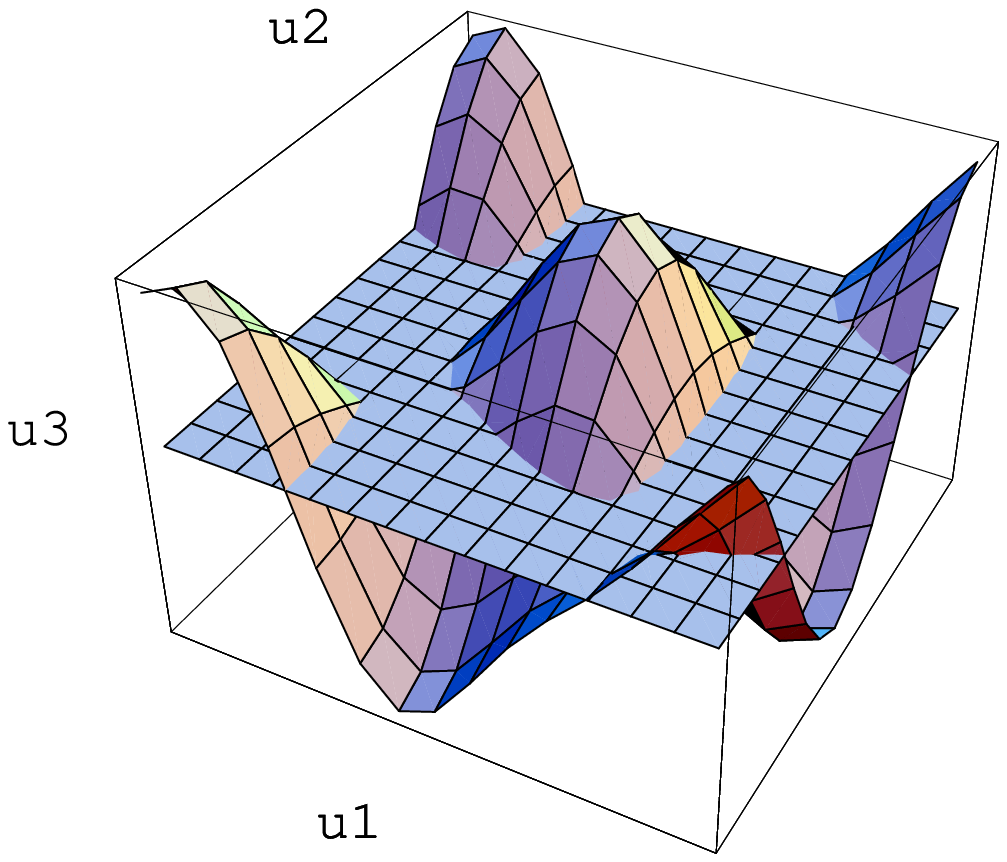}  \\
(a) & & (b)
\end{tabular}
 \caption{\label{fig:7-7}
(a) is a cartoon picture of $T^6$ with two stacks of
 D7-branes wrapping two different topological 4-cycles. 
Only real locus is described here, though.
One of the two stacks of D7-branes cannot be obtained by deforming 
continuously the configuration of the other stack of D7-branes. 
This is the typical situation one imagines for $S$ and $D'$.
In the D7-brane configuration in (b), on the other hand, one of the 
two stacks of D7-branes can be obtained by continuous deformation 
from D7-branes originally in the configuration of the other stack 
of D7-branes. This is an intuitive picture of $S''$ and $S'$ 
in (\ref{eq:cont-deform}) in Type IIB language. At the 7-brane intersection 
curves, however, there is no qualitative difference locally between 
the $S \cdot D'$ intersection in (a) and $S'' \cdot S'$ intersection in
 (b). This is why local physics of a system like (a) can be
 studied by local description of a system like (b), where 7-brane 
configuration is described by a globally defined field vev
 $\vev{\zeta}$.}
\end{center}
\end{figure}

Suppose that a zero mode exists in $\varphi$ for a given 
compactification; that is, $H^0(S; \mathfrak{g} \otimes K_S) \neq 0$.
Non-vanishing vacuum expectation value (vev) in $\varphi$ corresponds 
to deforming geometry of $X$, so that the irreducible decomposition of 
discriminant $\Delta$ in (\ref{eq:SandOthers}) becomes 
\begin{equation}
 \Delta = n'' S'' + S' + D',
\label{eq:cont-deform}
\end{equation}
with $n'' < n$ and $S'$ and $S''$ are topologically the same as $S$ 
in $B_3$.
Singularity along the irreducible discriminant locus $S$ is reduced 
from\footnote{Here, we abuse language and 
do not make a distinction between Lie algebra $\mathfrak{g}$
of $A$-$D$-$E$ type and singularity type of the corresponding 
one of $A$-$D$-$E$.} $\mathfrak{g}$ to the commutant $\mathfrak{g}''$ of 
$\vev{\varphi}$ in $\mathfrak{g}$. Suppose that 
$\vev{\varphi}$ takes its value in 
$\mathfrak{g}' \subset \mathfrak{g}$. Then, the irreducible
decomposition of $\mathfrak{g}$ is of the form 
\begin{equation}
 {\rm Res}^G_{G' \times G''} \mathfrak{g} = 
 ({\bf 1},{\bf adj.}) + ({\bf adj.},{\bf 1}) + \oplus_i (U_i, R_i).
\label{eq:irr-decmp}
\end{equation}
The hypermultiplets in the representation $R_i$ of unbroken symmetry $G''$
are localized along a complex curve $S'' \cdot S'$.
Chiral (and anti-chiral) matter multiplets exist in low-energy effective 
theory of 4-dimensions, if there are zero modes on the curve $S'' \cdot S'$.
Because the localization of these matter fields are solely due to 
the vev of $\varphi$, one can use the field theory on $S$ 
with non-vanishing $\vev{\varphi}$ to study 
behavior of localization of the $G''$-charged matter
on the curve $S'' \cdot S'$
and behavior of zero modes along the curve.\footnote{
Note that 
the field theory is formulated on a complex surface $S$, which is 
neither identified with the discriminant loci $S''$ nor $S'$ in the 
presence of $\vev{\varphi}$. To borrow an intuitive picture of Type IIB 
string in Figure~\ref{fig:7-7}, any points with the same $(u_1,u_2)$ 
coordinates on 7-branes are identified with a point with $(u_1,u_2)$ 
on ``$S$'', and the difference in the $u_3$ coordinate is ignored;  
configuration of 7-branes in the $u_3$ direction are encoded as 
$\vev{\varphi}$ that depend on $(u_1,u_2)$ in the field-theory 
formulation.}

Codimension-2 singularity loci $S \cdot D'$ in
(\ref{eq:SandOthers}), however, are not always obtained as 
$S'' \cdot S'$ associated with a non-trivial $\vev{\varphi}$ globally 
defined on ``$S$''.
Two cartoon pictures in 
Figure~\ref{fig:7-7} describe how D7--D7 intersection arises 
in Type IIB string theory. 
Not all D7--D7 intersection curves 
are obtained by turning on a vev $\vev{\zeta}$ or $\vev{\varphi}$ 
defined globally on $S$ (i.e., by deforming D7-brane configuration 
continuously), but local geometry around D7--D7 intersection curves 
is essentially the same in (a) and (b) of Figure~\ref{fig:7-7}.
Matter multiplets arise from open strings connecting two stacks 
of D7-branes, and are localized along the 7-brane intersection curve. 
They would not care about the global configuration of D7-branes, 
except that they interact at points where D7--D7 intersection curves 
meet one another.
Thus, the localization of massless matter multiplets will 
not be affected by the global configuration of irreducible 
discriminant loci (i.e. 7-branes) in F-theory. 
An important idea of \cite{BHV-1}, which essentially dates 
back to \cite{KV}, is that local geometry of F-theory compactification 
is approximately regarded as deformation of type $\mathfrak{g}$ 
singularity to type $\mathfrak{g}''$ for appropriate choice of 
$\mathfrak{g}$ and $\mathfrak{g}''$, and a corresponding gauge theory 
can be used as an (approximate) local model of local geometry to 
study physics issues such as the localization of matter multiplets 
and their interactions. 

Not all the aspects of low-energy effective theory in 4-dimensions 
depend on all the details of the elliptic fibered Calabi--Yau geometry
$X$, as we mentioned before. It is best to focus on relevant parts of
geometry to determine various aspects of the effective theory. 
Newton constant depends on the whole volume of $B_3$, whereas the gauge 
coupling of a non-Abelian gauge theory on $S$ depends only 
on ${\rm vol}(S)$. To study wavefunctions of localized zero-modes 
of charged matter, one needs to look at the entire compact matter 
curves (codimension-2 loci of $B_3$), but only a region around 
a codimension-3 singularity point in $B_3$ will be sufficient 
in studying individual contributions to Yukawa couplings. To study 
Yukawa couplings, local patches of $S$ and an appropriate Lie algebra 
$\mathfrak{g}$ can be chosen around codimension-3 singularity points, 
so that the local geometry is studied by a field theory local model 
with gauge group $G$, which is broken down to $G''$ by $\vev{\varphi}$. 
Such field-theory local models around codimension-3 singularity points 
are glued together along the matter curves, in order to obtain
low-energy effective theory below the Kaluza--Klein scale. 
We do not need an 8-dimensional field theory description that covers the 
entire discriminant locus $S$. This is a radical yet powerful idea.

In sections \ref{sec:warm-up} and \ref{sec:GUT}, we will construct 
field theory local models for various types of codimension-3 singularity 
points in F-theory and study the behavior of zero-mode wavefunctions.  
Although there are so many different choices of topology of Calabi--Yau 
4-fold and four-form flux on it, only a limited number of types of 
local geometry around codimension-3 singularity points. Thus, the local 
models are quite universal. Yukawa couplings generated in such 
local models are discussed in section \ref{sec:Yukawa}, and we will 
discuss how to glue the results of these local models together 
to obtain low-energy effective theory in sections \ref{sec:Higgs} 
and \ref{sec:Yukawa}.\footnote{Possible limitation of this approach 
of gluing local descriptions together is also discussed in section 6.2.
See discussion that follows (\ref{eq:ratio}).}

Reference \cite{BHV-1} determined the (leading order part of the) 
Lagrangian of the field theory formulation on 8-dimensions, 
and this Lagrangian is used in studying the field theory local models.
We quote the Lagrangian from \cite{BHV-1} 
in the appendix \ref{sec:action}, after 
clarifying details of conventions and correcting typos in signs, 
phases and coefficients.

\section{Generic Rank-2 Deformation of $A_{N}$ Singularity}
\label{sec:warm-up}

In order to construct field theory local models for 
a local geometry, we need to determined field vev $\vev{A_{\bar{m}}}$ 
and $\vev{\varphi}$ corresponding to local geometry of a Calabi--Yau
4-fold and a local configuration of 3-form potential. 
We begin by mapping the  complex structure of the Calabi--Yau 
4-fold into the vev of the field $\vev{\varphi}$ in sections 
\ref{sec:warm-up} and \ref{sec:GUT}. The 4-form flux background 
is implemented in field theory in section \ref{sec:Higgs}.
We only discuss local models until section \ref{sec:Yukawa}, 
and hence complex surfaces ``$S$'' on which field theory is defined 
is regarded as a non-compact space.\footnote{As we already mentioned in 
the previous section, ``$S$'' is not identified with a specific 
divisor in the base 3-fold $B_3$.}

The mapping between the Calabi--Yau geometry and the values of 
$\vev{\varphi}$ has been first used in \cite{KV}. The main idea 
is that the generic deformations of surface singularities 
of type $\mathfrak{g}$---one of $A_n$, $D_n$ or 
$E_{6,7,8}$---are parametrized by $\mathfrak{h} \otimes_\R \C/W$, 
where $\mathfrak{h}$ is a Cartan subalgebra of Lie algebra 
$\mathfrak{g}$, and $W$ its Weyl group. 
When we consider a local Calabi--Yau geometry 
that is a fibration (or family) of a deformation of 
a surface singularity of type $\mathfrak{g}$ over 
a base space $B$, the local geometry is described by a gauge theory 
with gauge group $G$.
The variation of the deformation parameters of the fiber
geometry over $B$ are encoded in 
$\vev{\varphi} \in \mathfrak{h} \otimes_\R \C$ varying over $B$ 
in the field-theory description.
The dictionary between the defining equation of deformation 
of singularity of type $\mathfrak{g}$ and the parametrization 
of deformation by $\mathfrak{h} \otimes_\R \C/W$ was well 
well-established in \cite{KM}. Thus, the dictionary can be 
used to translate the local geometry naively into 
$\vev{\varphi} \in \mathfrak{h} \otimes_{\R} \C$, if one 
ignores the difference between $\mathfrak{h} \otimes_\R \C /W$ 
and $\mathfrak{h} \otimes_\R \C$.

Most of the explicit studies in the literature essentially
deal with deformation over a non-compact complex curve $B$. 
The gauge group $\mathfrak{g}$ is often chosen to be 
just one-rank larger than the singularity (symmetry) 
$\mathfrak{g}''$ that remains over $B$. The consequence is simple:
the hypermultiplet is localized at points where $\mathfrak{g}''$ 
is enhanced to $\mathfrak{g}$, with a Gaussian wavefunctions around 
the enhancement points \cite{KV}.
But the idea itself is more general and we can apply it to for cases 
with a complex surface $B$. 
It is a natural guess to choose $\mathfrak{g}$ of a local
model to be two-rank larger 
than $\mathfrak{g}''$ left unbroken over $B$. This choice 
is proved to be correct for all the examples we study 
in this article. We will denote the base space as $S$ hereafter.

Calabi--Yau 4-fold geometry of F-theory compactification is much 
more general than the one coming from Calabi--Yau orientifold of 
Type IIB string compactification. In addition to the $\SL_2 \Z$ 
twist, and a twist that yields ``non-split'' singularity locus, 
there is another kind of twists generically in F-theory compactification, 
which corresponds to the difference between $\mathfrak{h} \otimes_\R \C$ 
and $\mathfrak{h} \otimes_\R \C/W$ mentioned above \cite{6authors}. 
To our knowledge, not much discussion is found in the literature 
how to deal with this Weyl-group twist of F-theory in the field theory 
description. We will construct local models for geometry with this 
twist in section \ref{ssec:AN-gen}, by using 8-dimensional field theory 
with branch cuts, and explain an useful idea is studying such field 
theory.

To begin with, let us study most generic deformation of 
$A_{N+1}$ singularity in F-theory. Since we are interested in 
application to study of local geometry around codimension-3
singularities, we only consider 2-parameter deformation of 
$A_{N+1}$ singularity. The most generic form of deformation 
of surface singularity $A_{N+1}$ to $A_{N-1}$ 
is given by two parameters, $s_1$ and $s_2$:
\begin{equation}
 Y^2 = X^2 + Z^N (Z^2 + s_1 Z + s_2).
\label{eq:AN+1--AN-1}
\end{equation}
When $s_1$ and $s_2$ are set to zero, $A_{N+1}$ singularity 
is restored. An alternative parametrization of deformation 
is\footnote{The deformation parameters are shifted from 
$\mathfrak{su}(N+2) \otimes \C$ to $\mathfrak{u}(N+2) \otimes \C$, 
but this overall ``center-of-mass'' shift is irrelevant because 
it is absorbed by the shift of the origin of the coordinate $Z$.
} 
\begin{equation}
 \diag(\overbrace{0,\cdots,0}^N, \tau_{N+1}, \tau_{N+2})
\end{equation}
in $(\mathfrak{su}(2)+\mathfrak{u}(1)) \otimes_\R \C \subset 
\mathfrak{su}(N+2) \otimes_\R \C$. The relation between 
$(s_1,s_2)$ and $(\tau_{N+1}, \tau_{N+2})$ is 
\begin{equation}
 s_1 = -(\tau_{N+1} + \tau_{N+2}), \qquad s_2 = \tau_{N+1} \tau_{N+2}.
\label{eq:relation-Atype}
\end{equation}
Both are invariant functions of Weyl group of $\mathfrak{su}(2)$, which 
is permutation $\mathfrak{S}_2$ of $\tau_{N+1}$ and $\tau_{N+2}$.

When the deformation parameters $s_1$ and $s_2$ 
in (\ref{eq:AN+1--AN-1}) are promoted to functions of local 
coordinates $(u_1,u_2)$ of a non-compact base space $S$, 
the equation (\ref{eq:AN+1--AN-1}) now defines a family (fibration) 
of surface singularity $A_{N+1}$ over $S$ deformed to $A_{N-1}$.
If a local geometry is given approximately by (\ref{eq:AN+1--AN-1}), 
then $\tau_{N+1}(u_1,u_2)$ and 
$\tau_{N+2}(u_1,u_2)$ are determined from (\ref{eq:relation-Atype}).
Local geometry given by (\ref{eq:AN+1--AN-1}) corresponds to 
a field-theory local model with non-vanishing vev\footnote{
``$\alpha$'' here is a constant associated with normalization of 
the field $\varphi$, and is largely unimportant in physics. See 
the appendix \ref{sec:action} for more details.} 
\begin{equation}
 2\alpha \vev{\varphi_{12}}(u_1,u_2) = 
\diag ( \overbrace{0, \cdots, 0}^N,  \tau_{N+1},\tau_{N+2});
\end{equation}
the field vev $\vev{\varphi}$ varies over the local coordinates 
$(u_1, u_2)$ of $S$, if $s_1, s_2$ do.
The gauge group of this field theory is $G = \SU(N+2)$, but 
only $\SU(N)$ symmetry\footnote{See discussion below for whether 
gauge fields of Abelian factors remain massless in the low-energy 
effective theory.} 
remains unbroken in effective theory, if $s_1$ and $s_2$ do not 
vanish at generic points of $S$.

\subsection{Three Intersecting D7-branes}
\label{ssec:warm-up-1}

\begin{figure}[t]
 \begin{center}
  \begin{tabular}{ccc}
  \includegraphics[width=.3\linewidth]{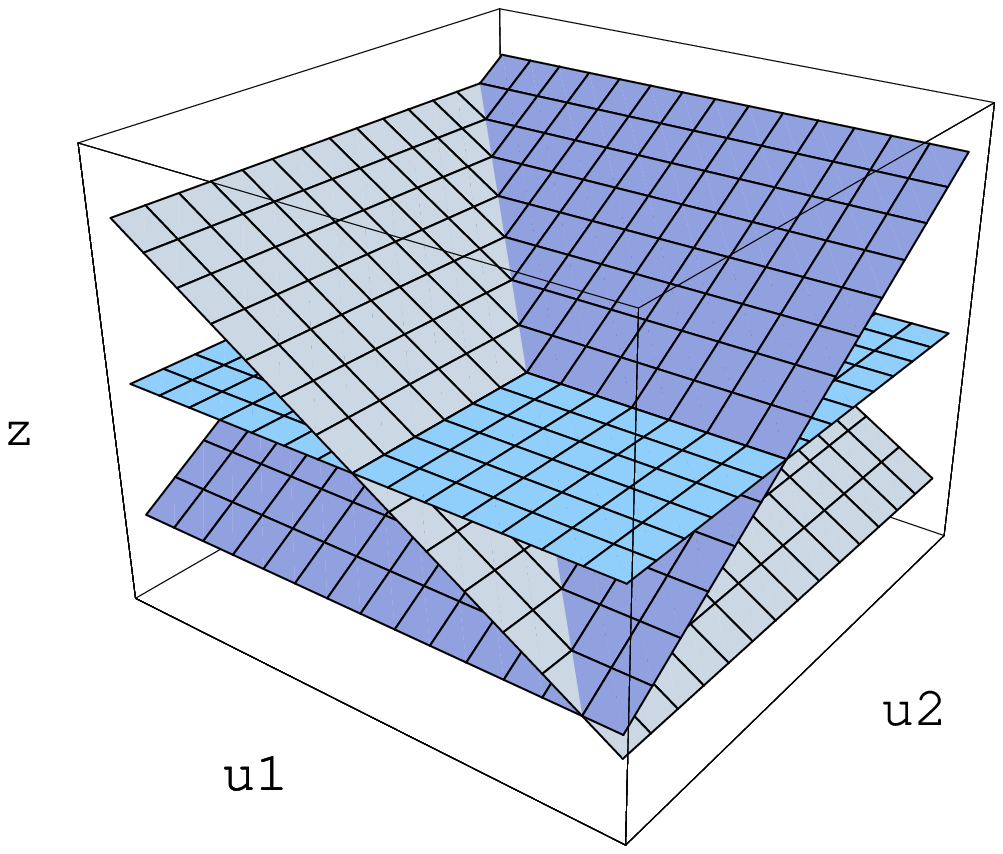} 
   & &
  \includegraphics[width=.3\linewidth]{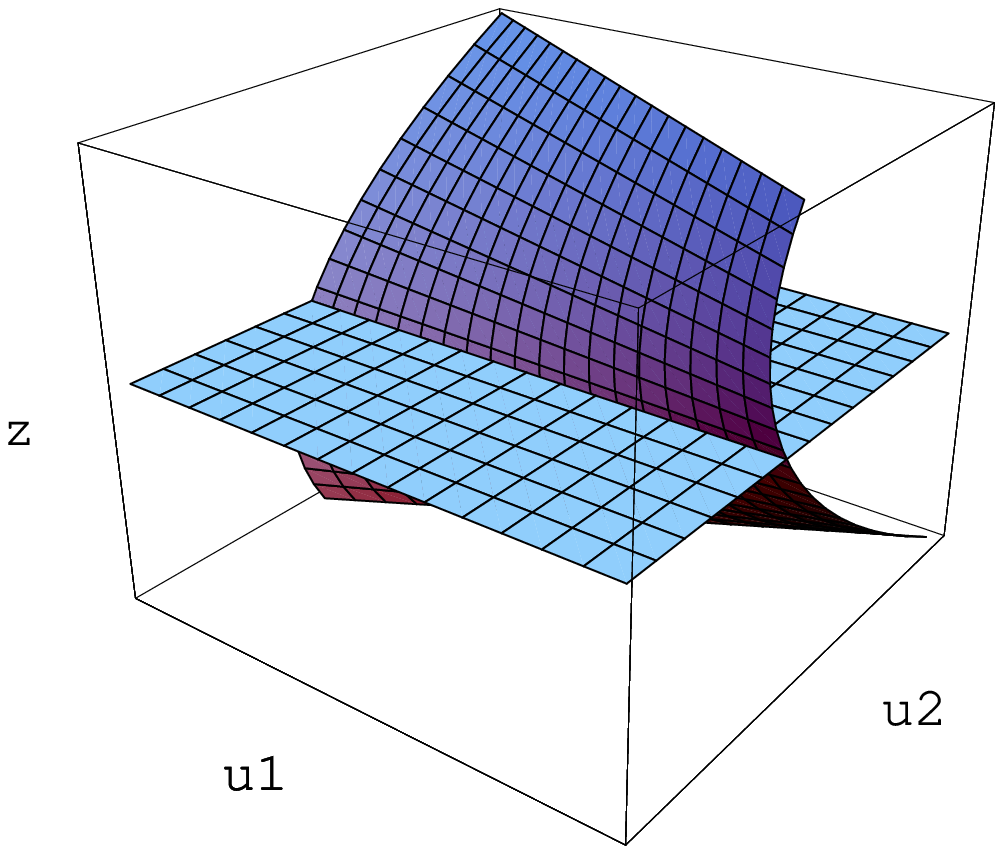}  \\
  (a) & & (b)
  \end{tabular}
\caption{\label{fig:Atype} Behavior of irreducible components of 
discriminant loci (7-branes) near codimension-3 singularities
associated with deformation of $A_{N+1}$ singularity down to $A_{N-1}$.
Only real locus is shown. (a) and (b) in this figure corresponds to 
the local behavior of $s_1$ and $s_2$ given in (\ref{eq:typeA-caseA}) 
and (\ref{eq:typeA-caseB}), respectively. Although there appears to be 
nothing singular at $(u_1,u_2) = (0,0)$ in the panel (b), the
discriminant $\Delta$ goes to $z^{N+2}$ there. 
}
 \end{center}
\end{figure}

We start with a choice for the values of $s_i$ as functions 
of $(u_1, u_2)$ which give rise to the well understood 
configuration of three intersecting D7-branes:
\begin{equation}
 s_1(u_1,u_2) = -(F_1 u_1 + F_2 u_2), \qquad 
 s_2(u_1,u_2) = F_1 F_2 u_1 u_2,
\label{eq:typeA-caseA}
\end{equation}
where $F_{1,2}$ are proportionality constants.
The irreducible components of discriminant locus,\footnote{
Although (\ref{eq:AN+1--AN-1}) does not define an elliptic fibration, 
it does contain a fibration of 1-cycle in $(X, Y)$ plane, and the 
``discriminant locus'' (\ref{eq:discr-AN}) is where the 1-cycle 
shrinks to zero size. }
\begin{equation}
 Z^N(Z^2 + s_1 Z + s_2) = 0 \quad \longrightarrow \quad 
 Z^N = 0 \quad {\rm and} \quad (Z^2 + s_1 Z + s_2)=0,  
\label{eq:discr-AN}
\end{equation}
behaves as in Figure~\ref{fig:Atype}~(a). 
Since the $(Z^2 + s_1 Z + s_2) = 0$ component of 
the discriminant locus (i.e., 7-brane) further factorizes into 
$(Z - F_1 u_1)= 0$ and $(Z - F_2 u_2) = 0$, 
this is nothing more than an F-theory 
description of a well-understood configuration in Type IIB 
string theory; three stacks of D7-branes intersect at angles, 
and one of the three stacks consists of $N$ D7-branes.
We begin with this almost trivial example, just to get 
started. The corresponding vev of $\varphi$ is 
\begin{equation}
2 \alpha \vev{\varphi_{12}} = \diag ( \overbrace{0,\cdots,0}^N,F_1 u_1,
			     F_2 u_2 ),
\label{eq:bg-AA}
\end{equation}
which is obtained by solving (\ref{eq:relation-Atype}) and 
substituting the explicit form of $s_1$ and $s_2$ in 
(\ref{eq:typeA-caseA}). 

The irreducible decomposition of $\mathfrak{su}(N+2)$ is\footnote{
There is no reason within this local model that the massless components 
of the vector fields for the two U(1) factors are gone from the
spectrum. Whether they remain in the massless spectrum of effective
theory below the Kaluza--Klein scale, however, is a more global issue. 
} 
\begin{equation}
 \mathfrak{su}(N+2)\mbox{-}{\bf adj.} \rightarrow 
 \mathfrak{su}(N)\mbox{-}{\bf adj} + \mathfrak{u}(1) + \mathfrak{u}(1) +
 \left[ N^{(-,0)}+N^{(0,-)}+ {\bf 1}^{(+,-)}\right] + {\rm h.c.}
\end{equation}
in this case, where the two signs on the shoulder indicate 
the charge under ${\rm ad}(\tau_{N+1})$ and ${\rm ad}(\tau_{N+2})$ 
respectively of a given irreducible component. Zero-mode equations,
\begin{eqnarray}
 {\rm e.o.m~of~}\eta : & & 
 2 \omega \wedge \partial \psi + |\alpha|^2
   [ \vev{\overline{\varphi}}, \chi] = 0, \label{eq:eomEta}\\
 {\rm e.o.m~of~}\chi : & & 
 \bar{\partial} \psi + \frac{i}{\sqrt{2}} \alpha^* 
 [\vev{\overline{\varphi}}, \eta] = 0, \label{eq:eomChi}\\
 {\rm e.o.m~of~} \psi : & & 
 \alpha \bar{\partial} \chi - \sqrt{2} i \omega \wedge \partial
 \eta - i \alpha [ \vev{\varphi}, \psi] = 0, \label{eq:eomPsi}
\end{eqnarray}
follow\footnote{(\ref{eq:eomEta}--\ref{eq:eomPsi}) are 
not precisely the equations of motions obtained from (\ref{eq:fermi-bilin}).
$(1/\sqrt{2}) \omega \wedge \omega \; \sigma^\mu D_\mu \bar{\eta}$ 
is missing from (\ref{eq:eomEta}), for example. True equations of
motions are solved with separation of variables, but the equations 
for the wavefunctions of zero modes are obtained by simply dropping 
derivatives in the Minkowski directions from the equations of motions.
} from the action\footnote{By taking the coordinates for 
orthonormal metric locally and promoting 
$i \; {\rm ad}(2 \alpha \varphi_{12})$ to 
a covariant derivative $\bar{\partial}_{\bar{3}}$ and 
$i \; {\rm ad}(2 \alpha \overline{\varphi}_{12})$ by 
$\partial_{3}$, one finds that these set of equations of motions 
have $\SU(3) \times \U(1) \subset \SU(4)$ R-symmetry of 16-SUSY 
Yang--Mills theories;
\begin{eqnarray}
\sum_{m=1}^3 \partial_m \psi_{\bar{m}} & = & 0, \\
 \partial_{m} \psi_{\bar{0}} + \epsilon^{\bar{m}\bar{k}\bar{l}}
  \bar{\partial}_{\bar{k}}\psi_{\bar{l}} & = & 0,
\end{eqnarray}
where $\psi_{\bar{0}} \equiv \eta/\sqrt{2}$ and 
$\psi_{\bar{3}} \equiv 2 \alpha \chi_{12}$ are left-handed 
spinors, just like $\psi_{\bar{1}, \bar{2}}$.
This serves as a check of various coefficients appearing in the
Lagrangian (\ref{eq:fermi-bilin}).} 
(\ref{eq:fermi-bilin}). 
These equations should be solved separately for each irreducible 
component, $N^{(-,0)}$, $N^{(0,-)}$ and their hermitian conjugates, 
in the presence of a common background such as (\ref{eq:bg-AA}). 
As long as the background preserves supersymmetry, the zero mode 
wavefunctions of $A_{\bar{m}}$ and $\varphi_{12}$ should be 
the same as those of $\psi_{\bar{m}}$ and $\chi_{12}$.

The zero mode equations can be written down more explicitly, 
if we assume that we can take\footnote{Although metric of $B_3$
should exhibit certain form of singularity along discriminant loci such
as $S$ in (\ref{eq:SandOthers}) (e.g., \cite{Sen}), 
one should note that the ``$S$'' here 
is a base space and is not a divisor of $B_3$. It is not clear what 
kind of metric should be used here. We just adopt the simplest
possibility here, which may be to ignore gravitational backreaction
and consider only the Yang--Mills sector. } 
a local coordinate system of $S$
so that the metric is locally approximately orthonormal, 
$g_{m\bar{n}} = \delta_{mn}$. 
\begin{eqnarray}
 & & \partial_2 \tilde{\psi}_{\bar{2}} + \partial_1
  \tilde{\psi}_{\bar{1}} - \lambda_i \bar{\tau_i}(\bar{u}_1,\bar{u}_2) 
  \tilde{\chi} = 0,   \label{eq:EOM1}\\
 & & \bar{\partial}_{\bar{1}} \tilde{\psi}_{\bar{2}} - 
         \bar{\partial}_{\bar{2}} \tilde{\psi}_{\bar{1}} = 0,
	 \label{eq:EOM2} \\
 & & \bar{\partial}_{\bar{1}} \tilde{\chi} - \lambda_i \tau_i(u_1,u_2)
  \tilde{\psi}_{\bar{1}} = 0, \label{eq:EOM3} \\
 & & \bar{\partial}_{\bar{2}} \tilde{\chi} - \lambda_i \tau_i(u_1,u_2)
  \tilde{\psi}_{\bar{2}} = 0, \label{eq:EOM4}
\end{eqnarray}
where $\tilde{\chi} \equiv 2 \alpha \chi_{12}$ and 
$\tilde{\psi}_{\bar{m}} = i \psi_{\bar{m}}$. 
$\lambda_i$ ($i=N+1, N+2$) corresponds to charges of a given 
irreducible component under the background $\tau_i$; 
$\lambda_i = \pm 1, 0$ and are read out from 
$N^{(\lambda_{N+1}, \lambda_{N+2})}$ or 
$\bar{N}^{(\lambda_{N+1}, \lambda_{N+2})}$.
We set $\eta = 0$, because we are interested in zero-modes that become 
chiral multiplets of ${\cal N} = 1$ supersymmetry in 4-dimensions.

The partial derivatives $\partial_{m},
\bar{\partial}_{\bar{m}}$ 
($m=1,2$) are 
covariant derivatives involving a gauge field that corresponds to 
four-form flux on a Calabi--Yau 4-fold. Before sections 
\ref{sec:Higgs} and \ref{sec:Yukawa}, we assume that this gauge field 
background has negligible effects in determining zero-mode
wavefunctions in a local region near points of codimension-3
singularity. Once they are ignored, then one can see that 
the zero mode equations for 
$(\tilde{\psi}_{\bar{m}}, \tilde{\chi})$ in the component 
$N^{(\lambda_{N+1},\lambda_{N+2})}$ are the same as those of 
$(- \tilde{\psi}_{\bar{m}}, \tilde{\chi})$ in the component 
$\bar{N}^{(-\lambda_{N+1},-\lambda_{N+2})}$. 
Chirality, the difference between 
the number of zero modes in $\SU(N)$-$N$ representation and 
$\bar{N}$ representation, comes purely from the gauge field. 
But, obtaining chiral spectrum at low-energy and generating 
Yukawa interactions among them are completely separate issues. The 
topological aspects of gauge field on the entire compact discriminant locus
is essential for the former, whereas only local geometry on $S$
is presumably relevant to the latter. As long as we discuss the 
Yukawa interactions, we ignore the gauge field background for now. 

It is not difficult to find a zero-mode wave function under the 
background (\ref{eq:bg-AA}). Since $\tau_{N+1}(u_1,u_2)$ 
[resp. $\tau_{N+2}(u_1,u_2)$] does not depend on $u_2$ [resp. $u_1$], 
one can seek for a wavefunction that does not depend on $u_2$
[resp. $u_1$] for the irreducible components $N^{(-,0)}$ and 
$\bar{N}^{(+,0)}$ [resp. $N^{(0,-)}$ and $\bar{N}^{(0,+)}$]. 
For the $N^{(-,0)}$ and $\bar{N}^{(+,0)}$ components,\footnote{
We assume that $F_1$ and $F_2$ are chosen real and positive, which 
is always possible without a loss of generality; one can redefine the 
phases of the coordinates $u_1$ and $u_2$, if $F_1$ and $F_2$ are not 
real and positive.} 
\begin{equation}
 \tilde{\chi}_{\mp} = c_{\mp} \exp \left[ - F_1 |u_1|^2 \right], \qquad 
 \tilde{\psi}_{\bar{1}\mp} = \pm c_{\mp}  
    \exp \left[ - F_1 |u_1|^2 \right], \qquad
    \tilde{\psi}_{\bar{2}} = 0.
\label{eq:Gaussian1}
\end{equation}
One can also see from the zero-mode equations that the coefficient 
$c_{\mp}$ 
can be holomorphic functions of $u_2$, yet the solutions above satisfy
all the equations; no $\bar{u}_2$ dependence is allowed,
however,  because of $\bar{\partial}_{\bar{2}}$ in 
(\ref{eq:EOM2}, \ref{eq:EOM4}). Thus, this zero mode is Gaussian 
in the $u_1$ direction, and is localized on a complex curve given 
by $\tau_{N+1} \propto u_1 = 0$. Along the curve parametrized 
by $u_2$, the zero mode should be a holomorphic function of $u_2$, 
which is not required to have any particular structure 
(such as pole or zero) at the codimension-3 singularity point 
$(U_1,u_2) = (0,0)$. In Type IIB language, these zero modes correspond 
to open strings connecting $N$ D7-branes and $(N+1)$-th D7-brane. 
Although another ($(N+2)$-th) D7-brane also passes through the 
point $(u_1,u_2) = (0,0)$,the  open strings in $N^{(-,0)}$ and 
$\bar{N}^{(+,0)}$ are not affected by the presence of 
the $N+2$-th D7-brane, quite a reasonable conclusion.

Similarly, for the $N^{(0,-)}$ component, 
\begin{equation}
 \tilde{\chi} = c(u_1) \exp \left[ - F_2 |u_2|^2 \right], \qquad 
 \tilde{\psi}_{\bar{2}} = c(u_1)  
    \exp \left[ - F_2 |u_2|^2 \right], \qquad
    \tilde{\psi}_{\bar{1}} = 0.
\label{eq:sol-AA-2}
\end{equation}
This is Gaussian in $u_2$ direction, and is localized along the curve 
$\tau_{N+2} \propto u_2 = 0$. The coefficient function $c$ can now 
be a holomorphic function of $u_1$, and the wavefunction in this local 
region is determined, once the behavior of $c(u_1)$ is determined.  
The holomorphic function $c(u_1)$ is not required to have any 
particular structure at $(u_1,u_2)=(0,0)$; this is an F-theory 
description. In type IIB interpretation, these zero modes correspond 
to open strings connecting $N$ D7-branes and $(N+2)$-th D7-brane.

\subsection{Generic Deformation of $A_{N+1}$ to $A_{N-1}$}
\label{ssec:AN-gen}

The first case (\ref{eq:typeA-caseA}) was a suitable choice 
in order to get three stacks of intersecting D7-branes 
in type IIB picture.
However, that choice may not be 
the most generic deformation of $A_{N+1}$ singularity 
down to $A_{N-1}$, because the zero locus of $s_2$ in (\ref{eq:typeA-caseA}) has 
a double point $(u_1,u_2) = (0,0)$ where $s_1$ also vanishes.
Not all the deformations of $A_{N+1}$ should be like that.
Instead, one can think of the following form of deformation:
\begin{equation}
 s_1 = 2 u_1,  \qquad s_2 = u_2,
\label{eq:typeA-caseB}
\end{equation}
where local coordinates $(u_1, u_2)$ are chosen so that 
the expressions above become simple. The essence of this
case is that there is a common zero of $s_1$ and $s_2$ 
on $S$, and no further assumptions are made.

The second case of the deformation of $A_{N+1}$ singularity 
is described by a field theory local model with 
$2\alpha \vev{\varphi_{12}}$ given by\footnote{It is not 
absolutely obvious whether the vev $2\alpha\vev{\varphi_{12}}$ 
can be diagonalized at $u_1^2 - u_2 =0$. 
The $2 \times 2$-matrix valued $2\alpha\vev{\varphi_{12}}$ 
may become a rank-2 Jordan block with a degenerate eigenvalue $-u_1$. 
We will be naive about anything around the branch locus 
($u_1^2 - u_2 = 0$) in the rest of this section, and similarly 
in section~\ref{sec:GUT}. We will return to this issue 
in section~\ref{sec:Higgs}, however, and develop an argument 
that gets around this issue.} 
\begin{equation}
 \tau_+ \equiv \tau_{N+1} = - u_1 + \sqrt{u_1^2 - u_2}, \qquad 
 \tau_- \equiv \tau_{N+2} = - u_1 - \sqrt{u_1^2 - u_2}.
\label{eq:bg-AB}
\end{equation}
This follows from solving (\ref{eq:relation-Atype}) and 
substituting (\ref{eq:typeA-caseB}). 
In this case, an irreducible decomposition\footnote{
The vector field for the $\mathfrak{su}(2)$ part does not remain
massless in the effective theory below the Kaluza--Klein scale; 
we know this only from this local model. The vector field for the 
U(1) factor, however, may or may not remain massless, depending on 
more global aspects of compactification.}
\begin{equation}
 \mathfrak{su}(N+2)\mbox{-}{\bf adj.} \rightarrow 
 \mathfrak{su}(N)\mbox{-}{\bf adj.} + 
 \mathfrak{su}(2)\mbox{-}{\bf adj.} + 
 \mathfrak{u}(1) + 
 (\bar{\bf 2}, N ) + ({\bf 2}, \bar{N})
\end{equation}
is more appropriate, because $\tau_{N+1}$ and $\tau_{N+2}$ turn into 
one another in the monodromy around $u_1^2 - u_2 = 0$.
The two components $\Psi_+ \equiv \Psi_{N+1} \equiv 
(\tilde{\psi}_{\bar{1}+},\tilde{\psi}_{\bar{2}+},\tilde{\chi}_+)$
and $\Psi_- \equiv \Psi_{N+2} \equiv 
(\tilde{\psi}_{\bar{1}-},\tilde{\psi}_{\bar{2}-},\tilde{\chi}_-)$ 
that form a doublet solution turn into one another, 
when they are traced along a path going around the branch locus 
$u_1^2-u_2 = 0$.

Here, we start with a gauge theory whose gauge group at each
point isi $\SU(N+2)$, because (\ref{eq:AN+1--AN-1}) defines a deformed 
$A_{N+1}$ surface singularity for each point $(u_1,u_2)$ in $S$.
$N-1$ independent vanishing 2-cycles are buried at 
$(X, Y, Z) = (0,0,0)$, 
and there are two other independent topological cycles that we define now. 
For a given point $(u_1,u_2)$, let the two roots of 
\begin{equation}
Z^2 + s_1(u_1,u_2) Z +s_2(u_1,u_2) = 0 
\label{eq:theothercmp}
\end{equation}
be $Z = z_+(u_1,u_2)$ and $Z = z_-(u_1,u_2)$.
Then, the two independent non-vanishing cycles can be chosen as 
\begin{eqnarray}
 C_{\pm}:& (X,Y,Z) = (r(Z) i \cos \theta, r(Z) \sin \theta, Z) & 
  Z \in [0, z_{\pm}] \quad \theta \in [0, 2\pi], \\
 & r(Z) \equiv \sqrt{Z^N(Z-z_-)(Z-z_+)} &
\end{eqnarray}
As $(u_1,u_2)$ varies over complex surface $S$, the value of 
$z_{\pm} = - u_1 \pm \sqrt{u_1^2 - u_2}$ changes and  $C_+$ turns 
into $C_-$ and vice versa around the branch locus $u_1^2 -u_2 = 0$.
The fields $\Psi_+$ and $\Psi_-$ are associated with membranes wrapped 
on $C_+$ and $C_-$, respectively, and this is why $\Psi_+$ becomes 
$\Psi_-$ and vice versa in a monodromy around the branch cut. 
At the same time, the $\mathfrak{su}(2)$-Cartan part of the 
vev of $2\alpha \varphi_{12}$, 
\begin{equation}
 \diag \left(+ \sqrt{u_1^2 - u_2}, - \sqrt{u_1^2 - u_2} \right) 
\end{equation}
becomes $\times (-1)$ of its own around the branch locus 
$u_1^2- u_2 = 0$. Overall, we need to introduce a branch cut 
extending out from the branch locus $u_1^2 - u_2 = 0$, and 
the $\mathfrak{su}(N+2)$-{\bf adj.} fields are glued to themselves
after twisting by $\mathfrak{S}_2$ transformation---Weyl group 
of $\mathfrak{su}(2) \subset \mathfrak{su}(N+2)$ algebra---across 
the branch cut. This is not a simple theory of fields in 
the $\mathfrak{su}(N+2)$-{\bf adj.} representation.
Since the branch locus $u_1^2 - u_2 = 0$ passes right 
through the codimension-3 singularity point $(u_1, u_2) = (0,0)$, 
we cannot take a local description for this codimension-3 singularity 
point that is free of this branch locus and $\mathfrak{S}_2$ twist.

Such a field theory description with branch cut and twist 
along the branch cut is quite common in generic configuration 
of F-theory compactification. The non-split $I_n$ singularity locus 
corresponding to $\mathfrak{su}(n)$-{\bf adj} gauge theory 
with a twist by outer automorphism of $\mathfrak{su}(n)$ 
algebra is considered in \cite{6authors}. Reference \cite{6authors} also 
discusses the twist by an element of Weyl group. The
necessity of the twist by the Weyl 
group above is traced back to the fact that the deformation 
of singularity of type $\mathfrak{g}$ is parametrized by 
$\mathfrak{h} \otimes_\R \C/W$, not by 
$\mathfrak{h} \otimes_\R \C$. 
In order to obtain an effective theory with an unbroken 
$G'' = \SU(N)$ gauge symmetry, one only has to maintain a family of 
undeformed untwisted (that is, split) $A_{N-1}$ surface singularity 
over $S$. The deformation parameters (varying over $S$) are in 
$\mathfrak{h}' \otimes_\R \C/W'$, where $\mathfrak{h}'$ and $W'$ 
are the Cartan subalgebra and Weyl group of $G'$ whose commutant 
in $G$ becomes $G''$. Generically, we should expect twists by 
$W'$ in field theory local descriptions in F-theory compactification.\footnote{
Charged matters are localized in codimension-1 loci of $S$, because 
that is where one of eigenvalues of $\varphi$ vanishes. The branch loci 
is where two eigenvalues of $\varphi$ become degenerate, and they are 
also codimension-1 loci in $S$. In F-theory 
compactification down to 5+1 dimensions, $S$ is a complex curve, and 
generically the matter loci and branch loci are isolated points on $S$, 
and one can take a local region around a matter locus, so that branch 
loci are not contained in it. 
In compactification down to 3+1 dimensions, however, $S$ is a complex 
surface, and codimension-1 matter curves and branch curves generically 
intersect. This is why the branch locus and Weyl-reflection twists 
are inevitable in the field-theory description for compactifications  
down to 4-dimensions.}

Aside from the branch locus where more than one eigenvalues are the same, 
$2\alpha \vev{\varphi}_{12} = \tau_+$ and 
$2\alpha \vev{\varphi}_{12} = \tau_-$ act separately on each of the 
two weights of $\mathfrak{su}(2)+\mathfrak{u}(1)$ doublets. 
Thus, we still have a zero-mode equations (\ref{eq:EOM1}--\ref{eq:EOM4})
separately for $\Psi_+$ and $\Psi_-$. Although it is not easy to find 
an exact solution of these zero-mode equations for general form of 
$\tau_\pm$, there is a common structure that any solutions of 
(\ref{eq:EOM1}--\ref{eq:EOM4}) satisfy. To see this, let us note 
first that $\tilde{\psi}_{\bar{m}}$ are given locally by 
\begin{equation}
 \tilde{\psi}_{\bar{m}} = \bar{\partial}_{\bar{m}} \; \tilde{\psi} 
\end{equation}
for some function $\tilde{\psi}$, because of (\ref{eq:EOM2}). 
Equations (\ref{eq:EOM3}, \ref{eq:EOM4}) now imply that 
\begin{equation}
 \tilde{\chi} = \tau \tilde{\psi} + f(u_1,u_2)
\end{equation}
for some (locally) holomorphic function $f$. 
Now (\ref{eq:EOM1}) is the only constraint we still have:
\begin{equation}
 \left(
 \partial_1 \bar{\partial}_{\bar{1}} + \partial_2 \bar{\partial}_{\bar{2}} 
 \right) \tilde{\psi} - \bar{\tau} \tau \tilde{\psi} - \bar{\tau} f = 0.
\end{equation}
For example, when $\tau = u_1$, 
\begin{equation}
 \tilde{\psi} = \frac{c(u_2)}{u_1}\left( \exp [ \pm |u_1|^2] - 1 \right), 
\qquad f(u_1,u_2)=c(u_2),
\end{equation}
and the Gaussian solution in (\ref{eq:Gaussian1}) corresponds to the one with 
$-$ sign here.

One can see that 
$\sum_{m=1}^2 \partial_m \bar{\partial}_{\bar{m}} - \tau\bar{\tau}$ 
is positive definite, does not have a zero mode, and hence its inverse exists. 
Thus, 
\begin{eqnarray}
 \tilde{\psi} & = & 
   \frac{1}{\partial_m \bar{\partial}_{\bar{m}} - \tau \bar{\tau}} 
  \bar{\tau} f, \label{eq:expr4psitot}\\
 \tilde{\chi} & = & \tau \; 
    \frac{1}{\partial_m \bar{\partial}_{\bar{m}}-\tau\bar{\tau}}
    \partial_n \bar{\partial}_{\bar{n}} \frac{1}{\tau} f.
\label{eq:expr4chi}
\end{eqnarray}
$\tilde{\psi}_{\bar{1},\bar{2}}$ are also obtained from $\tilde{\psi}$.
Thus, for any holomorphic functions $f(u_1,u_2)$, a solution exists 
for the zero-mode equation. However, the zero-mode solutions are not 
in one-to-one correspondence with the local holomorphic functions 
$f(u_1,u_2)$. 
$\partial_n \bar{\partial}_{\bar{n}} (f/\tau)$ vanishes 
wherever $(f/\tau)$ is holomorphic, and hence the behavior of $f$ 
at $\tau \neq 0$ is irrelevant to the zero-mode wavefunctions. 
Therefore, $\tilde{\chi}$ depends only on $f$ on the $\tau=0$ locus, 
and so do $\tilde{\psi}_{\bar{1},\bar{2}}$, 
because they are related to $\tilde{\chi}$ through 
(\ref{eq:EOM3}, \ref{eq:EOM4}).
Since $\tau=0$ locus is the matter curve, zero-mode solutions are
in one-to-one correspondence with local holomorphic functions 
(sections of a line bundle in general) on the matter curve.

Monodromy around the branch locus $u_1^2 - u_2 = 0$, or equivalently 
the gluing after $\mathfrak{S}_2$-twist, can be implemented in this 
set of partial differential equations. The following description 
leads to a key observation later on in this article. 
An idea is to replace the local coordinate system from $(u_1,u_2)$ 
to
\begin{equation}
  (u,x) \equiv \left( u_1, \sqrt{u_1^2 - u_2} \right).
\label{eq:coord-x}
\end{equation}
This is not a one-to-one map. 
\begin{equation}
 \pi_C: (u,x) \mapsto \pi_C((u,x)) \equiv (u_1,u_2) = (u, u^2 - x^2),
\end{equation}
because $(u,x)$ and $(u,-x)$ are mapped to the same point in $S$.
Let us introduce a new surface $C$, where local coordinates are 
$(u,x)$.
A loop around the branch locus $u_1^2 - u_2 = 0$ in $S$ 
is lifted to a path from $(u,x)$ to $(u,-x)$ in $C$. 
The new space $C$ 
is a 2-fold covering space of $S$. 
Now, we define $\Psi$ at $(u,x)$ as $\Psi_-(\pi_C((u,x)))$, 
$\Psi$ at $(u,-x)$ as $\Psi_+(\pi_C((u,-x)))$. 
$\tau(u,x) \equiv - u - x$ is $\tau_-(\pi_C((u,x)))$, and 
$\tau(u,-x) = -u+x$ becomes $\tau_+(\pi_C((u,-x)))$. 
The expressions (\ref{eq:expr4psitot}, \ref{eq:expr4chi}) are regarded as 
those on the covering space as well, with $\tau = -u - x$ and $f(u,x)$ now 
being single-valued functions on the covering space.
We can also use $(u,\tau)=(u,-u-x)$ as the local coordinates 
of this covering space. What really matters to the zero-mode 
wavefunctions is the residue of $f/\tau$ along the pole locus at $\tau=0$, 
or equivalently $f(u,\tau)$ modulo any functions generated by $\tau$.
 
\begin{figure}[t]
\begin{center}
    \includegraphics[width=.3\linewidth]{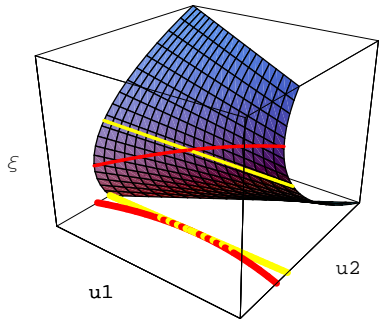} 
\caption{\label{fig:C-An}(color online) 
Covering surface of generic deformation 
of $A_{N+1} \rightarrow A_{N-1}$, which is identified with 
spectral surface of a $K_S$-valued rank-2 Higgs bundle 
in section \ref{sec:Higgs}.
This surface is given by $\xi^2 + 2 u_1 \xi + u_2 = 0$. 
Along the thin yellow curve on the surface, $\xi$ becomes zero. 
The field theory is formulated on a plane $S$ whose local coordinates 
are $(u_1, u_2)$, and the projection of the $\xi = 0$ curve
to the $(u_1,u_2)$ plane, thick yellow line in the figure, 
is the matter curve $u_2 = 0$ of matter multiplets in the $N$ and
$\bar{N}$ representations of unbroken $\SU(N)$ symmetry. 
Thin red curve on the surface $C$ is the ramification locus of 
$\pi_C: C \rightarrow S$, and its projection to the $S$ plane 
is the branch locus $u_1^2 - u_2 = 0$, denoted by a thick red 
curve in the figure.
}
\end{center}
\end{figure}

The covering surface $C$ is regarded as 
a space swept by all possible values of 
$\tau(u_1,u_2)$---eigenvalues of $2\alpha \vev{\varphi_{12}}$---for 
$(u_1, u_2) \in S$, because $\tau_+(u_1,u_2) \neq \tau_-(u_1,u_2)$ 
for a given point $(u_1, u_2) \in S$ are resolved in
$C$.
It is identified with a subspace in a space with 
coordinates $(u_1,u_2,\tau)$, given by 
\begin{equation}
 \tau^2 + 2 u_1 \tau + u_2 = 0.
\label{eq:C2-An}
\end{equation}
This defining equation of $C$ is the same 
as that of the $Z^2 + s_1 Z + s_2 = 0$ irreducible component 
of the discriminant locus, but they are different objects. 
The irreducible component of the discriminant is a divisor 
in $B_3$. On the other hand, $\varphi_{12}$ (and its eigenvalues)
should transform as sections of canonical bundle of $S$, $K_S$, 
and hence the space with coordinates $(u_1, u_2, \tau)$ is 
the total space of $K_S$, denoted by $\mathbb{K}_S$. 
Thus, the covering surface $C$ is regarded as a divisor 
of $\mathbb{K}_S$.
In the generic deformation of $A_{N+1}$ singularity 
to $A_{N-1}$, 
\begin{equation}
 \pi_C: C \rightarrow S, \qquad 
  (u, \tau) \mapsto (u_1, u_2) = (u, - (2u+\tau)\tau)
\end{equation}
is a degree-2 cover of $S$. $\tau+u=0$ is the ramification locus 
in $C$, which corresponds to the branch locus $u_1^2 - u_2 = 0$ in $S$.
The matter curve is $\tau=0$ on $C$, which is mapped to $u_2 = 0$ 
curve in $S$. See Figure~\ref{fig:C-An}.
The function $f$ on the covering space $C$ transforms as $\varphi_{12}$. 
Thus, $f$ is not just a function on the covering space $C$, but 
it is a section of a bundle containing\footnote{Since we 
have ignored gauge field background in this section, we cannot 
determine the other factor in the bundle associated with the gauge 
field background here. We will come back to this issue in 
section \ref{sec:Higgs}.} a factor
$\otimes \pi_C^* K_S$.

Let us now return to a particular case with the doublet 
background (\ref{eq:bg-AB}), before closing this section. 
Although we have not succeeded in solving this set of
partial differential equations analytically, it is still 
possible to discuss the asymptotic behavior of the zero-mode 
solution.  

In the large $|u_1|$ limit, $\tau_- \sim - 2u_1$ and 
\begin{equation}
 \tau_+  \sim - \frac{u_2}{2 u_1} - \frac{u_2^2}{8u_1^3} + \cdots.
\end{equation}
$|\tau_-|$ is large even along the matter curve, where $|u_2|$ is 
small, but $\tau_+$ remains small. The zero-mode wavefunction 
will have larger value in the $\Psi_+ = \Psi_{N+1}$ component, 
and it may decay as in Gaussian profile, $e^{-|u_2|^2}$, because 
$\tau_+$ is linear in the normal coordinate of the matter curve, $u_2$.

This Gaussian profile of zero modes across the matter curve 
$u_2 = 0$ will no longer be a good approximation near the 
codimension-3 singularity point $(u_1, u_2) = (0,0)$; 
the first term of the expansion of $\tau_+$ above simply does 
not give a good approximation to $\tau_+$. We can still study 
behavior of the zero mode wavefunction in a region with large 
$|u_2|$ and sufficiently small $|u_1|$. All the details are 
found in the appendix \ref{sec:0-mode}. Quoting 
(\ref{eq:chi-asymptotic}) here, 
\begin{equation}
  \tilde{\chi}_\pm \propto 
|u_2|^{- \frac{1}{4}} \exp\left(- \frac{4}{3} |u_2|^{\frac{3}{2}} \right).
\label{eq:chi-asymp-text}
\end{equation}
This solution is still localized along the matter curve, because 
it decays as $e^{-|u_2|^{3/2}}$ for large $|u_2|$. This wavefunction 
does not fall off as quickly as Gaussian $e^{-|u_2|^2}$. Such a 
profile of wavefunction has never been discussed in the literature, 
and this is interesting on its own.\footnote{Only the leading order 
(linear) terms were kept in (\ref{eq:typeA-caseB}), and one should 
be aware that this $e^{-|u_2|^{3/2}}$ profile follows only in a region 
where (\ref{eq:typeA-caseB}) is a good approximation.} It should 
also be noted that both $\tilde{\chi}_+$ and $\tilde{\chi}_-$
components have equally large value in this region, although 
only $\Psi_+$ is large and $\Psi_-$ is small in the large $|u_1|$
and small $|u_2|$ region.

The zero-mode wavefunction in this doublet background (\ref{eq:bg-AB})
has small $\tilde{\psi}_{\bar{1}\pm}$ components in this region, but 
$\tilde{\psi}_{\bar{2}\pm}$ components are comparable to 
$\tilde{\chi}_{\pm}$. We have\footnote{Here, we talk about the zero 
mode wavefunction in the ${\bf 2}$ representation of the background 
(\ref{eq:bg-AB}), and hence that of $({\bf 2}, \bar{N})$ irreducible 
component. The zero mode wavefunction of the $(\bar{\bf 2}, N)$
component is obtained by multiplying $(-1)$ to the $\psi_{\bar{m}\pm}$
components.}
\begin{equation}
 i \psi_{\pm} = i d\bar{u}_{\bar{2}} \psi_{\bar{2}\pm} \sim 
 \frac{d \bar{u}_{\bar{2}}}{\pm \sqrt{- \bar{u}_2}}|u_2|^{\frac{1}{4}} 
  \exp \left(- \frac{4}{3}|u_2|^{\frac{3}{2}} \right).
\label{eq:psi-asymp-text}
\end{equation}
This two-component (0,1)-form on $S$ $\psi_{\pm}$ can be expressed 
as a single component (0,1)-form on $C$. In the region with large 
$|u_2|$ and small $|u_1|$, $x \sim \pm \sqrt{- u_2}$, and 
\begin{equation}
 i \psi \sim -2 d \bar{x} \; |x|^{\frac{1}{2}} 
  \exp \left(- \frac{4}{3}|x|^3 \right).
\end{equation}
The monodromy around the branch locus $u_2 \sim 0$ in the zero-mode 
wavefunction can be traced by looking at 
$u_2 \rightarrow e^{2 \pi i} \times u_2$; $\psi_+$ becomes $\psi_-$
and vice versa in this process, just as we anticipated before. 
The same process corresponds to going from $x$ to $-x$ in 
the (0,1) form solution $\psi$ on the covering space $C$.

\section{Generic Deformation of Singularity for GUT Models}
\label{sec:GUT}

We have studied the field theory local model of generic deformation
of $A_{N+1}$ singularity down to $A_{N-1}$. 
Such a generic configuration around a point of codimension-3 
singularity in F-theory is much more complicated than just having 
intersecting D7-branes, but we now know how to deal with them. 

One of the most important advantages of using F-theory as 
opposed to Type IIB string theory is that all the Yukawa 
couplings can be generated even in SU(5) GUT gauge group. 
The same technique can be applied to F-theory compactification 
used for GUT model building, and this is the subject in this 
section. 

In order to preserve an unbroken SU(5) gauge symmetry, 
an elliptic fibered compact manifold should be given by 
an equation of the form \cite{6authors}\footnote{\label{fn:Tate}
When the elliptic
fibration is given by an equation 
\begin{equation*}
y^2 + A_1 x y + A_3 y  =  x^3 + A_2 x^2 + A_4 x + A_6, 
\end{equation*}
where $A_{1,3,2,4,6}$ are sections of certain line bundles on $B_3$, 
we need $A_i(u,z) = A_{i,(i-1)}(u) z^{i-1} + {\cal O}(z^i)$ for 
$i=1,2,3,4,6$, in order to have a locus of $A_4$ split 
singularity at $z=0$ \cite{6authors}; 
here $u$ denotes local coordinates on $S$ collectively, and $z$ the coordinate 
of $B_3$ normal to $S$. Local functions $A_{i,(i-1)}(u)$ here correspond 
to $(-1)^i a_{6-i}(u)$ in (\ref{eq:defeq}), and $f_0$ and $g_0$ can 
be regarded as higher-order terms in the expansion in $z$.
There is no rationale to maintain only $f_0$ and $g_0$ and drop all
other higher order terms purely in the context of F-theory, 
although it will be clear how to modify the rest of this article 
when such higher order terms are included in (\ref{eq:defeq}). 
Here, we adopted the form (\ref{eq:defeq}), because 
the duality map between F-theory and Heterotic string theory 
\cite{KMV-BM, CD, DW-1, HayashiEtAl} is to identify precisely 
$a_{6-i}(u) = (-1)^i A_{i,(i-1)}(u)$ with $a_{6-i}(u)$ 
in (\ref{eq:spectral-surface}), and $(f_0,g_0)$ in (\ref{eq:defeq}) 
with those in (\ref{eq:Weierstrass-Het}). 
We will discuss Heterotic--F theory duality in section \ref{sec:Higgs}.
}
\begin{eqnarray}
 y^2 & = & x^3 + a_5 y x + a_4 z x^2 + a_3 z^2 y + (a_2 z^3 + f_0 z^4) x 
  + (a_0 z^5 + g_0 z^6), \label{eq:defeq} \\
  & = & (x^3 + f_0 z^4 x + g_0 z^6) + 
  (a_5 x y + a_4 z x^2 + a_3 z^2 y + a_2 z^3 x + a_0 z^5), \nonumber 
\end{eqnarray}
where $(x,y)$ are coordinates for the elliptic fiber of 
the elliptic fibration in (\ref{eq:Weierstrass-F}), and 
$z$ is the coordinate normal to the discriminant locus $S$
of the GUT gauge group. $a_{0,2,3,4,5}$ are functions\footnote{
They are sections of appropriate line bundles on $S$, but we 
do not pay much attention to global issues here.} of 
local coordinates of $S$. Terms higher order in $z$ are 
omitted from this equation. Local geometry with $D_5$ singularity 
for SO(10) GUT models can be obtained by simply setting $a_5 = 0$.
For the purpose of studying geometry for SU(5) or SO(10) GUT 
models, therefore, it is very convenient to start from 
the defining equation (\ref{eq:defeq}).

\subsection{Codimension-3 Singularities in $\SU(5)$ GUT Models}
\label{ssec:SU(5)}

Let us begin with SU(5) GUT models. For this purpose, 
we do not need to make further assumptions on the sections 
$a_{0,2,3,4,5}$. 
The discriminant of this elliptic fibration is given by 
\begin{eqnarray}
 \Delta & \propto & 
z^5 \left( \frac{1}{16}a_5^4 P^{(5)} 
   + \frac{z}{16} a_5^2 
    \left(12 a_4 P^{(5)} - a_5^2 R^{(5)}\right)  \right.  \nonumber \\
   & & \qquad \qquad   \left. 
     + z^2 \left(a_3^2 a_4^3 + {\cal O}(a_5)\right) 
     + z^3 \left(\frac{27}{16}a_3^4 + {\cal O}(a_5)\right) 
     + {\cal O}(z^4)
\right). 
\label{eq:Det-E4}
\end{eqnarray}
$z=0$ is the locus of $\SU(5)$ GUT gauge fields (codimension-1
singularity in a base 3-fold), and there are two matter curves 
(codimension-2 singularities) specified by $a_5 = 0$ and 
$P^{(5)} = 0$. $a_5 = 0$ is where matter multiplets in the 
$\SU(5)$-${\bf 10}+\overline{\bf 10}$ representations are localized, 
while matter in the $\SU(5)$-${\bf 5}+\bar{\bf 5}$ representations 
are at the curve $P^{(5)} = 0$, with
\begin{equation}
 P^{(5)} \equiv a_0 a_5^2 - a_2 a_5 a_3 + a_4 a_3^2. 
\end{equation}
The $a_5 = 0$ curve is where the singularity in $(x,y,z)$-surface 
is enhanced from $A_4$ to $D_5$, while the enhanced singularity 
is $A_5$ along the curve $P^{(5)} = 0$.
The discriminant becomes $\Delta \propto z^7$ (as in $D_5$ singularity) 
when $a_5 = 0$, whereas $\Delta \propto z^6$ for $P^{(5)} = 0$ 
(as in $A_5$ singularity).

There are some isolated codimension-3 singularities along the matter 
curves. 
\begin{figure}[t]
 \begin{center}
  \includegraphics[width=.3\linewidth]{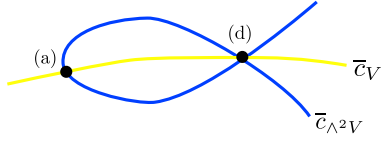}  
 \caption{\label{fig:curves-SU(5)} (color online) 
Structure of matter curves and codimension-3 singularities generically 
expected in F-theory compactification that has a locus of split $A_4$ 
singularity (SU(5) GUT models, in short). 
A blue curve $\bar{c}_V$ corresponds to $a_5 = 0$ curve, 
and a yellow curve $\bar{c}_{\wedge^2 V}$ to $P^{(5)} = 0$. There are 
two different kinds of intersection of these two curves, namely 
type (a) and type (d). The type (c1) codimension-3 singularity points 
are on the $\bar{c}_{\wedge^2 V}$ ($P^{(5)} = 0$) but not on 
$\bar{c}_V$ ($a_5 = 0$). This figure was recycled from
  \cite{HayashiEtAl} for readers' convenience.} 
 \end{center}
\end{figure}
On the $a_5 = 0$ curve, they are at \cite{AC, HayashiEtAl}\footnote{
In traditional literatures on F-theory like \cite{6authors, AC}, 
a section of a line bundle $h$ in such literature roughly corresponds 
to $a_5$, $H$ to $a_4$ and $q$ to $a_3$ in F-theory compactification 
that preserves an $A_4$ singularity locus. See \cite{HayashiEtAl} 
for the precise relation between them. In the traditional notation,
the type (a) singularity corresponds to the $h \cap H$, and type (d) 
to the $h \cap q$ loci. } 
\begin{itemize}
 \item type (a): common zero of $a_5$ and $a_4$,
 \item type (d): common zero of $a_5$ and $a_3$.
\end{itemize}
This is where the coefficient of the $z^7$ term in the discriminant 
vanishes.

On the $P^{(5)} = 0$ curve they are at:
\begin{itemize}
 \item type (c1): common zero of $P^{(5)}$ and $R^{(5)}$ but $a_5 \neq
       0$ \cite{HayashiEtAl}, with 
\end{itemize}
\begin{equation}
  R^{(5)} \equiv \left(a_2 - 2 \frac{a_3}{a_5}a_4 \right)^2 - a_5^2
  \left(\left(\frac{a_3}{a_5}\right)^3 + 
        f_0 \left(\frac{a_3}{a_5}\right) - g_0 \right)  
\end{equation}
for a local defining equation (\ref{eq:defeq}).
In what follows, we will see that the deformations of $E_6$, $D_6$ and 
$A_6$ singularity, respectively, to $A_4$ are good approximation 
of local geometry of each of the three types of codimension-3 
singularities above. Field theory local model with $E_6$, $\SO(12)$ and 
$\SU(7)$ gauge groups, respectively, can be used to analyze 
physics localized at these types of codimension-3 singularities.
All the three types of rank-2 deformation of $A$--$D$--$E$ singularity
to $A_4$ appear in F-theory compactification for SU(5) GUT models.
Using the local models, we will analyze the behavior of matter 
zero-mode wavefunctions around those codimension-3 singularities. 

\subsubsection{Type (a) Singularity: $E_6 \rightarrow A_4$}
\label{sssec:SU(5)-E6}

Let us begin with the type (a) codimension-3 singularities. 
The type (a) codimension-3 singularity 
is where Yukawa couplings of the form  
\begin{equation}
 \Delta W = {\bf 10}^{ab} {\bf 10}^{cd} {\bf 5}^e \epsilon_{abcde}
\label{eq:10105}
\end{equation}
is believed to be generated. 
This type of interaction becomes up-type Yukawa couplings 
in Georgi--Glashow $\SU(5)$ GUT's, and down-type Yukawa couplings and 
mass terms of colored Higgs multiplets in flipped SU(5) GUT's.
Thus, this type of local geometry of F-theory compactification 
is of a particular phenomenological interest.

The coefficient functions $a_r$ in (\ref{eq:defeq}) corresponds 
to those of defining equations of spectral surface of a vector 
bundle in Heterotic compactification. Thus, $a_5 = 0$ and $a_4 = 0$
at the type (a) codimension-3 singularities means that the 
structure group of vector bundle is reduced to $\SU(3)$ in 
Heterotic language, and the singularity is enhanced to $E_6$, 
the commutant of $\SU(3)$ in $E_8$. Even without a knowledge 
of Heterotic--F theory duality, the criterion in \cite{6authors} 
tell us that the singularity is enhanced to $E_6$, because 
${\rm ord} A_1 = 1$ when $a_5 = 0$ and ${\rm ord} A_2 = 2$ 
when $a_4 = 0$. Therefore, it is an obvious 
guess to use $E_6$ gauge theory for the field theory local model 
of the type (a) codimension-3 singularities, although not 
all F-theory vacua have Heterotic dual.

The most generic deformation of $E_6$ singularity can be given 
by a local equation 
\begin{equation}
 Y^2 = X^3 
 + X (\epsilon_2 Z^2 + \epsilon_5 Z + \epsilon_8)
 + \left(\frac{Z^4}{4} + \epsilon_6 Z^2 
      + \epsilon_9 Z + \epsilon_{12} \right).
\label{eq:E6-def-A}
\end{equation}
When all of $\epsilon_{2,5,8}$ and $\epsilon_{6,9,12}$ are set 
to zero, this equation defines a complex surface in a $(X,Y,Z)$ 
space with $E_6$ singularity. The coordinates $(X,Y)$ correspond 
to the direction of elliptic fiber, and $Z$ to a direction 
normal to $S$. The six deformation parameters $\epsilon_{2,5,8}$ 
and $\epsilon_{6,9,12}$ in (\ref{eq:E6-def-A}) are functions 
of local coordinates $u_m$ ($m=1,2$) on $S$. Since the rank of $E_6$ is six, 
the generically chosen $\epsilon_{2,5,8,6,9,12}$ will completely 
resolve the $E_6$ singularity.

Reference \cite{KM} shows that the most generic deformation 
of A-D-E singularity is parametrized by $\mathfrak{h} \otimes_\R \C/W$.
In the case of deformation of $E_6$ singularity, we can choose 
six complex numbers $t_i$ ($i=1,\cdots,6$) by choosing a basis 
in the Cartan subalgebra of $\mathfrak{e}_6$. The deformation 
parameters $\epsilon_{r=2,5,8,6,9,12}$ in (\ref{eq:E6-def-A}) 
are homogeneous functions of $t_i$'s of degree $r$ whose explicit 
forms are given in \cite{KM}. When we set the homogeneous degree 
of the three coordinates $(X,Y,Z)$ to be 4, 6 and 3, respectively, 
and the deformation parameters $t_i$'s as 1, the equation of 
$E_6$ singularity (and its deformation) is a homogeneous function 
of degree 12, the dual Coxeter number of $E_6$.
When a local geometry of F-theory is a family of surface with 
deformed $E_6$ singularity over a base space $S$, i.e. a local 
equation of geometry is approximately (\ref{eq:E6-def-A}) with 
some functions $\epsilon_{2,5,8,6,9,12}$ on $S$, 
the field theory description of \cite{KV, BHV-1} is to employ 
$E_6$ gauge theory with non-vanishing vev in the $\mathfrak{e}_6$-valued
complex scalar (or holomorphic 2-form) field given by corresponding 
$t_i$'s (so, actually the vev is in the Cartan subalgebra).

We are, however, interested in deformation of $E_6$ singularity 
to $A_4$, and want to preserve SU(5) unbroken symmetry. 
Thus, the deformation of interest is parametrized by two complex 
numbers for a given point on $S$.
In the local equation of geometry (\ref{eq:defeq}), $a_4$ and $a_5$ 
correspond to the two deformation parameters. To see only the local 
geometry around a family of deformed $E_6$ surface singularity, 
we consider the following scaling:
\begin{equation}
 (x,y,z) = (\lambda^4 x_0, \lambda^6 y_0, \lambda^3 z_0) , \qquad 
 \lambda \rightarrow 0,
\end{equation}  
and we zoom in to the singular locus as $\lambda \rightarrow 0$.
The deformation parameters $a_4$ and $a_5$ should also scale to zero 
toward a point of type (a) codimension-3 singularity, in order to
restore $E_6$ singularity in the fiber surface. 
\begin{equation}
 (a_4, a_5) = (\lambda a_{4,0}, \lambda^2 a_{5,0}), \qquad 
\lambda \rightarrow 0
\label{eq:scale-E6-a-para}
\end{equation}
turns out to be the appropriate scaling, as we will see. 
All other coefficients in (\ref{eq:defeq}), namely $a_{3,2,0}$, $f_0$
and $g_0$, remain finite and do not scale as $\lambda \rightarrow 0$. 
All the terms involving $a_{2,0}$, $f_0$ and $g_0$ scale as 
$\propto \lambda^{13}$, $\lambda^{15}$, $\lambda^{16}$ and
$\lambda^{18}$, and become relatively small compared with all other 
terms scaling as $\lambda^{12}$. Thus, those terms are irrelevant 
to the geometry close to the codimension-3 singularity. 
We therefore drop those terms in the following argument.

After a change in the coordinates from $(x,y,z)$ to 
\begin{equation}
 \tilde{z} = \frac{z}{a_3}, \qquad
 \tilde{y} = \frac{1}{a_3^3}\left(y - \frac{1}{2}(a_5 x + a_3 z^2)
			    \right), \qquad 
 \tilde{x} = \frac{1}{a_3^2} \left(x + \frac{1}{3}
  \left(\left(\frac{a_5}{2}\right)^2 + a_4 z \right) 
\right), 
\end{equation}
the local defining equation (\ref{eq:defeq}) becomes 
\begin{eqnarray}
 \tilde{y}^2 & \simeq & \tilde{x}^3 + 
 \tilde{x}\left[ 
 \left(- \frac{\tilde{a}_4^2}{3}+ \frac{\tilde{a}_5}{2} \right) \tilde{z}^2
 - \frac{2}{3} \left(\frac{\tilde{a}_5}{2}\right)^2 \tilde{a}_4 \tilde{z}
 - \frac{1}{3} \left(\frac{\tilde{a}_5}{2}\right)^4
          \right]  \label{eq:E6defB}\\
 & + & 
 \left[ \frac{1}{4} \tilde{z}^4 + 
 \left(\frac{2}{27}\tilde{a}_4^3 - \frac{1}{6}\tilde{a}_4 \tilde{a}_5 \right) 
 \tilde{z}^3
+ \frac{1}{3} \left(\frac{\tilde{a}_5}{2}\right)^2 
  \left( \frac{2}{3} \tilde{a}_4^2 - \frac{\tilde{a}_5}{2} \right) 
  \tilde{z}^2
+ \frac{2}{9} \left(\frac{\tilde{a}_5}{2}\right)^4 \tilde{a}_4 \tilde{z}
+ \frac{2}{27} \left(\frac{\tilde{a}_5}{2}\right)^6 
\right]. \nonumber 
\end{eqnarray}
Here, $\tilde{a}_4 \equiv a_4 / a_3$ and $\tilde{a}_5 \equiv a_5/ a_3$.
Irrelevant terms have been dropped. This equation is ready to be 
identified with (\ref{eq:E6-def-A}).

In order to determine $t_i$'s and hence the vevs of $\varphi$ 
in terms of $\tilde{a}_4$ and $\tilde{a}_5$, 
we need a little more preparation. Not all six degrees of freedom 
of $t_i$'s are necessary, because the geometry (\ref{eq:defeq}) 
preserves $\SU(5) \subset E_6$ symmetry. Some two dimensional subspace 
for unbroken $\SU(5)$ symmetry has to be identified.
To this end, we start off with a review of a (rather common) 
convention of choice of basis in describing the Cartan subalgebra 
of $E_6$. Consider a 7-dimensional vector space spanned by 
a basis vectors $e_{0,\cdots,6}$, and introduce a bilinear 
form $\diag(+1, -1,\cdots,-1)$ in this basis. 
A special element $k$ of this vector space is defined 
by $k = -3 e_0 + (e_1 + \cdots + e_6)$. The orthogonal 
complement of $k$, which is a 6-dimensional space, is 
identified with the dual space of the Cartan subalgebra 
of $\mathfrak{e}_6$. Six independent vectors 
\begin{equation}
 v_i = e_i - e_{i+1} \quad (i=1,\cdots,5), \qquad 
 v_0 = e_0 - (e_1 + e_2 + e_3)
\end{equation}
can be chosen as the simple roots. The maximal root 
becomes $\theta = 2e_0 - (e_1 + \cdots + e_6)$, 
and the intersection form of 
$v_{0,\cdots,5}$ and $-\theta$ is described by 
the extended Dynkin diagram of $E_6$.

The Cartan subalgebra of $\mathfrak{e}_6$ is a dual space of 
the six dimensional space given above. Using the bilinear form, 
however, the Cartan subalgebra can be identified with 
the same six-dimensional vector space. Thus, an arbitrary 
element in the Cartan subalgebra can be written as 
\begin{equation}
 H = \xi_0 e_0 + \sum_{i=1}^6 \xi_i e_i, 
\label{eq:dP-parametrizationA}
\end{equation}
where the coefficients satisfy 
$3\xi_0 + (\xi_1 + \cdots + \xi_6) = 0$. 
It is conventional (at least in \cite{KM}) that 
the independent six numbers $t_i$ ($i=1,\cdots,6$) 
are chosen as 
\begin{equation}
 t_i = \frac{1}{3} \xi_0 + \xi_i \qquad ({\rm for~} i=1,\cdots,6).
\label{eq:dP-parametrizationB}
\end{equation}

When the $\mathfrak{e}_6$ Cartan subalgebra 
is parametrized by $t_i$, functions $s_{1,2,3,4,5,6}$ 
symmetric under any permutations of $t_i$'s are defined by 
\begin{equation}
 s_j = \sum_{i_1 < i_2 < \cdots < i_j} t_{i_1} t_{i_2} \cdots t_{i_j}.
\label{eq:def-s}
\end{equation}
The permutation group of six elements $\mathfrak{S}_6$ is the Weyl 
group of $\mathfrak{su}(6)$ generated by $v_{1,\cdots,5}$. 
Functions of $t_i$'s symmetric under the full Weyl group of 
$\mathfrak{e}_6$---$\epsilon_{2,5,8,6,9,12}$---are constructed 
from the $\mathfrak{S}_6$-invariant functions $s_{1,2,3,4,5,6}$; 
explicit form of $\epsilon_{2,5,8,6,9,12}$ are found 
in \cite{KM}. The deformation parameters $\epsilon$'s  
in (\ref{eq:E6-def-A}) are related to the Cartan parameters 
$t_i$'s in this way. 

An $\mathfrak{so}(10)$ subalgebra can be generated by 
simple roots $v_{1,2,3,4}$ and $v_0$. Since 
\begin{equation}
 H_\sigma \sigma =(-3 e_0 + (e_1 + \cdots + e_5) + 4 e_6) \sigma
\label{eq:H4SO(10)-A}
\end{equation}
satisfies $\vev{H_\sigma,v_{1,2,3,4}} = 0$ and $\vev{H_\sigma,v_0}=0$, 
vev of $\varphi$ can be introduced in this rank-1 subspace of 
the Cartan subalgebra of $\mathfrak{e}_6$ in order to deform 
the $E_6$ singularity so that $D_5$ singularity remains. 
This choice corresponds to 
\begin{equation}
 (t_1,\cdots,t_6) = (\overbrace{0,\cdots,0}^5,3\sigma).
\end{equation}
One can also see that 
\begin{equation}
 H = (-3e_0 + 2(e_1 + \cdots + e_5) - e_6) \tau
\label{eq:H4SO(10)-B}
\end{equation}
satisfies $\vev{H,v_{1,2,3,4}} = 0$ and $\vev{H,v_0+v_5}=0$. 
Roots $v_{1,2,3,4}$ and $v_0+v_5$ can also generate another 
$\mathfrak{so}(10)$ subalgebra of $\mathfrak{e}_6$. 
This choice corresponds to 
\begin{equation}
 (t_1,\cdots,t_6) = (\overbrace{\tau,\cdots,\tau}^5,-2\tau).
\end{equation}
Thus, when the vev of $\varphi$ is in the direction parametrized 
by $\tau$ as above, then the singularity of $E_6$ is deformed 
and $D_5$ singularity remains.

An $\mathfrak{su}(6)$ subalgebra of $\mathfrak{e}_6$ can be 
generated by $v_{1,\cdots,5}$. Thus, 
\begin{equation}
 H_\tau \tau = (-6 e_0 + 3(e_1 + \cdots + e_6)) \tau
\label{eq:H4SU(6)}
\end{equation}
satisfies $\vev{H_\tau,v_{1,\cdots,5}} = 0$, and the vev of $\varphi$
should be introduced in this rank-1 subspace of the Cartan 
of $\mathfrak{e}_6$, if $A_5$ singularity is to remain.
This choice corresponds to\footnote{Weyl reflection in the 
direction of $v_0$ takes $H_\tau \tau$ in (\ref{eq:H4SU(6)}) 
to $(-e_0 + (e_4 + e_5 + e_6))3\tau$, and 
$(t_1,\cdots,t_6)$ to 
$(-\tau,-\tau,-\tau,2\tau,2\tau,2\tau)$. $v_{1,2,4,5}$ and $v_0 + v_3$
are the simple roots of $\mathfrak{su}(6)$, then.} 
\begin{equation}
 (t_1 ,\cdots,t_6) = (\tau, \cdots, \tau).
\end{equation}

The three generators (\ref{eq:H4SO(10)-A}, \ref{eq:H4SO(10)-B}, 
\ref{eq:H4SU(6)}) of the Cartan subalgebra of $\mathfrak{e}_6$
span a two dimensional subspace. Two generators 
(\ref{eq:H4SO(10)-A}) and (\ref{eq:H4SU(6)}) can be chosen 
as independent, 
\begin{equation}
 H = H_\sigma \sigma + H_\tau \tau, 
\label{eq:2dim-sub-E6}
\end{equation}
and (\ref{eq:H4SO(10)-B}) can be obtained 
as a linear combination of both with $\sigma = - \tau$.
All the roots $v_{1,2,3,4}$ are neutral under the two-dimensional 
subspace of the Cartan, and vev of $\varphi$ in this subspace 
does not break the SU(5) symmetry (or deform the $A_4$ singularity)
generated by $v_{1,2,3,4}$. The $(\sigma, \tau)$-parametrization 
of the two-dimensional subspace corresponds to 
\begin{equation}
 (t_1,\cdots, t_6) = (\overbrace{\tau,\cdots,\tau}^5, \tau+3\sigma).
\label{eq:2para-E6}
\end{equation}

This parametrization is plugged into the expressions in the appendix 
of \cite{KM} to obtain $\epsilon_{2,5,8,6,9,12}$. 
After a shift\footnote{This shift does not change the homogeneous
nature.} of coordinate $Z$ \cite{KV, BHV-1}, 
\begin{equation}
 Z' \equiv Z - \frac{1}{2} \sigma (9 \tau(\tau+\sigma) + 4 \sigma^2),
\end{equation}
the equation (\ref{eq:E6-def-A}) becomes 
\begin{eqnarray}
 Y^2 & = & X^3  \nonumber \\
 & + & X \left[
 \left(- \frac{(3\sigma)^2}{3} - \frac{9\tau(\tau+\sigma)}{2}\right)Z^{'2}
-\frac{2}{3}\left(\frac{9\tau(\tau+\sigma)}{2}\right)^2(3\sigma) Z'
- \frac{1}{3}\left(\frac{9\tau(\tau+\sigma)}{2}\right)^4
\right] \nonumber \\
 & + & \left[ \frac{1}{4} Z^{'4} 
 +\left(\frac{2}{27}(3\sigma)^3 + \frac{1}{6}(3\sigma)9\tau(\tau+\sigma)
  \right) Z^{'3} \right. \nonumber \\
  &  &  \left. \quad 
 + \frac{1}{3} \left(\frac{9\tau(\tau+\sigma)}{2}\right)^2 
  \left( \frac{2}{3}(3\sigma)^2 + \frac{9\tau(\tau+\sigma)}{2} \right)
  Z^{'2} \right. \nonumber \\
& & \left. \quad 
 + \frac{2}{9}\left(\frac{9\tau(\tau+\sigma)}{2}\right)^4 3\sigma Z'
+\frac{2}{27}\left(\frac{9\tau(\tau+\sigma)}{2}\right)^6 
  \right].  \label{eq:E6defA}
\end{eqnarray}
Now (\ref{eq:E6defB}) and (\ref{eq:E6defA}) have exactly the same form. 
This proves that we can choose $E_6$ as the gauge group of field theory 
local model for the geometry (\ref{eq:defeq}) around the type (a)
codimension-3 singularity, and the vev of $\varphi$ field can be chosen 
in the rank-2 subspace specified above. 
Three homogeneous coordinates $(\tilde{x}, \tilde{y}, \tilde{z})$
in (\ref{eq:E6defB}) are identified with $(c^4 X,c^6 Y, c^3 Z')$ 
in (\ref{eq:E6defA}), where undetermined scaling factor $c$ is inserted.
The deformation parameter of geometry corresponds to Cartan vev of 
$\varphi$ through
\begin{equation}
 \left(\frac{a_4}{a_3}, \frac{a_5}{a_3} \right)= 
(3\sigma c, -3\tau (3\tau +3\sigma)c^2).
\label{eq:correspondence-E6}
\end{equation}
We ignore the scaling factor $c$ in the rest of the argument. 

Let us now proceed to study how zero-modes behave around 
the type (a) codimension-3 singularities, using this 
field-theory local model.
Deformation by $(a_4,a_5)$, or equivalently the rank-2 
parametrization of $\mathfrak{e}_6$ Cartan subspace, 
$H = H_\tau \tau+ H_\sigma \sigma$ in (\ref{eq:2dim-sub-E6}), is designed to 
break $E_6$ symmetry to $\SU(5)$. $\mathfrak{e}_6$ contains 
subalgebra 
\begin{equation}
 \mathfrak{su}(2)+\mathfrak{su}(6) \supset 
\mathfrak{su}(2) + \mathfrak{u}(1) + \mathfrak{su}(5),
\end{equation}
and the 2-parameter subspace (\ref{eq:2dim-sub-E6}) is regarded as 
the Cartan part of $\mathfrak{su}(2)+\mathfrak{u}(1)$ above.
The irreducible decomposition of $\mathfrak{e}_6$-{\bf adj.} 
under the $\mathfrak{su}(2)+\mathfrak{u}(1)+\mathfrak{su}(5)$ subalgebra 
is given by
\begin{eqnarray}
 \mathfrak{e}_6\mbox{-}{\bf adj.} & \rightarrow & 
  ({\bf adj.}, {\bf 1})+({\bf 1},{\bf adj.}) +({\bf 1},{\bf 1}) \nonumber \\
& & \quad  + ({\bf 1},{\bf 5})+({\bf 1},\bar{\bf 5})
 +({\bf 2},{\bf 10})+({\bf 2},\overline{\bf 10}),
\label{eq:e6-decomp}
\end{eqnarray}
and the components in the second line, which we refer to as 
off-diagonal components,\footnote{In some other literatures, 
a word ``bi-fundamental components'' is used for the same thing. 
Too much intuition of Type IIB string theory tends to be carried on 
by this word, however, and we do not use this word in this article.}
are where charged matter multiplets of $\SU(5)$ come from.
They correspond to the sum $\oplus_i (U_i, R_i)$ in (\ref{eq:irr-decmp}).
\begin{table}[tb]
\begin{center}
\begin{tabular}{c|c|c}
irr. comp. & roots & $\vev{H_\sigma \sigma + H_\tau \tau, \bullet}$ \\
\hline
$({\bf 1}_+,{\bf 10})$ & $e_0-(e_k+e_l+e_m)$ & $3\tau$ \\
$({\bf 1}_-,{\bf 10})$ & $-e_0+(e_i+e_j)+L_6$ & $-3(\tau+\sigma)$ \\
\hline
$\overline{({\bf 1}_+,{\bf 10})}$ & $-e_0+(e_k+e_l+e_m)$ & $-3\tau$ \\
$\overline{({\bf 1}_-,{\bf 10})}$ & $e_0-(e_i+e_j)-e_6$ &
 $3(\tau+\sigma)$ \\
\hline
$({\bf 1},{\bf 5})=(\overline{{\bf 1}_+} \otimes \overline{{\bf 1}_-},{\bf 5})$ &
   $e_i-e_6$ & $3\sigma$ \\
\hline
$({\bf 1},\bar{\bf 5})=({\bf 1}_+ \otimes {\bf 1}_-,\bar{\bf 5})$ &  $-e_i+e_6$ & $-3\sigma$
\end{tabular} 
\caption{\label{tab:e6-decomposition} ${\bf 1}_+$ and ${\bf 1}_-$ 
form a doublet representation ${\bf 2}$ of
 $\mathfrak{su}(2)+\mathfrak{u}(1)$, and hence the first and second rows
 [resp. third and fourth] form an irreducible component 
$({\bf 2},{\bf 10})$ [resp. $(\bar{\bf 2},\overline{\bf 10})$] together.
The irreducible component in the fifth [resp. sixth] row
is regarded as $(\wedge^2 \bar{\bf 2}, {\bf 5})$ [resp. 
$(\wedge^2 {\bf 2}, \bar{\bf 5})$] representation of 
$\mathfrak{u}(2)+\mathfrak{su}(5)$. Indices $i,j,k,l,m$ are mutually 
different and can take values within $1, \cdots, 5$. }
\end{center}
\end{table}
The zero-mode wavefunctions of $\SU(5)$-${\bf 10}+\overline{\bf 10}$
representations are determined by 
\begin{equation}
 \rho_{U={\bf 2}}(2\alpha \vev{\varphi_{12}}) 
  = \diag(3\tau, -3(\tau+\sigma)) 
 \quad {\rm and} \quad 
 \rho_{U = \bar{\bf 2}}(2\alpha \vev{\varphi_{12}}) 
  = \diag(-3\tau, 3(\tau+\sigma)),
\end{equation}
respectively, and 
\begin{equation}
 \rho_{U = \wedge^2 {\bf 2}}(2\alpha \vev{\varphi_{12}})
  = -3\sigma, 
  \quad  {\rm and} \quad 
 \rho_{U = \wedge^2 \bar{\bf 2}} (2\alpha\vev{\varphi_{12}}) 
  = + 3\sigma
\end{equation}
should be used for the $\lambda_i \tau_i$ in the zero-mode 
equation (\ref{eq:EOM1}--\ref{eq:EOM4}) for the matter in 
$\SU(5)$-$\bar{\bf 5}+{\bf 5}$ representations. 
See Table~\ref{tab:e6-decomposition}.

Matter curves are where some of eigenvalues of 
$\rho_U(\vev{\varphi})$ vanish. Thus, $\sigma = 0$ 
is the matter curve of $\bar{\bf 5}+{\bf 5}$ representations, 
which is equivalent to $\tilde{a}_4 = 0$ from
(\ref{eq:correspondence-E6}). This agrees with the conventional 
understanding in F-theory compactification that $P^{(5)} = 0$ is 
the matter curve of $\bar{\bf 5}+{\bf 5}$ representations; under 
the scaling (\ref{eq:scale-E6-a-para}), $P^{(5)}\simeq \vev{a_3}^2 a_4$.
The matter curve of $\SU(5)$-${\bf 10}+\overline{\bf 10}$ 
representations is $\tau(\sigma+\tau) = 0$, which is equivalent to 
$\tilde{a}_5 = 0$ from (\ref{eq:correspondence-E6}). 
It is also a conventional understanding in F-theory compactification 
that $a_5 = 0$ [$h = 0$ in the notation of \cite{6authors, AC}]
is the matter curve for these representations. 

\begin{figure}[t]
\begin{center}
 \begin{tabular}{ccc}
  \includegraphics[width=.3\linewidth]{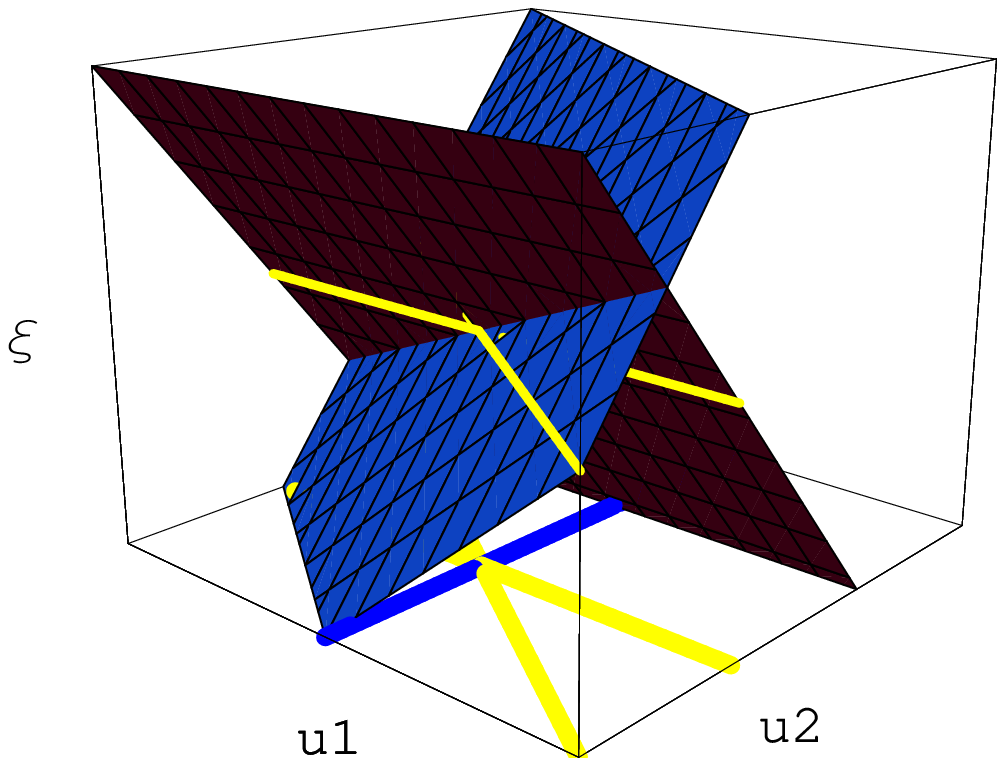} & 
  \includegraphics[width=.3\linewidth]{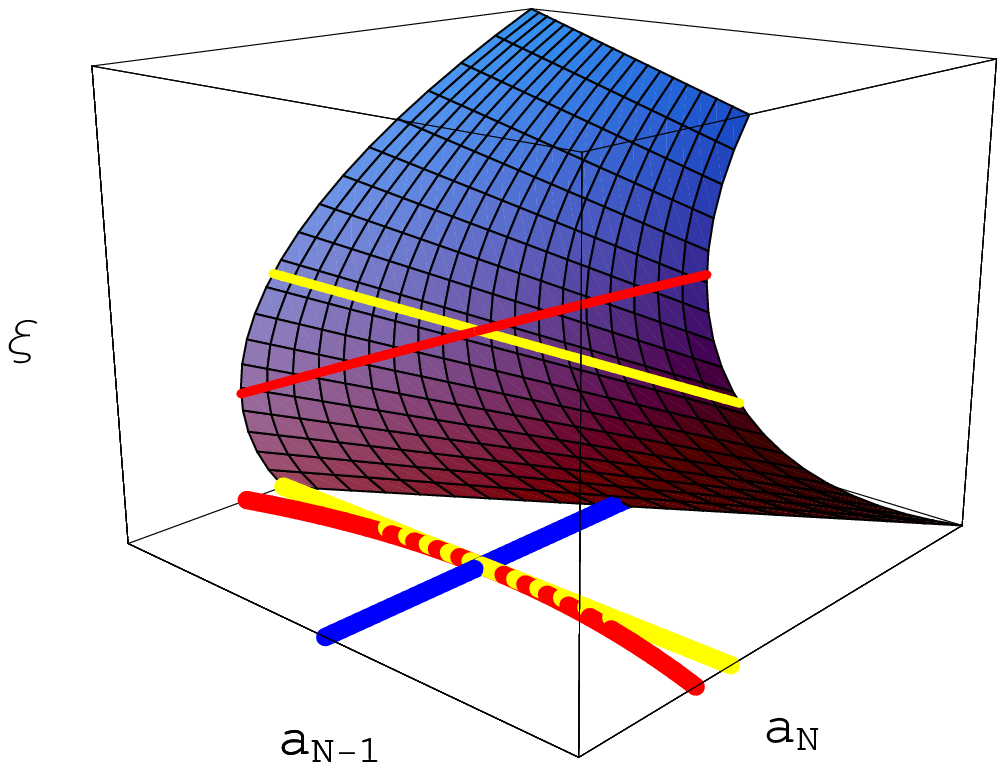} & 
  \includegraphics[width=.3\linewidth]{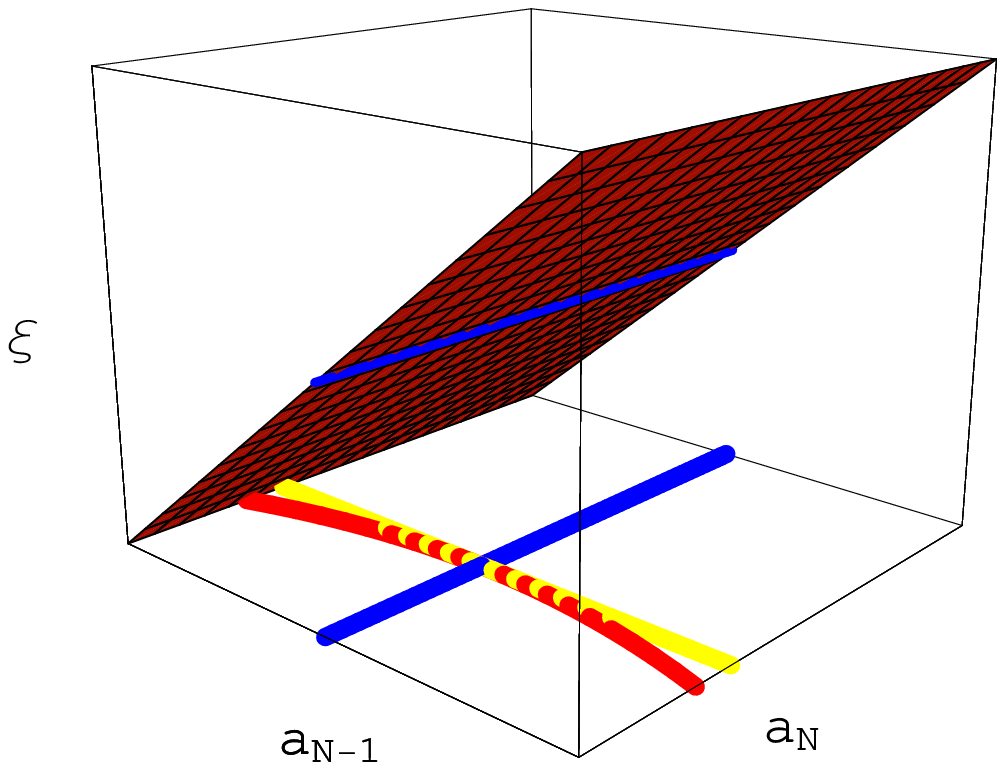} \\
 (ia): $\widetilde{C}_{{\bf 2} = {\bf 1}_+ + {\bf 1}_-} = 
    C_{{\bf 1}_+} \coprod C_{{\bf 1}_-}$ & (ib): $C_{{\bf 2}}$ & 
 (ii): $C_{\wedge^2 {\bf 2}}$   
 \end{tabular}
\caption{\label{fig:C-E67-a} (color online) 
Covering (spectral) surface for various irreducible components
around a type (a) codimension-3 singularity point of SU(5) and SO(10) 
GUT models. The panel (ia) and (ib) are for the $({\bf 2},{\bf 10})$ 
component [resp. $({\bf 2},{\bf 16})$ component] for the SU(5) GUT 
[resp. SO(10) GUT] models, with special (ia) and generic (ib) choices 
of complex structure moduli. Thin yellow curves on the covering surfaces
are the matter curves of these components, and thick yellow curves are
 their projection to $S$. A thin red curve in (ib) is the ramification 
divisor of $\pi_C: C \rightarrow S$, and its projection to $S$---branch 
locus---is denoted by a thick red curve. The panel (ii) is the 
covering (spectral) surface for the $(\wedge^2 {\bf 2}, \bar{\bf 5})$ 
[resp. $(\wedge^2 {\bf 2},{\bf vect.})$] component of SU(5) GUT [resp. 
SO(10) GUT] models. Thin and thick blue curves are the matter curves 
of these components and their projection to $S$. The coordinates 
$a_N$ and $a_{N-1}$ in the panels (ib) and (ii) should be read as 
$\tilde{a}_5$ and $\tilde{a}_4$ [resp. $\tilde{a}_4$ and $\tilde{a}_3$]
in SU(5) GUT models [resp. SO(10) GUT models].
}
\end{center} 
\end{figure}

If one assumes that the Cartan vev parameters $\sigma$ and $\tau$
vary linearly over the local coordinates of $S$, then we can set the
local coordinates $(u_1,u_2)$, so that\footnote{The choice of local
coordinates for (\ref{eq:bg-E6-A}) is not, in general, the same as 
the choice to make the metric simple, $g_{m\bar{n}} = \delta_{mn}$.
We do not try to be make the presentation more generic in this respect, 
at the cost of losing simplicity and clarity of discussion.} 
\begin{equation}
 \sigma \sim u_1, \qquad \tau \sim u_2. \label{eq:bg-E6-A}
\end{equation}
The matter curve of the 
${\bf 10}+\overline{\bf 10}$ representations have two irreducible 
branches in this case: $\tau \sim u_2 =0$ and 
$(\sigma+\tau) \sim (u_1 + u_2)= 0$. This is 
the case studied in \cite{BHV-1}. The two irreducible branches 
here, as well as the curve $u_1 = 0$ of $\bar{\bf 5}+{\bf 5}$ representations, 
all lie within the $A_4$ singularity surface and all pass through 
this type (a) codimension-3 singularity point \cite{BHV-1}.
See Figure~\ref{fig:C-E67-a}~(ia). In the choice of the background 
(\ref{eq:bg-E6-A}), the codimension-3 singularity point 
$(u_1, u_2) = (0,0)$ is also a double point singularity of the 
$a_5 = 0$ curve. 

For a generic choice of complex structure of a Calabi--Yau 4-fold, 
however, there is no guarantee that the Cartan vev parameters 
$\sigma$ and $\tau$ of the field theory-local model depend linearly 
on local coordinates. The type (a) codimension-3 singularity points 
are characterized as a common subset of $a_4 = 0$ and $a_5 = 0$. 
There is no reason to believe generically that the $a_5 = 0$ curve 
has a double-point singularity right at such a point. The type (a) 
codimension-3 singularity points are generically simple zero of 
both $a_4$ (i.e., $P^{(5)}$) and $a_5$, and the two matter curves 
intersect there normally. See Figure~\ref{fig:curves-SU(5)}.
For a generic choice of complex structure of a Calabi--Yau 4-fold, 
it is more appropriate to take $(\tilde{a}_4, \tilde{a}_5)$ as a 
set of local coordinates on $S$ in the field theory local model.
From the relation (\ref{eq:correspondence-E6}), we can now determine 
how the Cartan vev parameters of $\vev{\varphi_{12}}$ depend on 
local coordinates:
\begin{eqnarray}
 3\tau & = & -(\tilde{a}_4/2) + \sqrt{(\tilde{a}_4/2)^2 - \tilde{a}_5}, 
  \label{eq:bg-E6-B1}\\
 -(3\tau+3\sigma) & = &
  -(\tilde{a}_4/2) - \sqrt{(\tilde{a}_4/2)^2 - \tilde{a}_5}. 
  \label{eq:bg-E6-B2}
\end{eqnarray}
One can see from this expression explicitly that the condition of 
the matter curve $\tau(\sigma+\tau) = 0$ for the ${\bf 10}+\overline{\bf 10}$
representations become $\tilde{a}_5 = 0$, which now defines a smooth
curve without a double-point singularity. See Figure~\ref{fig:C-E67-a}
(ib).

The wavefunctions of zero-modes in the 
$({\bf 2},{\bf 10})+(\bar{\bf 2},\overline{\bf 10})$ components behave 
quite differently, depending upon whether the complex structure of a 
Calabi--Yau 4-fold is special and $\vev{\varphi_{12}}$ varies linearly 
on local coordinates, or the complex structure is generic. 
The field-theory local model for the former case does not need a branch 
cut and twists, but it has a branch locus at 
$(\tilde{a}_4/2)^2 - \tilde{a}_5 = 0$ for generic choice of complex
structure. The vev $\vev{\varphi}$ is in the $\U(2)$ subgroup 
of $E_6$, and the $E_6$-{\bf adj.} fields are twisted by the Weyl group 
transformation $\mathfrak{S}_2$ of the $\SU(2) \subset E_6$ structure 
group.

For a special choice of complex structure corresponding to the
background (\ref{eq:bg-E6-A}), the zero-mode equations can be studied 
separately for the $({\bf 1}_+, {\bf 10})$ component and 
$({\bf 1}_-,{\bf 10})$ component separately. The equations are
completely the same as those in section \ref{ssec:warm-up-1}. 
Zero modes from the $({\bf 1}_+, {\bf 10})$ component are localized 
along the curve $u_2 = 0$, and their wavefunctions are in Gaussian 
profile in the normal $u_2$ direction. Its profile along the curve is
determined by a holomorphic function $c(u_1)$ on the curve. 
Zero modes from the $({\bf 1}_-, {\bf 10})$ component are localized 
along the curve $(u_1 +u_2) = 0$, with a Gaussian profile in the normal
direction. Holomorphic functions on the curve determine the behavior 
of the zero modes. 
This is basically a known story \cite{BHV-1, BHV-2}.
Note also, although it will be obvious from the field theory formulation 
with $E_6$ gauge group without any twists, that the fields (and hence 
the holomorphic functions on the matter curves) for the two 
components $({\bf 1}_+, {\bf 10})$ and $({\bf 1}_-, {\bf 10})$ should 
remain independent within the local model of a given codimension-3
singularity point. Put another way, matter zero-modes are identified 
with locally free holomorphic functions (sections) on the covering 
matter curve (the one obtained by resolving the double point 
$u_2(u_1+u_2)=0$). 
The zero-mode wavefunctions in the $({\bf 1}_+,{\bf 10})+
({\bf 1}_-,{\bf 10})$ components are 
\begin{eqnarray}
& &
\left( \begin{array}{c}
 \tilde{\psi}_{\bar{1}; {\bf 1}_+} \\ \tilde{\psi}_{\bar{1}; {\bf 1}_- } 
\end{array}\right) = 
\left( \begin{array}{c}
  0 \\ \frac{f_{{\bf 1}_-}(u')}{\sqrt{2}} e^{-3|u_1+u_2|^2/\sqrt{2}} 
\end{array} \right), \quad 
\left( \begin{array}{c}
 \tilde{\psi}_{\bar{2}; {\bf 1}_+} \\ \tilde{\psi}_{\bar{2}; {\bf 1}_- } 
\end{array} \right) = 
\left( \begin{array}{c}
 - f_{{\bf 1}_+}(u_1) e^{-3|u_2|^2} \\ 
 \frac{f_{{\bf 1}_-}(u')}{\sqrt{2}} e^{-3|u_1+u_2|^2/\sqrt{2}}
\end{array} \right),  \nonumber \\
 & & 
\left( \begin{array}{c}
 \tilde{\chi}_{{\bf 1}_+} \\ \tilde{\chi}_{{\bf 1}_-} 
\end{array} \right) = 
\left( \begin{array}{c}
 f_{{\bf 1}_+}(u_1) e^{-3|u_2|^2} \\ 
 f_{{\bf 1}_-}(u') e^{-3|u_1+u_2|^2/\sqrt{2}}
\end{array} \right).
\end{eqnarray}
$f_{{\bf 1}_+}(u_1)$ is a holomorphic function on the $u_2=0$ branch, 
and $f_{{\bf 1}_-}(u')$ one of the $(u_1+u_2)=0$ branch; $u'$ is a local 
coordinate on the $(u_1+u_2)=0$ branch, e.g. $u_1-u_2$. 
The two functions are free within this field theory local model. 
\label{page:covering4tripleintersect}

For a generic choice of complex structure, results of 
section \ref{ssec:AN-gen} can be used instead, to determine 
the behavior of zero-mode wavefunctions in the $\SU(5)$-${\bf 10}$ 
representation. One will notice that (\ref{eq:bg-E6-B1}, \ref{eq:bg-E6-B2})
are exactly the same as (\ref{eq:bg-AB}), after identifying the local
coordinates $(\tilde{a}_4, \tilde{a}_5)$ with $(2u_1,u_2)$.
The wavefunctions are localized along the matter curve, which is now 
$u_2 = 0$. In a region where $|u_1|$ is large, the wavefunctions are 
large in $(\tilde{\psi}_{+\bar{2}}, \tilde{\chi}_+)$ components, 
and small in others. The $(\tilde{\psi}_{+\bar{2}}, \tilde{\chi}_+)$ 
decays in a Gaussian profile in the normal $u_2$ direction. In a 
region where $|u_2|$ is large and $|u_1|$ is small instead,
wavefunctions for the $\SU(5)$-${\bf 10}$ representation are 
large in all the $(\tilde{\psi}_{\pm \bar{2}}, \tilde{\chi}_{\pm})$ 
components (but not in $\tilde{\psi}_{\pm \bar{1}}$ components), and 
their asymptotic behavior is already obtained in
(\ref{eq:chi-asymp-text}, \ref{eq:psi-asymp-text}); they do not decay 
as fast as in Gaussian profile, but still they become small
exponentially off the matter curve. 

In section \ref{ssec:AN-gen}, we introduced a covering surface $C$ 
in a study of zero-mode wavefunctions of 
$({\bf 2},\bar{N}) + (\bar{\bf 2}, N)$ components in the local model 
of generic deformation of $A_{N+1}$ to $A_{N-1}$ singularity.
We argued that the zero-modes and zero-mode wavefunctions are better 
understood and better behave on the covering surface $C$, rather than 
on the base space $S$. Similarly, we introduce a notion of spectral
surface (covering surface) for each irreducible component $(U_i, R_i)$
appearing in the decomposition (\ref{eq:irr-decmp}, \ref{eq:e6-decomp}).
For a representation $U_i$ such as ${\bf 2}$, $\bar{\bf 2}$, 
$\wedge^2 {\bf 2}$ and $\wedge^2 \bar{\bf 2}$, spectral surface
$C_{U_i}$ is determined by zero locus of 
\begin{equation}
 {\rm det}\left(\xi {\bf 1} - \rho_{U_i} (2\alpha\vev{\varphi_{12}})\right).
\end{equation}
The spectral surface for a representation $U_i$ describes the behavior 
of eigenvalues of $\rho_{U_i}(2\alpha\vev{\varphi_{12}})$.
Just like in the case we discussed in the previous section, 
the spectral surfaces are divisors of $\mathbb{K}_S$, total space 
of canonical bundle of $S$. For the representation $U_i = {\bf 2}$, 
it is given by 
\begin{equation}
 a_3 (\xi - 3\tau)(\xi + (3\tau+3\sigma)) = a_3 \xi^2 + a_4 \xi + a_5 =
  0, 
\label{eq:C-E6-A4}
\end{equation}
and $\pi_{C_{\bf 2}}: C_{U_i={\bf 2}} \rightarrow S$ is a degree-2 
cover. $\tilde{a}_4^2 - 4 \tilde{a}_5 = 0$ is the branch locus 
of this covering.\footnote{
For a special choice of complex structure of a Calabi--Yau 4-fold, 
the $\varphi$ vev becomes linear in local coordinates, and the spectral 
surface becomes a sum of two irreducible pieces, as 
in Figure~\ref{fig:C-E67-a} (ia), each one of the irreducible pieces 
is not ramified over $S$.} See Figure~\ref{fig:C-E67-a}~(ib).
The zero-mode solutions for the $\SU(5)$-{\bf 10} representation
do not have a singular behavior on the spectral surface 
$C_{U_i = {\bf 2}}$, as we saw at the end of section \ref{ssec:AN-gen}.
Applying the same argument as in section \ref{ssec:AN-gen} to the 
irreducible component $({\bf 2}, {\bf 10})$ and its spectral surface, 
we learn that the zero modes of $\SU(5)$-${\bf 10}$ representation 
are ultimately characterized by holomorphic functions on the spectral 
surface $C_{U_i = {\bf 2}}$ modulo those that vanish along the matter curve 
$\bar{c}_{U_i = {\bf 2}}$. 

The zero  modes of matter in the $(\wedge^2 \bar{\bf 2},{\bf 5})$ and 
$(\wedge^2 {\bf 2},\bar{\bf 5})$ representations always have Gaussian 
wavefunctions in the direction normal to the matter curves.
Even in case of general unfolding of $E_6$ singularity, 
$\rho_{\wedge^2 {\bf 2} [\bar{\bf 2}]}(2\alpha \vev{\varphi_{12}})$ is 
$\mp 3\sigma = \mp \tilde{a}_4 = \mp (2 u_1)$ and is linear in local 
coordinates for this pair of representations.
This is sufficient to conclude that the zero-mode solutions behave 
locally as 
\begin{eqnarray}
 \tilde{\chi}_{{\bf 5}} = f_{{\bf 5}}(u_2) 
   \exp \left[ - 2|u_1|^2 \right], \qquad 
 \tilde{\psi}_{\bar{1}; {\bf 5}} = - f_{{\bf 5}}(u_2) 
  \exp \left[ - 2|u_1|^2 \right], \qquad 
 \tilde{\psi}_{\bar{2}; {\bf 5}} = 0, \\
 \tilde{\chi}_{\bar{\bf 5}} = f_{\bar{\bf 5}}(u_2) 
   \exp \left[ - 2|u_1|^2 \right], \qquad 
 \tilde{\psi}_{\bar{1}; \bar{\bf 5}} = + f_{\bar{\bf 5}}(u_2) 
  \exp \left[ - 2|u_1|^2 \right], \qquad 
 \tilde{\psi}_{\bar{2}; \bar{\bf 5}} = 0,
\end{eqnarray}
where the first line is for the matter in the $\SU(5)$-${\bf 5}$
representation, and the second line for those in the $\bar{\bf 5}$ 
representation.
Zero mode wavefunctions are determined, once the holomorphic functions 
$f_{{\bf 5}[\bar{\bf 5}]}(u_2)$ on the matter curve $u_2 = 0$ 
for zero modes are determined. 
They are in a simple Gaussian form in the direction normal to the 
matter curve $\tilde{a}_4 \propto u_1 = 0$.

There is not a strong need to introduce a spectral surface 
to study the zero modes in the ${\bf 5}+\bar{\bf 5}$ representations. 
As a preparation for discussion in section \ref{sec:Higgs}, and for 
an illustrative purpose for section \ref{ssec:discr-spec}, however, 
let us see how the spectral surface of 
$(U_i, R_i) = (\wedge^2 {\bf 2}, \bar{\bf 5})$ looks like. 
It is a divisor of $\mathbb{K}_S$, and is defined by 
\begin{equation}
 \xi -(-3\sigma) = \xi + \tilde{a}_4 = 0.
\end{equation}
See Figure~\ref{fig:C-E67-a} (ii). 
This spectral surface $C_{\wedge^2 {\bf 2}}$ is ramified over $S$, 
and branch cut is not necessary for the study of zero modes 
in the $\bar{\bf 5}$ representation. The zero modes are characterized 
by holomorphic functions on the matter curve, $\tilde{a}_4 = 0$, which 
is equivalent to holomorphic functions on the spectral surface 
$C_{\wedge^2 {\bf 2}}$ in Figure~\ref{fig:C-E67-a}, modulo those 
that vanish along the matter curve, which it denoted by a thin blue 
curve in it.

\subsubsection{Type (d) Singularity: $D_6 \rightarrow A_4$}
\label{sssec:SU(5)-D6}

In F-theory compactifications that leave $\SU(5)$ gauge symmetry 
in low-energy effective theory, there is another type of 
codimension-3 singularity, type (d), where Yukawa couplings of the form 
\begin{equation}
 \Delta W = \bar{\bf 5}_a \; {\bf 10}^{ab} \; \bar{\bf 5}_b
\label{eq:5105}
\end{equation}
are generated. In Georgi--Glashow SU(5) GUT theories, the down-type
quark and charged leptons Yukawa couplings come from this type of 
interactions. In flipped SU(5) models, they become up-type and 
neutrino Yukawa couplings. The type (d) codimension-3 singularity 
is where $a_5$ and $a_3$ in (\ref{eq:defeq}) vanish simultaneously. 

It is almost obvious from the pattern of deformation of singularity, 
$D_6 \rightarrow A_4$, that local geometry of this type of 
codimension-3 singularity allows for interpretation in terms of 
D7-branes and an O7-plane in Type IIB string theory. In the rest 
of this section \ref{sssec:SU(5)-D6}, we confirm that this 
intuition is correct. We will do so by constructing a field-theory 
local model for this local geometry, and then, see that the background 
assumed in the local model is interpreted as D7-branes and an O7-plane 
intersecting at angles. 

The behavior of the zero-mode wavefunctions are determined 
by using the 8-dimensional field-theory formulation of \cite{DW-1, 
BHV-1}, and we find that the results agree with what was obtained 
in \cite{HayashiEtAl} by using Heterotic--F theory duality. 
The results also agree with well known intuitive picture of 
open strings in Type IIB string theory.

An 8-dimensional field theory with a gauge group $D_6 = \SO(12)$ 
is used in describing the local geometry of this codimension-3 
singularity. We need to determine the background configuration 
$\vev{\varphi}_{12}(u_1,u_2)$ of the local model in terms 
of coefficients of the defining equation of the local geometry.

A most generic deformation of $D_n$ surface singularity is described by 
\begin{equation}
 Y^2 = - Z X^2 - 2 \gamma_n X + Z^{n-1} + \delta_{2} Z^{n-2} + \cdots
 + \delta_{2(n-1)},
\label{eq:Dn-def-A}
\end{equation}
where $(X,Y,Z)$ are coordinates and $\delta_{2,4,\cdots,2(n-1)}$ and 
$\gamma_n$ are deformation parameters. The deformation parameters 
vary over a non-compact base space $S$ in the context of F-theory
geometry, and (\ref{eq:Dn-def-A}) is regarded as a defining equation 
of local geometry of a Calabi--Yau 4-fold that is a fibration of
deformed $D_n$ singularity surface.
The Cartan subalgebra ($\otimes_{\R} \C$) of $D_n$ is parametrized by 
$n$ complex numbers, $t_i$ ($i=1,\cdots,n$). If we take a basis of 
$2n \times 2n$ matrix so that the Cartan subalgebra is diagonalized, 
the Cartan parameters $t_i$'s are the diagonal entries of  the 
$2n \times 2n$ matrix:
\begin{equation}
H = \diag(t_1,\cdots,t_n,-t_1,\cdots,-t_n). 
\end{equation}
The symmetric functions under the full Weyl group of $\SO(2n)$---just like 
$\epsilon_{2,5,8,6,9,12}$ for the deformation of $E_6$---are given by 
\begin{equation}
 \delta_{2j} = \sum_{i_1 < i_2 < \cdots < i_j} 
  t_{i_1}^2 \, t_{i_2}^2 \cdots t_{i_j}^2, \quad {\rm and} \quad 
 \gamma_n = t_{i_1} \cdots t_{i_n}.
\label{eq:symmetric-Dn}
\end{equation}
If all the Cartan parameters $t_i$'s are assigned homogeneous 
degree 1, the deformation parameters $\delta_{2i}$ and $\gamma_n$ 
are homogeneous functions of degree $2i$ and $n$, respectively. 
If the coordinates $(X,Y,Z)$ have degree $n-2$, $n-1$ and $2$,
respectively, then the local equation (\ref{eq:Dn-def-A}) of 
$D_n$ singularity and its deformation are given by a homogeneous 
function of degree $2n - 2$, the dual Coxeter number of $D_n$.

Here, we are interested in studying a deformation of $D_6$ to $A_4$. 
We thus take $n=6$ and the vev of $\varphi$ is turned on only in a rank-2 
subspace of the Cartan subalgebra of $\mathfrak{so}(12)$.
We choose the 2-dimensional subspace to be 
\begin{eqnarray}
 H_\sigma & = & \diag (\overbrace{1,\cdots,1}^6,
  \overbrace{-1,\cdots,-1}^6),  \\
 H_\tau & = & \diag (\overbrace{0,\cdots,0}^5, 1, 
  \overbrace{0,\cdots,0}^5,-1), \\
 H & = & H_\sigma \sigma + H_\tau \tau = 
  \diag(\overbrace{\sigma,\cdots,\sigma}^5,\sigma+\tau, 
        \overbrace{-\sigma,\cdots,-\sigma}^5,-(\sigma+\tau)),  
  \label{eq:D6-2para}
\end{eqnarray}
where $(\sigma,\tau) \in \C^2$ is a parametrization of this subspace.
It is straightforward to see that the equation (\ref{eq:Dn-def-A}) becomes
\begin{eqnarray}
 Y^2 & = & - Z X^2 - 2 \sigma^5 (\sigma+\tau) X  \nonumber \\
 &  & + Z^5 + (5\sigma^2 + (\sigma+\tau)^2) Z^4 
 + (10 \sigma^4 + 5 \sigma^2(\sigma+\tau)^2) Z^3  \label{eq:D6-def-A}\\
 &  & +  (10 \sigma^6 + 10 \sigma^4(\sigma+\tau)^2) Z^2
 + (5 \sigma^8 + 10 \sigma^6 (\sigma+\tau)^2) Z
 + (\sigma^{10} + 5 \sigma^8 (\sigma+\tau)^2). \nonumber 
\end{eqnarray}

The local equations of the deformations (\ref{eq:Dn-def-A}) and 
(\ref{eq:D6-def-A}) are not particularly useful, however, 
in identifying independent deformations that are supposed to be 
smooth in local coordinates of $S$. It is hard to pick up 
the two independent deformation parameters among the Weyl-group 
invariant deformation parameters $\delta_{2,4,\cdots,10}$ and
$\gamma_6$ in (\ref{eq:Dn-def-A}). The parameters 
$t_i = \sigma,(\sigma+\tau)$ in (\ref{eq:D6-def-A}) are not Weyl-group 
invariant quantities, and are not necessarily required to be smooth 
in the local coordinates of $S$. The deformation equation
(\ref{eq:defeq}) is instead suitable for this purpose. 
Thus, we will rewrite (\ref{eq:defeq}) for small $a_3$ and $a_5$---this 
is the characterisation of type (d) codimension-3 
singularity---in exactly the same form as (\ref{eq:D6-def-A}).

To start with, we need to specify how to zoom in into the codimension-3 
singularity. The scaling of the coordinates $(x,y,z)$ is 
\begin{equation}
 (x,y,z) = (\lambda^{n-2} x_0, \lambda^{n-1} y_0, \lambda^2 z_0)
\qquad \lambda \rightarrow 0
\end{equation}
with $n=6$, the same as the scaling of coordinates $(X,Y,Z)$ 
in (\ref{eq:Dn-def-A}). The deformation parameters $a_3$ and $a_5$ 
also become small around the type (d) codimension-3 singularities, 
and we assign the scaling 
\begin{equation}
 (a_3, a_5) = (\lambda a_{3,0}, \lambda a_{5,0}).
\end{equation}
All other deformation parameters are not supposed to vanish 
around the codimension-3 singularity generically, and hence 
we assign zero weight in the scaling by $\lambda$. 
Terms with weight greater than 10 become irrelevant near 
the codimension-3 singularity point, and the local equation 
(\ref{eq:defeq}) becomes 
\begin{equation}
 \left(y - \frac{1}{2}(a_5 x + a_3 z^2) \right)^2 \simeq 
 \left(\frac{a_5^2}{4} + a_4 z \right) x^2  + 
 \left(a_2 z^3 + \frac{1}{2} a_5 a_3 z^4 \right) x + 
 \left(\frac{1}{4} a_3^2 z^4 + a_0 z^5 \right).
\label{eq:tempD6}
\end{equation}
By introducing a shorthand notation
\begin{equation}
 \tilde{s} \equiv \left(\frac{a_5}{2a_4}\right)^2, \qquad 
 \tilde{t} \equiv \frac{1}{\tilde{D}} 
   \left(\frac{a_3}{2a_4}- \frac{a_2}{4a_4}\frac{a_5}{a_4}\right)^2, \qquad 
 \tilde{D} \equiv \frac{a_2^2 - 4 a_4 a_0}{4 a_4^2} \quad {\rm and}
 \quad
 \tilde{a}_{2,3,5} \equiv \frac{a_{2,3,5}}{a_4},
\end{equation}
and new coordinates 
\begin{eqnarray}
 \tilde{z} & = & - \left(\frac{z}{a_4}+\tilde{s}\right), \\
 \tilde{y} & = & \frac{1}{\sqrt{\tilde{D}}} \frac{1}{a_4^3}
   \left(y - \frac{1}{2}(a_5 x + a_3 z^2)\right), \\
 \tilde{x} & = & \frac{-1}{\sqrt{\tilde{D}}}\left( \frac{x}{a_4^2} 
  - \frac{1}{8}\tilde{a}_5^3 
    \left(\tilde{a}_3 -\frac{3}{4}\tilde{a}_2 \tilde{a}_5\right)
  - \frac{\tilde{a}_5}{2} 
    \left( \frac{\tilde{a}_3}{2} - \frac{3}{4}\tilde{a}_2 \tilde{a}_5 \right)
    \tilde{z}
  + \frac{\tilde{a}_2}{2}\tilde{z}^2
\right), 
\end{eqnarray}
we find that (\ref{eq:tempD6}) becomes 
\begin{eqnarray}
 \tilde{y}^2 & \simeq & - \tilde{z} \tilde{x}^2 
   - 2 (\tilde{s}^5 \tilde{t}) \tilde{x} 
  + \tilde{z}^5 + \left[5 \tilde{s} + \tilde{t} \right] \tilde{z}^4  
  \nonumber \\
& &  + \left[10 \tilde{s}^2 + 5 \tilde{s} \tilde{t} \right] \tilde{z}^3
  + \left[10 \tilde{s}^3 + 10 \tilde{s}^2 \tilde{t} \right] \tilde{z}^2
  + \left[5 \tilde{s}^4 + 10 \tilde{s}^3 \tilde{t} \right] \tilde{z}
  + \left[ \tilde{s}^5 + 5 \tilde{s}^4 \tilde{t} \right].
\label{eq:D6defB}
\end{eqnarray}
Now (\ref{eq:D6-def-A}) and (\ref{eq:D6defB}) have exactly the same
form. This means that an $\SO(12)$ gauge theory can be used for 
the local model of the type (d) codimension-3 singularities, and 
the rank-2 parametrization of the background $\vev{\varphi}$ 
in (\ref{eq:D6-2para}) is sufficient for the most generic complex 
structure.

We now only need to determine the background $\tau$ and $\sigma$ 
in terms of the coefficients of the defining equation of local geometry 
(that is, $a_3$ and $a_5$) in order to fix the field-theory local model.
Coordinates $(\tilde{x}, \tilde{y}, \tilde{z})$ in (\ref{eq:D6defB}) 
are identified with $(c^4 X, c^5 Y, c^2 Z)$ in (\ref{eq:D6-def-A}), 
and the Cartan vev parameters $\sigma$ and $\tau$ are identified as follows:
\begin{equation}
  c \left(\sigma, (\sigma+\tau) \right) = 
 \left( \frac{a_5}{2a_4},  \frac{1}{\sqrt{\tilde{D}}}
   \left(\frac{a_3}{2a_4} - \frac{a_2}{4a_4}\frac{a_5}{a_4}\right)
 \right). 
\end{equation}
We will ignore the factor $c \in \C^\times$ in the following.

\begin{figure}[t]
 \begin{center}
  \begin{tabular}{ccc}
  \includegraphics[width=.3\linewidth]{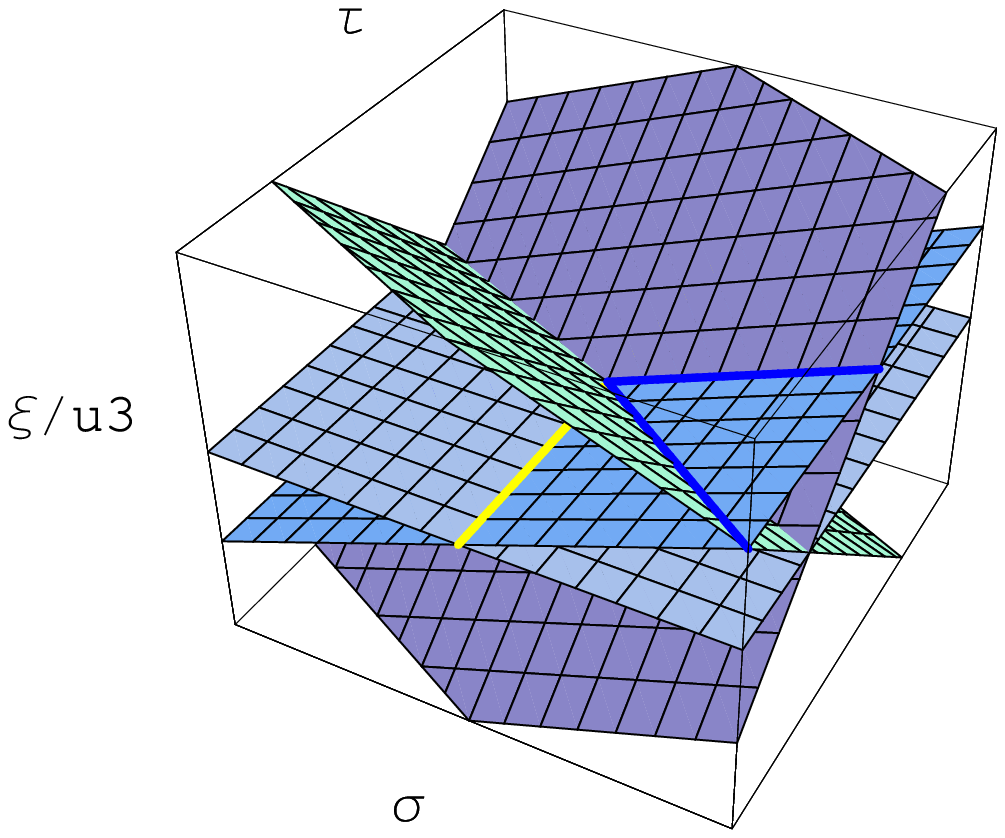} & $\qquad$ & 
  \includegraphics[width=.3\linewidth]{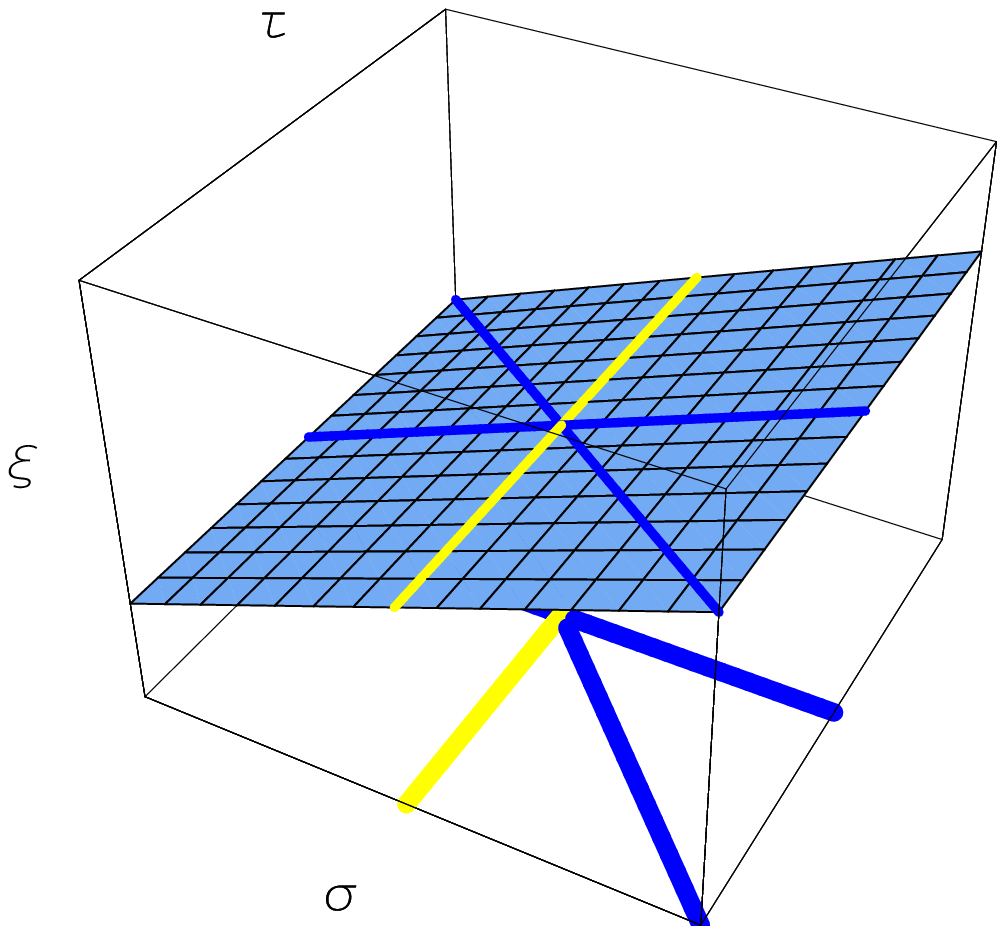}  \\
(a) & & (b)
  \end{tabular}
\caption{\label{fig:C-D6} (color online)
Configuration of D7-branes and an O7-plane corresponding 
to the local geometry of type (d) codimension-3 singularity.
An O7-plane (light blue) is placed at $u_3 = 0$. 
A stack of five D7-branes for unbroken SU(5) symmetry 
(blue) and the O7-plane intersect at angle along 
the $\sigma=0$ locus (yellow). The 6th D7-brane and 
12th D7-brane are colored purple and green, respectively 
in (a). Their intersection curve is right on the O7-plane at 
$u_3 = 0$, showing that they are the orientifold image of 
each other. The 6th D7-brane and the SU(5) D7-branes intersect 
along a curve at $\tau=0$, and the intersection curve of 
the 12th D7-brane and the SU(5) GUT D7-branes is 
at $(2\sigma + \tau) = 0$; they are shown by dark blue curves 
in the figure. The right panel shows only the SU(5) GUT 
branes and matter curves among them, which is close to the picture 
in F-theory description. The three matter curves within the 
SU(5) GUT D7-branes all pass through the type (d) codimension-3 
singularity points.
}
 \end{center}
\end{figure}

We can use $(\tilde{a}_3, \tilde{a}_5) = (a_3/a_4, a_5/a_4)$ 
as the local coordinates on $S$ around a point of type (d) codimension-3
singularity. Thus, the vev of $\varphi$ field 
in the $\mathfrak{so}(12)$ field theory on $S$
\begin{equation}
 2\alpha \vev{\varphi_{12}} = H_\sigma \sigma(\tilde{a}_3,\tilde{a}_5) 
 + H_\tau \tau(\tilde{a}_3,\tilde{a}_5)
\label{eq:SO(12)vev}
\end{equation}
are linear in local coordinates. There is no branch locus
in $\sigma(\tilde{a}_3, \tilde{a}_5)$ or 
$\tau(\tilde{a}_3, \tilde{a}_5)$ for generic unfolding of 
$D_6$ singularity to $A_4$.
In Type IIB Calabi--Yau orientifold language, 
$\vev{\varphi_{12}}$ is regarded as the coordinate 
of a given D7-brane in the 3rd complex dimension, ``$u_3$''.
Linear behavior of $\sigma(u_1,u_2)$ and $\tau(u_1,u_2)$ means 
that the local geometry of this codimension-3 singularity is 
interpreted as (locally almost) flat D7-branes intersecting 
at angles. See Figure~\ref{fig:C-D6}.

The $\SO(12)$ symmetry is broken by $\vev{\varphi}$ in a 
$\U(1) \times \U(1)$ subgroup of $\SO(12)$ specified by 
(\ref{eq:SO(12)vev}). The off-diagonal components in the 
irreducible decomposition of $\mathfrak{so}(12)$-{\bf adj.} 
under $\mathfrak{u}(1)+\mathfrak{u}(1)+\mathfrak{su}(5)$
are 
\begin{equation}
  [{\bf 10}^{2\sigma}+{\bf 5}^{-\tau}+{\bf 5}^{2\sigma+\tau}] 
 + {\rm h.c.},
\label{eq:SO(12)-irr}
\end{equation}
where $2\sigma$, $-\tau$ and $(2\sigma+\tau)$ show how 
${\rm ad}(\vev{\varphi})$ acts on these irreducible components.
Thus, the matter curve of $\SU(5)$-${\bf 10}+\overline{\bf 10}$ 
representations is $\sigma = 0$, which is equivalent to $a_5 = 0$.
The matter curve of $\SU(5)$-${\bf 5}+\bar{\bf 5}$ representations 
is either $\tau= 0$ (equivalently $\sigma = + (\sigma+\tau)$) or 
$(2\sigma+\tau) = 0$ (equivalently $\sigma = - (\sigma+\tau)$).
Both of these two branches of curves pass through the point of 
type (d) codimension-3 singularity. The condition 
$\sigma^2 = (\sigma+\tau)^2$ for the entire matter curve of 
${\bf 5}+\bar{\bf 5}$ representations is equivalent to 
\begin{equation}
 0 = (\sigma+\tau)^2 - \sigma^2 = 
 \frac{a_4 a_3^2 - a_2 a_3 a_5 + a_0 a_5^2}{a_4 (a_2^2 - 4 a_4 a_0)}
 \propto P^{(5)}.
\end{equation}
Thus, the locus of vanishing ${\rm ad}(\vev{\varphi})$ reproduces 
the matter curves, $a_5 = 0$ and $P^{(5)}=0$, a result we are familiar 
with in F-theory compactification in general. 
In Type IIB language,  $\sigma= 0$ is the matter curve of 
{\bf 10} representation as this is where 5 D7-branes intersect 
an O7-plane. $\sigma = + (\sigma+\tau)$ [$\sigma = - (\sigma+\tau)$] 
is where the first 1--5 D7-branes intersect the 6th [12-th respectively]
D7-brane, and this is why they are matter curves for the ${\bf 5}$ 
representation. It is now easy to see that the three matter curves all 
pass through the type (d) codimension-3 singularity points, essentially 
because an O7-plane requires the 12-th D7-brane as a mirror image 
of the 6th D7-brane. This is purely a Type IIB phenomenon.

The zero-mode wavefunctions are Gaussian for all the irreducible components 
in (\ref{eq:SO(12)-irr}) in directions normal to their matter curves, 
because ${\rm ad}(\vev{\varphi})$ is linear in local coordinates 
for all these irreducible components. The behavior of zero mode
wavefunctions along the matter curves is in one-to-one correspondence 
with holomorphic functions $f$ (introduced 
in section~\ref{sec:warm-up}) 
for the matter curves, $2\sigma=0$, $-\tau=0$ and $2\sigma+\tau=0$, 
respectively. Note that the holomorphic functions $f$'s, especially 
their values at $(\tilde{a}_3,\tilde{a}_5)=(0,0)$, are defined 
on the two branches of $P^{(5)} = 0$ curve separately \cite{HayashiEtAl}.
This statement is almost trivial in this field-theory formulation 
(or in Type IIB D-brane picture), because they come from different 
irreducible components of $\mathfrak{so}(12)$-{\bf adj.} fields on $S$. 

The spectral surface, 
${\rm det} (\xi {\bf 1} - \rho_U(2\alpha\vev{\varphi_{12}})) = 0$, 
becomes a plane $\xi - 2\sigma = 0$ for ${\bf 10}^{2\sigma}$ component. 
The spectral surface for 
the $\bar{\bf 5}^\tau + \bar{\bf 5}^{-(2\sigma+\tau)}$ components 
is 
\begin{equation}
 C_{6+12}: (\xi - \tau)(\xi + (2\sigma + \tau)) = 0,
\label{eq:def-CV2-d}
\end{equation}
but it is more natural to define
\begin{equation}
 \widetilde{C}_{6+12} = C_6 \coprod C_{12} = 
 (\xi - \tau = 0) \coprod (\xi + (2\sigma + \tau) = 0). 
\label{eq:def-CV2-d-resolve}
\end{equation}
The $C_6$ component corresponds to the system of the 6th D7-brane and 
a stack of 1--5th D7-branes intersecting at $\tau=0$, and 
$C_{12}$ to that of the 12th D7-brane and the stack of five D7-branes 
intersecting at $-(2\sigma+\tau) = 0$.
There is no double-curve singularity in $\widetilde{C}_{6+12}$, which 
was present in $C_{6+12}$. 
The covering matter curve is characterized by $\xi = 0$ 
on $\widetilde{C}_{6+12}$, and the double point singularity of 
the matter curve $P^{(5)} = 0$ is resolved. The same should be 
applied to the spectral surface and matter curve on it for the 
$({\bf 1}_+ \oplus {\bf 1}_-,{\bf 10})+{\rm h.c.}$ components for 
the special choice 
of complex structure moduli corresponding to (\ref{eq:bg-E6-A}) and 
Figure~\ref{fig:C-E67-a}~(ia).
This desingularization of the spectral surface 
$\nu: \widetilde{C}_{6+12} \rightarrow C_{6+12}$ is a 
natural and useful concept as we see in section~\ref{sec:Higgs}.
\begin{figure}[t]
 \begin{center}
  \begin{tabular}{ccc}
  \includegraphics[width=.3\linewidth]{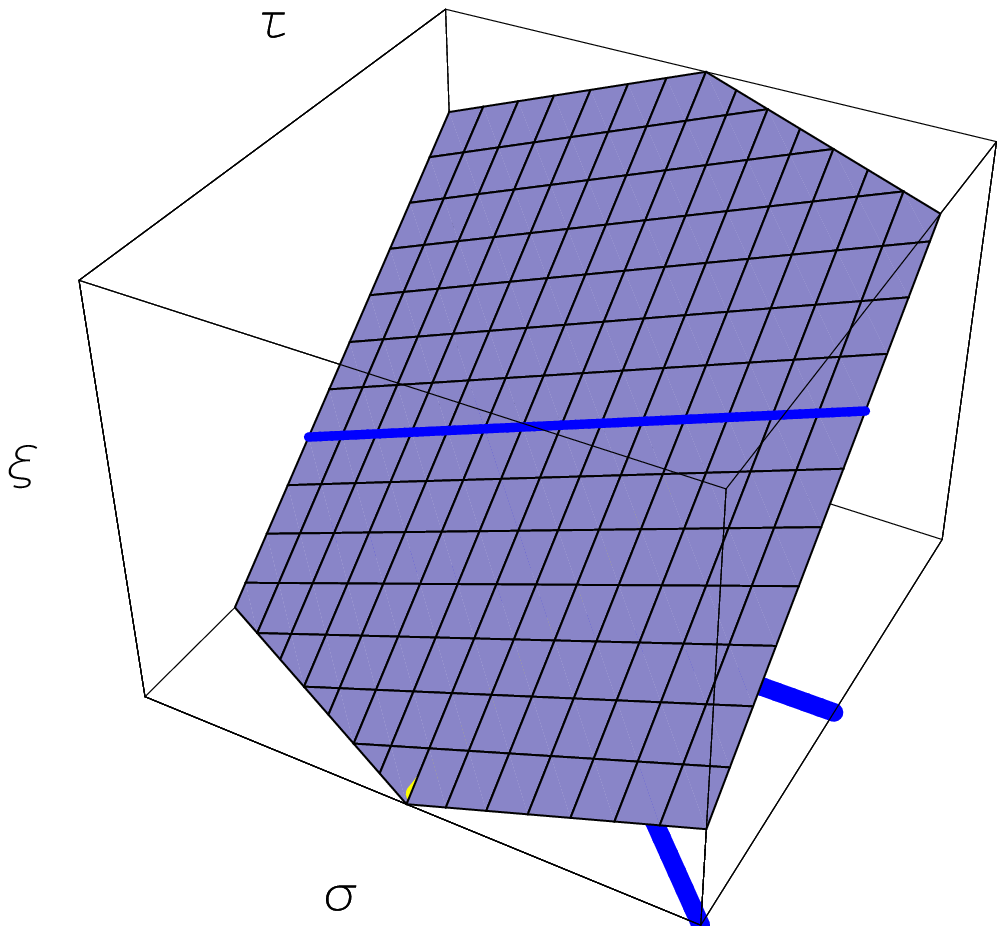} & $\coprod$ & 
  \includegraphics[width=.3\linewidth]{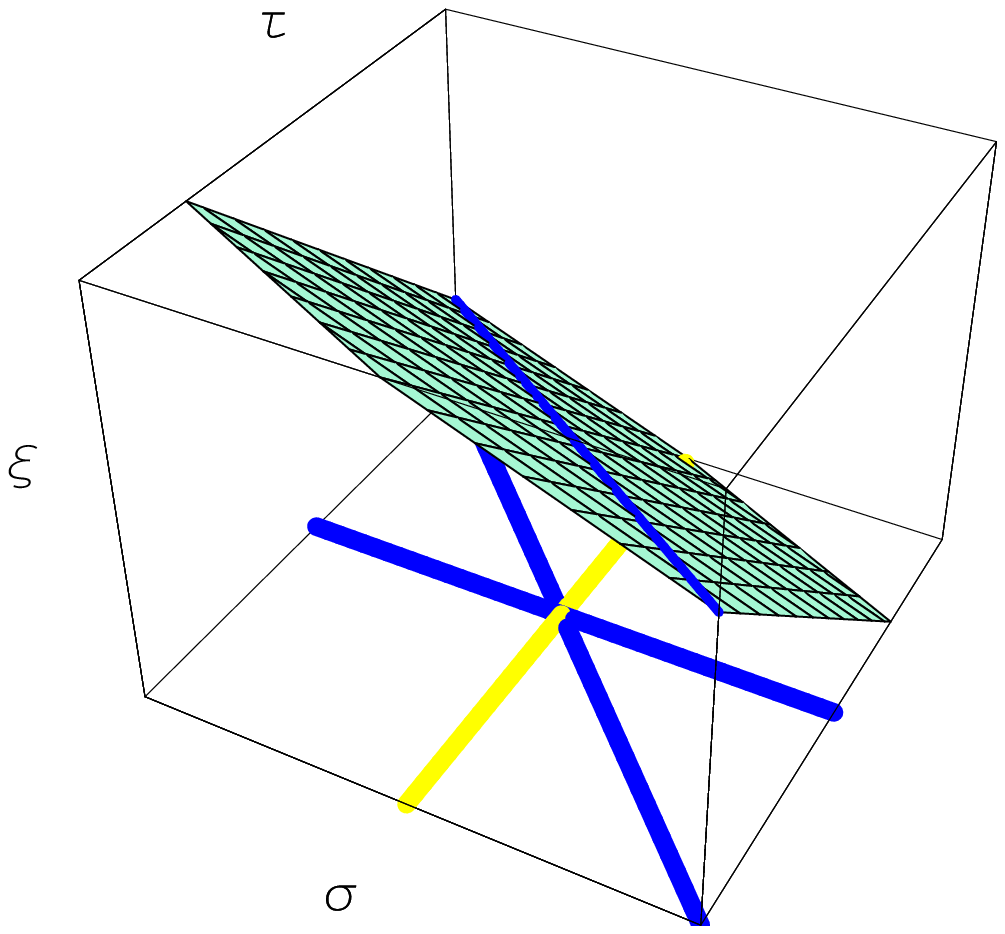}  \\
(a) & & (b)
  \end{tabular}
\caption{\label{fig:C2-D6} The 6th (a) and 12th (b) D7-branes are 
extracted from Figure~\ref{fig:C-D6}~(a), instead. The intersection 
curve with the SU(5) GUT D7-branes are marked by blue lines on them.
Their projection to the plane with local coordinates $(\sigma,\tau)$ 
becomes the two branches of the matter curve $R^{(5)} = 0$. Their 
disjoint union corresponds (after subtracting the $\xi$ coordinate by 
$\sigma$) to the desingularization
$\widetilde{C}_{6+12}$ of the spectral surface $C_{6+12}$.
}
 \end{center}
\end{figure}

\subsubsection{Type (c1) Singularity: $A_6 \rightarrow A_4$}
\label{sssec:SU(5)-A6}

There are isolated codimension-3 singularities characterized 
by $P^{(5)}=0$ and $R^{(5)}=0$ (but $a_5 \neq 0$).
Using new coordinates $\tilde{y} = y - (a_5 x + a_3 z^2)/2$ and 
\begin{equation}
  \tilde{x}  =  x + \frac{a_3}{a_5} z^2 
   + \left(\frac{2}{a_5}\right)^2 
     \left(\frac{a_2}{2}-\frac{a_3 a_4}{a_5}\right) z^3
   + \frac{2}{a_5^2} 
     \left(f_0 + 3 \left(\frac{a_3}{a_5}\right)^2 
           - \left(\frac{2}{a_5}\right)^2 
                \left(a_2 - 2 \frac{a_3 a_4}{a_5}\right)
     \right) z^4,  \nonumber
\end{equation}
the defining equation of local geometry (\ref{eq:defeq}) 
can be rewritten as 
\begin{equation}
 \tilde{y}^2 =
  \frac{a_5^2}{4} \tilde{x}^2 
+ \frac{P^{(5)}}{a_5^2} z^5 - \frac{R^{(5)}}{a_5^2} z^6 +{\cal O}(z^7)
 + {\cal O}(\tilde{x}^3)  + {\cal O}(z) \tilde{x}^2
 + {\cal O}(z^5)\tilde{x} + {\cal O}(z^8).
\label{eq:A6-def-B}
\end{equation}
By zooming into the codimension-3 singularity with a scaling 
\begin{equation}
 (\tilde{x},\tilde{y},z) = (\lambda^{7/2} \tilde{x}_0, 
  \lambda^{7/2} \tilde{y}_0, \lambda^7 z_0 ) \qquad \lambda \rightarrow 0,
\end{equation}
we see that the last four terms of (\ref{eq:A6-def-B}) become
irrelevant. We have  an $A_6$ singularity deformed by parameters 
$P^{(5)}$ and $R^{(5)}$ to $A_4$, which is the $N=5$ case of 
the deformations of $A_{N+1}$ to $A_{N-1}$ discussed by us 
in section \ref{sec:warm-up}.
Therefore, the local geometry of type (c1) codimension-3 singularities 
can be studied with an  $\SU(7)$ gauge theory.\footnote{The group
$\SU(7)$ was already identified in \cite{DW-2}.} $R^{(5)}$ and $P^{(5)}$ play 
the role of the variables $s_1$ and $s_2$ in section \ref{sec:warm-up}.

For a generic choice of complex structure moduli of a compact 
Calabi--Yau 4-fold, we do not expect that a common zero of 
$P^{(5)}$ and $R^{(5)}$ is a double point singularity of 
$P^{(5)} = 0$ away from $a_5 = a_3 = 0$ locus. Thus, for such 
generic unfolding of $A_6$ singularity to $A_4$, only one 
smooth matter curve $P^{(5)} = 0$ pass through a type (c1)
codimension-3 singularity, and the $\SU(7)$ gauge theory on $S$ 
should have a branch locus and a twist by Weyl reflection of 
$\SU(2) \subset \SU(7)$ that commutes with unbroken $\SU(5)$.
Zero mode wavefunctions of multiplets in the $\SU(5)$-${\bf 5}$ 
and $\bar{\bf 5}$ representations are determined by a doublet 
vev of $\varphi$ as in (\ref{eq:bg-AB}), where now local coordinates 
$u_1$ and $u_2$ are set in directions where $R^{(5)}$ and $P^{(5)}$ 
increase. The matter wavefunctions are largely localized along 
the curve $P^{(5)} \sim u_2 = 0$.

The type (a) and type (c1) codimension-3 singularity points are
characterized in a totally different way. The former is where 
two matter curves for different $\SU(5)$ representations intersect, 
and the singularity is enhanced to the $E_n$ type. On the other hand, 
only one type of matter curve just passes through the type (c1) 
codimension-3 singularity points, and the singularity is enhanced 
only to $A_n$ type. The defining equations $P^{(5)} = 0$ and 
$R^{(5)} = 0$ involve very complicated functions of the parameters 
$a_{0,2,3,4,5}$. Yet surprisingly, the rank-2 $\varphi$ field vev 
for the matter in the 
$\SU(5)$-${\bf 10}+\overline{\bf 10}$-representations at the type (a) 
points and for the matter in the $\SU(5)$-$\bar{\bf 5}+{\bf 5}$ 
representations at the type (c1) points are exactly the same. 
The spectral surfaces for these irreducible components in these 
local models are exactly the same, and so are the behavior of the 
zero-mode wavefunctions. 

\subsection{Codimension-3 Singularities in SO(10) GUT Models}
\label{ssec:SO(10)}

If only the global holomorphic section $a_5$ vanishes, 
and $a_{0,2,3,4}$ are generic, then we have a locus of 
$D_5$ singularity. The discriminant is given by\footnote{
A few typos in \cite{HayashiEtAl} are corrected.} 
\begin{equation}
 \Delta = z^7 \left(a_4^3 a_3^2 + 
   z \left(\frac{27}{16} a_3^4 - \frac{9}{2} a_3^2 a_2 a_4
             - a_4^2 (a_2^2 - 4 a_0 a_4) \right)
   + z^2 \left(4 a_4^3 y_*^2 + {\cal O}(a_3, (a_2^2 - 4 a_4 a_0))
	 \right)
  \right).
\label{eq:Det-E5}
\end{equation}
Here, $y_*^2 = x_*^3 + f_0 x_* + g_0$ and $x_* = - (a_2/(2a_4))$.
The singularity is $D_5$ along the codimension-1 $z = 0$ locus, and 
$\Delta \sim {\cal O}(z^7)$. 
Along codimension-2 loci $a_4 = 0$ ($\bar{c}_V$) and $a_3 = 0$ 
($\bar{c}_{\wedge^2 V}$), $\Delta \sim {\cal O}(z^8)$, and 
the singularity is enhanced to $E_6$ and $D_6$, respectively.

In F-theory compactification that leaves $\SO(10)$ gauge symmetry, 
there are two types of codimension-3 singularity.
Here is a list \cite{AC, HayashiEtAl}:\footnote{In the traditional 
literature on F-theory like \cite{6authors, Het-F-4D, AC}, sections 
$a_4$ and $a_3$ roughly correspond to sections $h$ and $q$ of the same 
line bundles. Reference \cite{HayashiEtAl} will be useful in figuring
out the precise relations between them including coefficients.}
\begin{itemize}
 \item type (a): common zero of $a_4$ and $a_3$, 
 \item type (c): common zero of $a_3$ and $(a_2^2 - 4 a_4 a_0)$.
\end{itemize}
Two matter curves $a_4 = 0$ for ${\bf 16}+\overline{\bf 16}$
representations and $a_3 = 0$ for ${\bf vect.}$ representation 
pass through the type (a) codimension-3 singularities. On the 
other hand, only the $a_3 = 0$ curve passes through the type (c) 
points.

In this subsection, we will see that local geometry of type (a) 
and type (c) codimension-3 singularities are approximated by 
deformation of $E_7$ and $D_7$ singularity, respectively, to $D_5$.
Field theory of $E_7$ and $\SO(14)$ gauge group can be used for 
the field-theory local model of these codimension-3 singularities. 
The two types of codimension-3 singularity exhaust both possibilities 
of rank-2 deformation of $A$--$D$--$E$ singularity to $D_5$.

\subsubsection{Type (a) Singularity: $E_{7}\rightarrow D_{5}$}
\label{sssec:SO(10)-E7}

From the algebra of topological 2-cycles that vanish simultaneously 
at the type (a) codimension-3 singularities of SO(10) GUT models of 
F-theory, it is believed that the Yukawa couplings of the form 
\begin{equation}
 \Delta W = \lambda {\bf 16} \; {\bf 16}\; {\bf vect}
\label{eq:1616vect}
\end{equation}
are generated. With a field theory local model of this singularity, 
one will be able to calculate the Yukawa couplings of this form. 

Let us first see that rank-2 deformation of $E_7$ 
can be a good approximation of local geometry of type (a) 
codimension-3 singularity. An equation of $E_7$ surface singularity 
and its most generic rank-7 deformation is given by 
\begin{equation}
Y^{2}=Z^{3}-16XZ^{3}-\epsilon_{2}X^{2}Z-\epsilon_{6}Z^{2}-\epsilon_{8}XZ
-\epsilon_{10}Z^{2}-\epsilon_{12}X-\epsilon_{14}Z-\epsilon_{18},
\label{eq:def-E7-A}
\end{equation}
where $\epsilon_{i}$ ($i=2,6,8,10,12,14,18$) are deformation 
parameters. Those deformation parameters are homogeneous functions 
of degree $i$ of 7-dimensional space $\mathfrak{h} \otimes_\R \C$, 
where $\mathfrak{h}$ is the Cartan subalgebra of $\mathfrak{e}_7$.
Explicit expressions for $\epsilon_i$ are found in \cite{KM}.

We are not interested in full deformation of $E_7$ singularity, but 
in a partial deformation that leaves $D_5$ singularity unresolved, 
so that such geometry can be used for SO(10) GUT models. We thus 
start off with identifying an appropriate rank-2 subspace of 
the complexified Cartan subalgebra $\mathfrak{h} \otimes_\R \C$ of 
$\mathfrak{e}_7$, just like we did in section \ref{sssec:SU(5)-E6}
for the deformation of $E_6$ to $A_4$. The 7-dimensional space 
$\mathfrak{h} \otimes_\R \C$ is identified with the orthogonal 
complement of $k \equiv -3 e_0 + (e_1 + \cdots + e_7)$ in an 
8-dimensional space $H = \xi_0 e_0 + \sum_{i=1}^7 \xi_i e_i$ 
($\xi_0, \xi_i \in \C$) under an intersection form 
$\diag (+1,-1,\cdots,-1)$ in the $\{e_0,\cdots, e_7\}$ basis.
The 7-dimensional subspace is parametrized by $t_i = \xi_0/3 + \xi_i$ 
($i=1,\cdots,7$) just like in (\ref{eq:dP-parametrizationA},
\ref{eq:dP-parametrizationB}). The root space $\mathfrak{h}^*$ 
is also identified with $\mathfrak{h}$ using the intersection 
form, and the simple roots of $E_7$ can be chosen as 
\begin{equation}
v_{i}=e_{i}-e_{i+1}\;(i=1, \cdots, 6), \;\;\;\;v_{0}=e_{0}-(e_{1}+e_{2}+e_{3}).
\end{equation}
The maximal root becomes $v_{\theta}=2e_{0}-(e_{2}+\cdots+e_{7})$. 

The deformation parameter in the direction 
\begin{equation}
 H_1 \sigma = (-3 e_0 + (e_1 + \cdots + e_6) + 3 e_7) \sigma \qquad 
(\sigma \in \C)
\end{equation}
leaves $E_6$ singularity undeformed, because 
$\vev{H_1,v_{0,1,\cdots,5}} = 0$. 
In other words, $\varphi$ field vev specified in this direction 
does not break $E_6$ symmetry. This choice of $H_1$ 
corresponds to 
%
%
\begin{equation}
\vec{t}_{1}=(0, 0, 0, 0, 0, 0, 2\sigma).
\end{equation}
Another direction 
\begin{equation}
H_2 \tau = (-6 e_0 + 3(e_2 + \cdots + e_7))\tau \qquad (\tau \in \C)
\end{equation}
leaves $D_6$ singularity undeformed, because 
$\vev{H_2,v_{0,2,\cdots,6}} = 0$. This choice corresponds to 
\begin{equation}
\vec{t}_{2}=(-2, 1, 1, 1, 1, 1, 1)\tau.
\end{equation}
The common subset of the $E_6$ and $D_6$ symmetries is 
$D_5 = \SO(10)$ generated by simple roots $v_0, v_{2,\cdots,5}$.
\begin{figure}[t]
 \begin{center}
   \includegraphics[width=.6\linewidth]{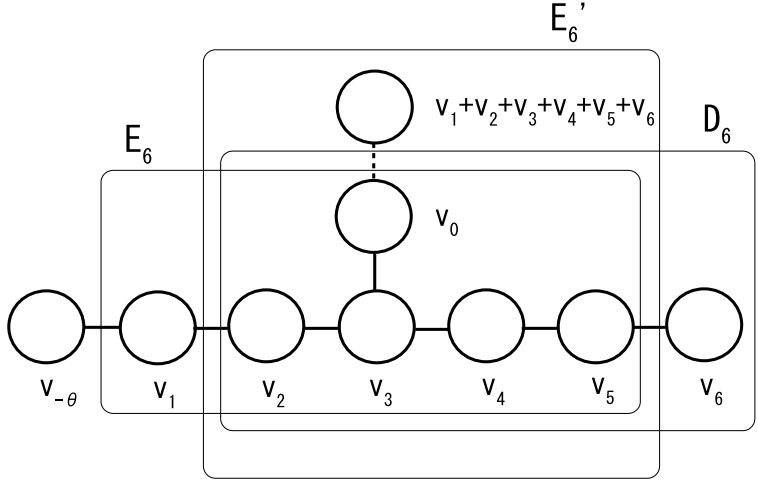}  
\caption{The Dynkin diagram which describes the deformation of $E_{7}$ singularity to $E_{6}\;(\tau=0), D_{6}\;(\sigma=0)$ and $E_{6}^{\prime}\;(2\sigma+3\tau=0)$.}
\label{fig:E7}
 \end{center}
\end{figure}

Thus, the $\varphi$ field vev can be oriented in an arbitrary 
linear combination
of the directions specified by $\vec{\xi}_1$ and $\vec{\xi}_2$.
Therefore, the rank-2 subspace of $\mathfrak{h} \otimes_\R \C$ 
for SO(10) GUT models is given by 
\begin{equation}
\vec{t}_{1+2}=(-2\tau, \tau, \tau, \tau, \tau, \tau, 2\sigma+\tau),
\label{eq:cartan}
\end{equation}
and the deformation parameters $\epsilon_i$'s in (\ref{eq:def-E7-A}) 
are calculated for $t_i$ of this form.

Within the rank-2 subspace of deformation, $\sigma = 0$ is the condition 
for unbroken $D_6 = \SO(12)$ symmetry. Although $E_6$ symmetry is
enhanced for $\tau = 0$, there is another 1-dimensional subspace for 
unbroken $E_6$, namely, a linear combination satisfying 
\begin{equation}
 2\sigma + 3\tau = 0.
\end{equation}
This is because $H' \propto (-e_0 -e_1+(e_2 + \cdots + e_6)-e_7)$ 
in this case, and $\vev{H', v_1+\cdots + v_6}= 0$ as well as
$\vev{H',v_{0,2,\cdots,5}}=0$. It is not hard to see that the
intersection form of $v_{0,2,\cdots,5}$ and $v_1+\cdots+v_6$ is 
equal to the Cartan matrix of $E_6$.
If we are to consider a geometry of F-theory compactification where 
$\sigma$ and $\tau$ vary linearly on the local coordinates $(u_1,u_2)$ 
of $S$, then three matter curves on $S$ intersect at the type (a) 
codimension-3 singularity $(\sigma,\tau) = (0,0)$, where the singularity 
is enhanced to $E_7$. $\sigma = 0$ is the curve for $\SO(10)$-{\bf
vect.} representation, and both $\tau=0$ and $(2\sigma+3\tau)=0$ curves 
form the matter curve for spinor representations of $\SO(10)$ \cite{BHV-1}.

The Cartan vev of $\varphi$-field 
\begin{equation}
 2 \alpha \vev{\varphi}_{12} = H_1 \sigma + H_2 \tau
\label{eq:vev-phi-E7}
\end{equation}
does not have to be linear around 
the codimension-3 singularity points as we have already seen 
such examples in the precedent sections. The most generic
characterization of type (a) codimension-3 singularity point is 
the common zero of sections of certain bundles $a_4$ and $a_3$.
Generically, curves $a_4 = 0$ or $a_3 = 0$ do not have singularity 
at their intersection, and $(\tilde{a}_4, \tilde{a}_3) \equiv 
(a_4/a_2, a_3/a_2)$ can be chosen as a set of local coordinates around 
a generic type (a) codimension-3 singularity point. Thus, we will now 
rewrite (\ref{eq:defeq}) and compare it with (\ref{eq:def-E7-A}), so
that we can determine the most generic behavior of the Cartan vev 
parameters $\sigma$ and $\tau$ as functions of local coordinates 
$\tilde{a}_4$ and $\tilde{a}_3$.

Using a new set of coordinates 
\begin{eqnarray}
 \tilde{x} & = & \frac{(-16)^2}{a_2^2}\left(x+\frac{1}{3}a_4 z \right), \\
 \tilde{y} & = & \frac{(-16)^3}{a_2^3}\left(y -\frac{1}{2}a_3 z^2 \right), \\
 \tilde{z} & = & \frac{-16}{a_2} z,
\end{eqnarray}
we find that the equation of local geometry (\ref{eq:defeq}) is
approximately 
\begin{equation}
 \tilde{y}^2  \simeq  \tilde{x}^{3}
 + \left( - 16 \tilde{z}^{3} 
          - \frac{(-16)^{2}}{3}\tilde{a}_{4}^{2}\tilde{z}^{2}
   \right) \tilde{x}
 + \left(   \frac{(-16)^{2}}{4}\tilde{a}_{3}^{2}
          - \frac{16^{2}}{3}\tilde{a}_{4} \right) \tilde{z}^{4}
 + \frac{2(-16)^{3}}{27}\tilde{a}_{4}^{3}\tilde{z}^{3}.
\label{eq:def-E7-B}
\end{equation}
Here, we have dropped higher-order terms that become irrelevant in 
zooming up to the codimension-3 singularity through 
\begin{equation}
 (\tilde{x},\tilde{y},\tilde{z},\tilde{a}_3, \tilde{a}_4) = 
  (\lambda^6 \tilde{x}_0, \lambda^9 \tilde{y}_0, \lambda^4 \tilde{z}_0,
   \lambda \tilde{a}_{3;0}, \lambda^2 \tilde{a}_{4;0})  \qquad 
  \lambda \rightarrow 0.
\end{equation}
When the standard form of deformation (\ref{eq:def-E7-A}) is rewritten 
with a new set of coordinate $(X,Y,Z')$ with 
\begin{equation}
 Z' = Z + \sigma^2 (\sigma^{2}+4\sigma \tau+6\tau^{2}),
\end{equation}
one notices that (\ref{eq:def-E7-A}) and (\ref{eq:def-E7-B}) 
look exactly the same. 
Identifying the two sets of local coordinates as 
$(\tilde{x},\tilde{y},\tilde{z}) = (c^6 X, c^9 Y, c^4 Z')$, 
we find that the Cartan vev parameters $(\sigma,\tau)$ of 
deformation of $E_7$ to $D_5$ correspond to  
\begin{equation}
\left( 2 \sigma \times \left(\frac{c}{2i}\right), \;  
     - 3 \tau (2\sigma+3\tau) \times \left(\frac{c}{2i}\right)^2 
\right) = 
\left( \tilde{a}_3, \tilde{a}_4 \right).
\end{equation}
We will set $c = 2i$.
$\sigma$ always depends linearly on the local coordinates 
$(\tilde{a}_3, \tilde{a}_4)$, but there is no guarantee that 
$\tau$ is linear in the local coordinates in general. Only 
when the curve $a_4 = 0$ has a double point singularity at 
the codimension-3 singularity point, the two branches of 
the $a_4 = 0$ curve are identified with $\tau=0$ and 
$(2\sigma+3\tau)=0$ and all of $\sigma, \tau$ and $(2\sigma+3\tau)$ 
become the normal coordinates of the three curves passing through 
the codimension-3 singularity point. This happens only for 
special choice of complex structure moduli of F-theory 
compactification.

Now, we are ready to determine the $\varphi$ field vev of the 
field-theory local model of this type of codimension-3 singularity.
For a given complex structure of a Calabi--Yau 4-fold, $a_{0,2,3,4}$
etc. in (\ref{eq:defeq}) are determined, and from this data, 
we can determine $\vev{\varphi}$ through 
\begin{eqnarray}
3\tau & = & - (\tilde{a}_3/2) + \sqrt{(\tilde{a}_3/2)^2 - \tilde{a}_4},
 \\
-(2\sigma + 3\tau) & = & 
 - (\tilde{a}_3/2) - \sqrt{(\tilde{a}_3/2)^2 - \tilde{a}_4}, \\
 2\sigma & = & \tilde{a}_3.
\end{eqnarray}
This field-theory local model has an $E_7$ gauge group, and 
the structure group of the field vev is 
\begin{equation}
  \U(2) = \SU(2) \times \U(1) \subset \SU(2) \times \SO(12) \subset E_7. 
\end{equation}
For a generic complex structure, $\sqrt{(\tilde{a}_3/2)^2 - \tilde{a}_4}$
in the Cartan vev parameters above introduces branch locus 
$\tilde{a}_3^2 - 4 \tilde{a}_4 = 0$ and a branch cut. 
We will see below that the $E_7$-{\bf adj.} fields are twisted 
by the $\SU(2)$ Weyl group reflection at the branch cut in this local model.

The irreducible decomposition of $\mathfrak{e}_7$-{\bf adj.}
representation under $\mathfrak{su}(2) + \mathfrak{u}(1) 
+ \mathfrak{so}(10)$ is 
\begin{eqnarray}
{\rm Res}^{E_7}_{\SU(2) \times \U(1) \times \SO(10)} {\bf 133}
 & = & ({\bf 1}, {\bf 45})+({\bf 3}, {\bf 1})
 +({\bf 1}, {\bf 1})  \nonumber \\
 & & + ({\bf 2},{\bf 16}) + (\bar{\bf 2}, \overline{\bf 16})
   + (\wedge^2 {\bf 2},{\bf vec.}) + (\wedge^2 \bar{\bf 2},{\bf vec.}).
\end{eqnarray}
We have worked out in Table \ref{tb:E7} how the background field 
configuration $\vev{\varphi}$ enters into the zero-mode equation 
(\ref{eq:EOM1}--\ref{eq:EOM4}) of each irreducible component.
\begin{table}[t]
\begin{center}
\begin{tabular}{|c|c|c|}
\hline
irr. comp.& roots $\omega$ & ${\rm ad}(H_1\sigma + H_2 \tau)$ on $\omega$\\
\hline
$({\bf 2},{\bf 16})$ & $-e_{i}+e_{1}$, $e_{0}-e_{i}-e_{j}-e_{k}$, $2e_{0}-(e_{1}+\cdots+e_{6})$& $3\tau$\\
& $-2e_{0}+(e_{1}+\cdots+e_{7})-e_{i}$, $-e_{0}+e_{i}+e_{j}+e_{7}$, $-e_{1}+e_{7}$&$-(2\sigma+3\tau)$\\
\hline
$(\bar{\bf 2},\overline{\bf 16})$ 
& $e_{i}-e_{1}$, $-e_{0}+e_{i}+e_{j}+e_{k}$, $-2e_{0}+(e_{1}+\cdots+e_{6})$& $-3\tau$\\
& $2e_{0}-(e_{1}+\cdots+e_{7})+e_{i}$, $+e_{0}-e_{i}-e_{j}-e_{7}$, $+e_{1}-e_{7}$&$(2\sigma+3\tau)$\\
\hline
$(\wedge^2 \bar{\bf 2},{\bf vec.}_+)$ 
&$e_{i}-e_{7}$, $e_{0}-e_{i}-e_{1}-e_{7}$&$2\sigma$\\
$(\wedge^2 {\bf 2},{\bf vec.}_-)$ &$-e_{i}+e_{7}$, $-e_{0}+e_{i}+e_{1}+e_{7}$&$-2\sigma$\\
\hline
\end{tabular}
\caption{\label{tb:E7}
The roots and U(1) charge of the irreducible representation from the decomposition of $E_{7}$, where $i, j, k=2,\cdots, 6$.
}
\end{center}
\end{table}
Eigenvalues are given by 
\begin{equation}
 \rho_{U={\bf 2}}(2\alpha\vev{\varphi_{12}}) = \diag(3\tau, -
  (2\sigma+3\tau)), \qquad 
 \rho_{U=\bar{\bf 2}}(2\alpha\vev{\varphi_{12}}) = \diag(-3\tau, 
 (2\sigma+3\tau))
\end{equation}
for the $({\bf 2},{\bf 16})$ and $(\bar{\bf 2},\overline{\bf 16})$ 
components, and 
\begin{equation}
 \rho_{\wedge^2 {\bf 2}}(2\alpha \vev{\varphi_{12}}) = - 2\sigma, \qquad 
 \rho_{\wedge^2 \bar{\bf 2}}(2\alpha \vev{\varphi_{12}}) = 2\sigma
\end{equation}
for the $(\wedge^2 {\bf 2},{\bf vect.}_-)$ and 
$(\wedge^2 \bar{\bf 2},{\bf vect.}_+)$ irreducible components.

The $\varphi$ field background for the $({\bf 2},{\bf 16})$ component 
is exactly the same as the one for the $({\bf 2}, {\bf 10})$ component 
at the type (a) points in the deformation of $E_6$ to $A_4 = \SU(5)$, 
and as the one for the $({\bf 2}, \bar{N})$ component 
in the deformation of $A_{N+1}$ to $A_{N-1}$. The local coordinates 
$(\tilde{a}_3, \tilde{a}_4)$ in this local model correspond to 
$(\tilde{a}_4, \tilde{a}_5)$ and $(2u_1,u_2) = (s_1,s_2)$ in those local
models, respectively.
We thus know the profile of the zero-mode wavefunctions of 
matter in the $\SO(10)$-{\bf 16} representation. We will not repeat it 
here. 

The $\varphi$ field background for either one of the irreducible
component in the $\SO(10)$-{\bf vect} representation is linear 
in the local coordinates. Thus, their wavefunctions are Gaussian 
in the direction of $\tilde{a}_3$, the normal direction of the matter 
curve $\tilde{a}_3 = 0$. 

In sections \ref{sec:warm-up} and \ref{ssec:SU(5)}, we introduced 
a covering surface $C_{U_i}$ of $S$ for each irreducible component 
$(U_i, R_i)$. It was a surface that eigenvalues of 
$\rho_{U_i}(\vev{\varphi})$ scan over $S$. 
Zero mode wavefunctions of matter multiplets in representation $R_i$ 
become smooth and single-valued on the spectral surface $C_{U_i}$, and 
the zero-modes themselves are characterized (locally) as holomorphic 
functions on $C_{U_i}$ modulo those that vanish on the $\xi = 0$ locus 
(matter curve).
The defining equation for the $({\bf 2},{\bf 16})$ component is 
\begin{equation}
  0 =  a_2 {\rm det} \left(\xi - \diag( 3\tau, -(2\sigma+3\tau))\right)
  = a_2 (\xi^2 + \tilde{a}_3 \xi + \tilde{a}_4)
  = a_2 \xi^2 + a_3 \xi + a_4.
\label{eq:C2-E7-D5}
\end{equation}
$\pi_{C_{{\bf 2}}}: C_{{\bf 2}} \rightarrow S$ is a degree-2 cover, and 
becomes a branched cover at $\tilde{a}_3^2 - 4 \tilde{a}_4 = 0$. 
This spectral surface behaves exactly the same as (\ref{eq:C-E6-A4}), 
when $a_{2,3,4}$ are replaced by $a_{3,4,5}$. Thus, 
Figure~\ref{fig:C-E67-a}~(ib) shows the spectral surface of 
the $({\bf 2},{\bf 16})$ component. 
The spectral surface of the $(\wedge^2 \bar{\bf 2},{\bf vect})$
component is 
\begin{equation}
 0 =  \xi + (2\sigma) = \xi + \tilde{a}_3.
\label{eq:C1-E7-D5}
\end{equation}
$\pi_{C_{\wedge^2 \bar{\bf 2}}}: C_{\wedge^2 \bar{\bf 2}} \rightarrow S$ 
is a one-to-one correspondence, and looks as in Figure~\ref{fig:C-E67-a}~(ii).

\subsubsection{Type (c) Singularity: $D_{7}\rightarrow D_{5}$}
\label{sssec:SO(10)-D7}

The type (c) codimension-3 singularity is characterized by $a_3 = 0$ 
and $(a_2^2 - 4 a_4 a_0)= 0$ in (\ref{eq:defeq}). What would the field 
theory local model be for the local geometry around this type of 
codimension-3 singularity? The most naive guess (and the most minimal
choice, if possible) is to take the gauge group larger than 
the unbroken $D_5 = \SO(10)$ by rank 2. In the following, we show 
indeed that the local geometry is approximately regarded as 
a family of deformed $D_7$ surface singularity.

The most general form of local equation of $D_n$ singularity and 
its deformation are given in (\ref{eq:Dn-def-A}--\ref{eq:symmetric-Dn}), 
and we use the case of $n=7$ here. Since we are interested only 
in rank-2 deformation of $D_7$ singularity that leaves $\SO(10) = D_5$ 
symmetry unbroken, the deformation parameters (and the vev of 
$\varphi$) in the $\mathfrak{h} \otimes_\R \C$ should be of the following 
form:
\begin{equation}
H={\rm diag}(0^{5}, \tau_+, \tau_-, 0^{5}, -\tau_+, -\tau_-)
\label{eq:cartan_D7}
\end{equation}
for some complex numbers ($\C$-valued functions) $\tau_+$ and $\tau_-$.
Thus, (\ref{eq:Dn-def-A}) becomes 
\begin{equation}
Y^{2} = -X^{2}Z+Z^{6}+(\tau_+^2 + \tau_-^2) Z^{5}+(\tau_+^2 \tau_-^2)Z^{4}.
\label{eq:D7_1}
\end{equation}

Certainly we can think of local geometry where $\tau_+$ and $\tau_-$ 
behave linearly on the local coordinates around type (c) codimension-3 
singularity points, but we have already seen that such configuration 
is not necessarily the most generic one. The equation of local geometry 
(\ref{eq:defeq}) with vanishing $a_3$ and $(a_2^2 - 4 a_4 a_0)$ 
at some point is the most generic definition of the type (c) codimension-3 
singularity. To see the relation between (\ref{eq:defeq}) and 
(\ref{eq:D7_1}), we introduce a new set of local coordinates 
\begin{equation}
 \tilde{x} \equiv \frac{1}{a_4^2}\frac{1}{y_*}
  \left(x + \frac{a_2}{2a_4}z^2\right), \quad 
 \tilde{y} \equiv \frac{1}{a_4^3}\frac{1}{y_*}
  \left(y - \frac{1}{2} a_3 z^2 \right), \quad 
 \tilde{z} \equiv - \frac{z}{a_4},
\end{equation}
where $(y_*^2) = x_*^3 + f_0 x_* + g_0$, and $x_* = - a_2 /(2a_4)$.
In this new coordinates, (\ref{eq:defeq}) becomes 
\begin{equation}
 \tilde{y}^2 \simeq - \tilde{z}\tilde{x}^2 + \tilde{z}^6 
   + \frac{\tilde{D}}{y_*^2} \tilde{z}^5 + \frac{(\tilde{a}_3/2)^2}{y_*^2},
\label{eq:D7_2}
\end{equation}
where higher-order terms in the scaling of 
\begin{equation}
 \left( \tilde{x}, \tilde{y}, \tilde{z}, \tilde{D}, \tilde{a}_3 \right)
 = \left(\lambda^5 \tilde{x}_0, \lambda^6 \tilde{y}_0, \lambda^2
    \tilde{z}_0, \lambda^2 \tilde{D}_0, \lambda^2
    \tilde{a}_{3,0}\right), \qquad \lambda \rightarrow 0
\label{eq:D7-scaling}
\end{equation}
are ignored because they become irrelevant to the local geometry 
close to the codimension-3 singularity. 
Now (\ref{eq:D7_1}) and (\ref{eq:D7_2}) have exactly the same form, 
and this proves that the local geometry around the type (c)
codimension-3 singularities is approximately a family of $D_7$ surface 
singularity deformed to $D_5$.

Comparing (\ref{eq:D7_1}) with (\ref{eq:D7_2}), we find the relation 
between $(\tilde{a}_3, \tilde{D})$ and $(\tau_+, \tau_-)$:
\begin{equation}
 \left(c^2 (\tau_+^2 + \tau_-^2), c^4 \tau_+^2 \tau_-^2 \right) = 
 \left(\frac{\tilde{D}}{y_*^2}, \frac{(\tilde{a}_3/2)^2}{y_*^2}\right).
\end{equation}
For general type (c) codimension-3 singularity points, the  zero loci of 
$\tilde{a}_3$ and $\tilde{D}$ intersect transversely, and they can be 
chosen as local coordinates. The Cartan vev of $\varphi$ can be solved 
in terms of local coordinates $(\tilde{a}_3, \tilde{D})$, and 
\begin{eqnarray}
 \pm \tau_+ & = & \pm \left(
  \frac{ \sqrt{ \tilde{D} + \tilde{a}_3 y_*} +
         \sqrt{ \tilde{D} - \tilde{a}_3 y_*} 
       }{2y_*} \right), \\
 \pm \tau_- & = & \pm \left(
  \frac{ \sqrt{ \tilde{D} + \tilde{a}_3 y_*} -
         \sqrt{ \tilde{D} - \tilde{a}_3 y_*} 
       }{2y_*} \right). 
\end{eqnarray}

The Cartan vev of $\varphi$ is in the $\mathfrak{so}(4)$ subalgebra of 
$\mathfrak{so}(14)$. Thus, the charged matter multiplets arise from the 
off-diagonal components of the irreducible decomposition
\begin{equation}
 \mathfrak{so}(14)\mbox{-}{\bf adj.} \rightarrow
  \mathfrak{so}(4)\mbox{-}{\bf adj.} + \mathfrak{so}(10)\mbox{-}{\bf adj.}
 + ({\bf vect.}, {\bf vect.}). 
\end{equation}
Only the last one is the off-diagonal component. The zero-mode 
wavefunctions of matter multiplets in $\SO(10)$-${\bf vect.}$ 
representation are determined by 
\begin{equation}
 \rho_{U_i = {\bf vect.}}(2\alpha \vev{\varphi_{12}}) = 
\diag(+\tau_+, + \tau_-, - \tau_+, - \tau_-).
\end{equation}
Because of the expressions of $\tau_+$ and $\tau_-$ obtained above, 
the field theory local model of this codimension-3 singularity also 
involves branch cuts and twists by the Weyl group. 
Fields are twisted by a Weyl reflection associated with a root 
$(e_6-e_7)$ around the branch locus $\tilde{D}=\tilde{a}_3 y_*$, 
and by a reflection associated with a root $(e_6+e_7)$ around 
the other branch locus $\tilde{D}=-\tilde{a}_3 y_*$. 

The zero-mode wavefunctions of the $\SO(10)$-{\bf vect.} representation 
are localized along the $\tilde{a}_3 = 0$ curve, and mainly take values 
in the second and fourth components, where $\pm \tau_-$ act; apart from 
the codimension-3 singularity point at 
$(\tilde{a}_3, \tilde{D}) = (0,0)$, $\tau_-$ can be expanded in terms of 
$\tilde{a}_3 y_* / \tilde{D}$, and $\tau_-$ begins with a term linear 
in $\tilde{a}_3$ in this expansion. The wavefunction has a Gaussian
profile in the $\tilde{a}_3$ direction. It is not easy, however, to 
guess how the wavefunctions behave around the codimension-3 singularity 
point, because of very complicated behavior of the four eigenvalues, 
$\pm \tau_+$ and $\pm \tau_-$. 

We introduced spectral surfaces in earlier sections, which is a surface 
that eigenvalues of $\rho_{U_i}(2\alpha\vev{\varphi_{12}})$ scan. 
We found that the zero-mode wavefunctions become smooth and single
valued when expressed on the spectral surface, even when they do not 
when expressed on the base space $S$. The spectral surface 
$C_{\bf vect.}$ of this local model is given by 
\begin{equation}
 (\xi^2 - \tau_+^2) (\xi^2 - \tau_-^2) = 
 \xi^4 - \frac{\tilde{D}}{y_*^2} \xi^2 +
 \frac{(\tilde{a}_3/2)^2}{y_*^2}=0,
\label{eq:def-CV2}
\end{equation}
which is a divisor of $\mathbb{K}_S$ with coordinates 
$(\xi, \tilde{a}_3, \tilde{D})$.
\begin{figure}[t]
 \begin{center}
  \begin{tabular}{ccc}
  \includegraphics[width=.3\linewidth]{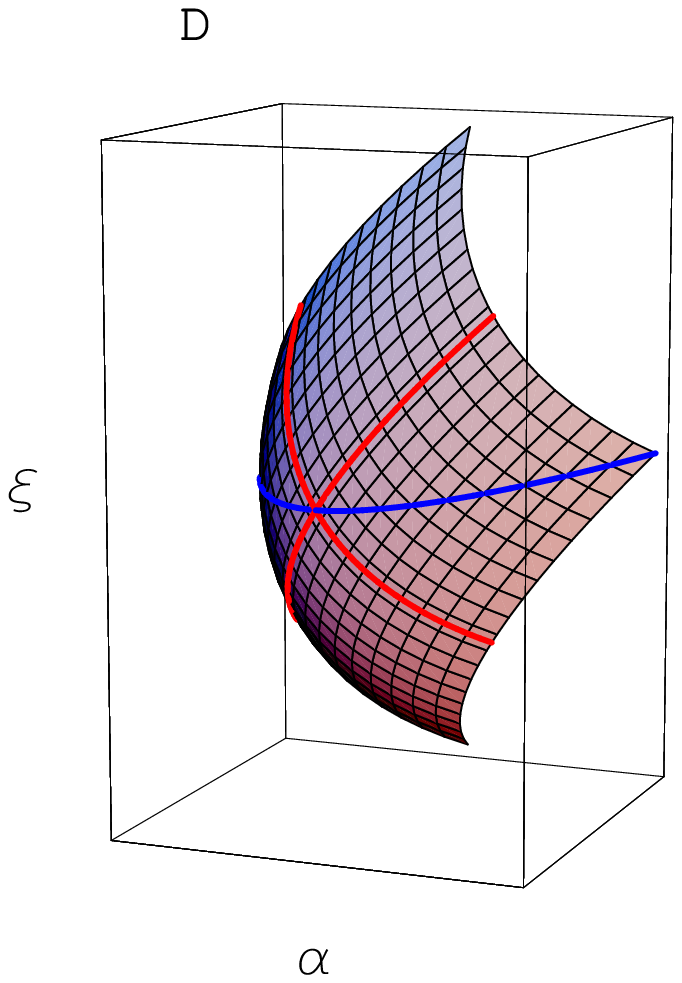}  &   & 
  \includegraphics[width=.3\linewidth]{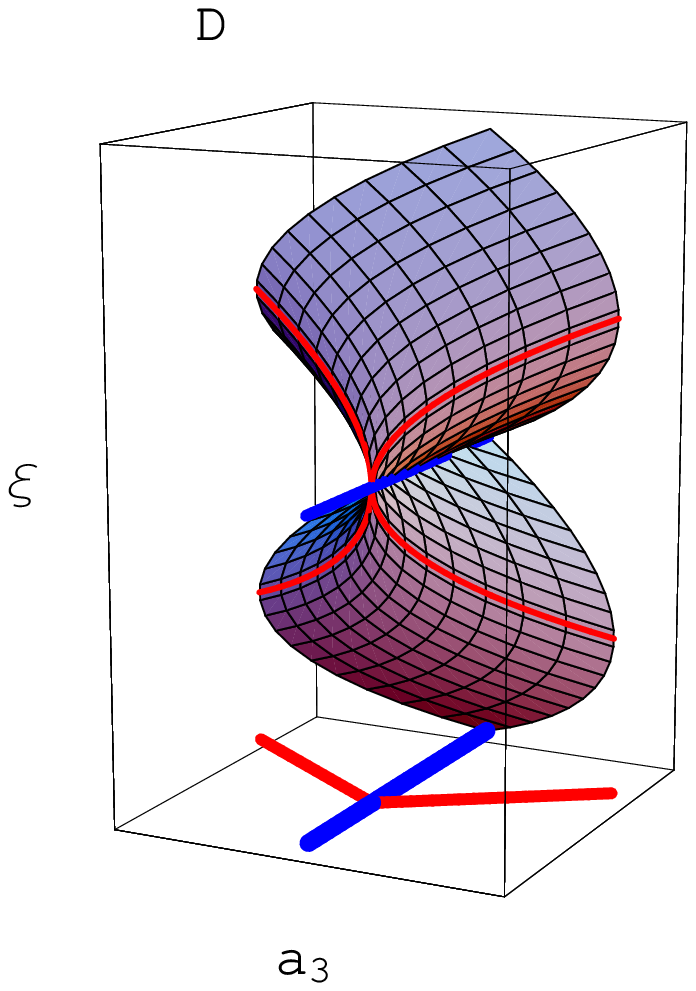}  \\
(a) & & (b)
  \end{tabular}
\caption{\label{fig:C2-D7} (color online) 
Spectral surface $C_{U={\bf vect.}}$ (b) and its desingularization 
$\widetilde{C}_{\bf vect.}$ (a). The coordinates $a$ and $D$ in the panel 
(b) means $\tilde{a}_3$ and $\tilde{D}$ in the text, respectively. 
The map (\ref{eq:resolve-pinch}) from $\widetilde{C}_{\bf vect}$ to 
$C_{\bf vect.}$ is given by $(\xi, \alpha, \tilde{D}) \mapsto 
(\xi , \tilde{a}_3, \tilde{D}) = (\xi, \xi \alpha, \tilde{D})$. 
Blue curves in the panel (b) are the matter curve of the 
$\SO(10)$-{\bf vect.} representation, and the blue curve in the panel 
(a) is their inverse image of the map. The grid curves on the surface 
$\widetilde{C}_{\bf vect.}$ correspond to the contours of $w_+$ and 
$w_-$ coordinates. The two red curves correspond to the $w_\pm = 0$
curves, the ramification divisor of 
$\pi_C \circ \nu: \widetilde{C}_{\bf vect.} \rightarrow S$. Their 
images in $C_{\bf vect.}$ are also shown in the panel (b), and it is 
clear that they are ramification divisors indeed. The fact that 
the {\it two} irreducible red curves intersect the blue curve in 
the panel (a) explains the coefficient $2 \times 1/2 = 1$ of 
the $\tilde{b}^{(c)}$ divisor in \cite{HayashiEtAl}.}
 \end{center}
\end{figure}
This spectral surface has a double-curve singularity and a pinch point, 
as in Figure~\ref{fig:C2-D7}. More natural extension of the idea 
in section \ref{sec:warm-up}, however, is to consider a space 
$\widetilde{C}_{{\bf vect.}}$ with a set of coordinate 
\begin{equation}
 (w_+, w_-) = \left(\sqrt{\tilde{D}+\tilde{a}_3 y_*}, 
     \sqrt{\tilde{D}-\tilde{a}_3 y_*} \right).
\end{equation}
To be more precise, the new space $\widetilde{C}_{\bf vect.}$ has 
a set of local coordinates $(w_+, w_-)$, and a map 
$\nu: \widetilde{C}_{\bf vect.} \rightarrow C_{\bf vect.}$ is given by
\begin{equation}
 \nu: (w_+,w_-) \mapsto (\xi,\tilde{a}_3,\tilde{D}) = 
  \left(\frac{w_+ + w_-}{2y_*}, \frac{w_+^2 - w_-^2}{2y_*}, 
    \frac{w_+^2 + w_-^2}{2} \right) \in C_{\bf vect.}.
\label{eq:resolve-pinch}
\end{equation}
This surface does not have a double-curve singularity or 
pinch point singularity. It is smooth, as in Figure~\ref{fig:C2-D7}~(a).
The map $\pi_C \circ \nu: \widetilde{C}_{\bf vect.} \rightarrow S$ 
is ramified at 
\begin{equation}
 r = {\rm div} (w_+ w_-).
\end{equation}
This ramification divisor on $\widetilde{C}_{\bf vect.}$ is 
projected onto the branch loci $\tilde{D} = \pm \tilde{a}_3 y_*$.
Repeating the same argument as in section~\ref{sec:warm-up}, we 
expect that the zero-mode solution of the $\SO(10)$-${\bf vect.}$ 
multiplet is obtained as a single-component field $\Psi$ on the 
desingularized spectral surface $\widetilde{C}_{\bf vect.}$, and 
that the zero-modes are in one-to-one correspondence with holomorphic 
functions on $\widetilde{C}_{\bf vect.}$ modulo those that vanish 
on the covering matter curve $\tilde{\bar{c}}_{\bf vect.}$, 
defined as the inverse image of $\xi= 0$ from $C_{\bf vect.}$, 
$\nu^* (\xi)=(w_+ + w_-)/(2y_*) = 0$.
It is very natural to consider that this covering matter curve 
$\tilde{\bar{c}}_{\bf vect.}$ is the same thing as the covering 
matter curve introduced in \cite{HayashiEtAl}. 
We will show in the next section that they are indeed the same object.

\subsection{Discriminant Locus and Spectral Surface}
\label{ssec:discr-spec}

Let us take a moment here to study relation between the discriminant 
locus and (desingularized) spectral surface. Behavior of spectral
surfaces around codimension-3 singularities has been analyzed in 
detail in the preceding sections. We will now take a brief look 
at the behavior of discriminant locus around codimension-3 
singularities. 

Discriminant locus is, by definition, where the elliptic fiber 
of F-theory compactification degenerates. Depending upon which 
1-cycle of the elliptic fiber shrinks, a label $(p,q)$ is assigned 
to a discriminant locus (remember the $\SL_2 \Z$ monodromy, though) \cite{Vafa-Het-F}.
 Degrees of freedom corresponding to various roots of 
$E_n$-${\bf adj.}$ representations are obtained as string junctions \cite{Gaberdiel:1997ud} 
.  When a pair of irreducible discriminant locus approach 
each other, therefore, an extra set of degrees of freedom corresponding 
to the junctions stretching between the two discriminant locus 
becomes massless, and charged matter fields are localized 
at the intersection---codimension-2 locus---called matter 
curves \cite{6authors}. 
Intuitive picture of D7-branes in Type IIB string theory almost 
seems to hold for the discriminant locus in F-theory, except 
that the gauge group (including its deformation) can be $E_n$ type.
To what extent is this intuition correct, when we look at the behavior 
of discriminant locus around codimension-3 singularities? 
To what extent is the converse true, or put differently, are there 
irreducible pieces of discriminant locus for individual matter curves 
(and matter multiplets) in different representations? We will see this 
in the following.\footnote{We will pay attention only to the local 
geometry of Calabi--Yau 4-fold around the (deformed) singularity locus, 
but we do not study how the overall elliptic fiber becomes around the 
codimension-3 singularities. This is because the geometry around the 
singularity locus is directly relevant to physics of gauge theory. 
} 

Let us begin with a codimension-3 singularity around which 
$A_{N+1}$ singularity surface is deformed to surface with $A_{N-1}$ 
singularity. Section~\ref{sec:warm-up} was dedicated for the 
study of this case. The non-$z=0$ component of the discriminant 
locus, denoted by $D'$ in (\ref{eq:SandOthers}), is locally 
given by $(z^2 + s_1 z + s_2) = 0$, and its behavior is shown 
in Figure~\ref{fig:Atype}. For the case with a complex structure 
given in (\ref{eq:typeA-caseA}), the discriminant locus $D'$ consists 
of two irreducible pieces, and both of them are interpreted as
D7-branes. Even in the case of generic complex structure, 
(\ref{eq:typeA-caseB}), $D'$ may be understood as a recombination of 
two D7-branes. (See \cite{BHV-1} for related discussion.) There is 
nothing particularly surprising or interesting, in the case $A_{N-1}$ 
type singularity is deformed to $A_{M-1}$ type. It is worth noting, 
the following, however. There is only one Hermitian conjugate pair 
of off-diagonal components in the irreducible decomposition of 
$\mathfrak{su}(N)$ under the $\SU(M)$ subgroup, which is 
a pair of $\SU(M)$ fundamental and anti-fundamental representations.
A spectral surface is associated with this pair of
representations.\footnote{The spectral surface of
$\SU(M)$-anti-fundamental representation is obtained by multiplying 
$(-1)$ to the coordinate $\xi$ of that of the $\SU(M)$-fundamental 
representation.} Local behavior of the discriminant locus $D'$ in $B_3$ 
(see Figure~\ref{fig:Atype}~(b)) and the spectral surface $C$ 
in $\mathbb{K}_S$ (see Figure~\ref{fig:C-An}) is exactly the same 
for this type of deformation.
 
\begin{figure}[t]
\begin{center}
 \begin{tabular}{ccc}
  \includegraphics[width=.27\linewidth]{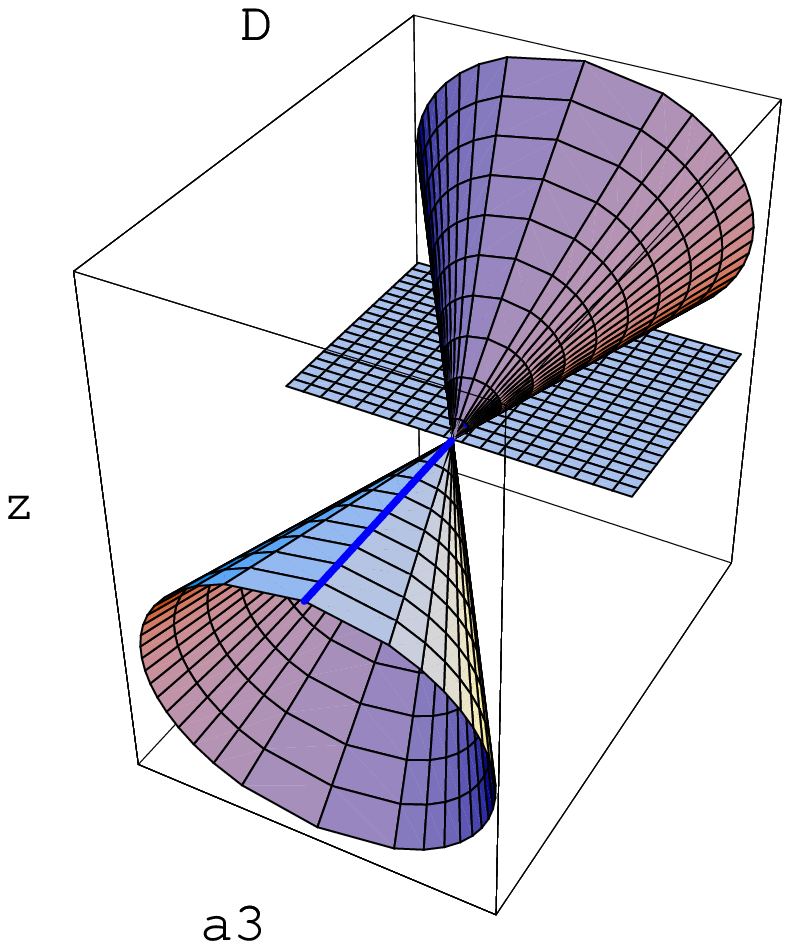}  
  & &
  \includegraphics[width=.38\linewidth]{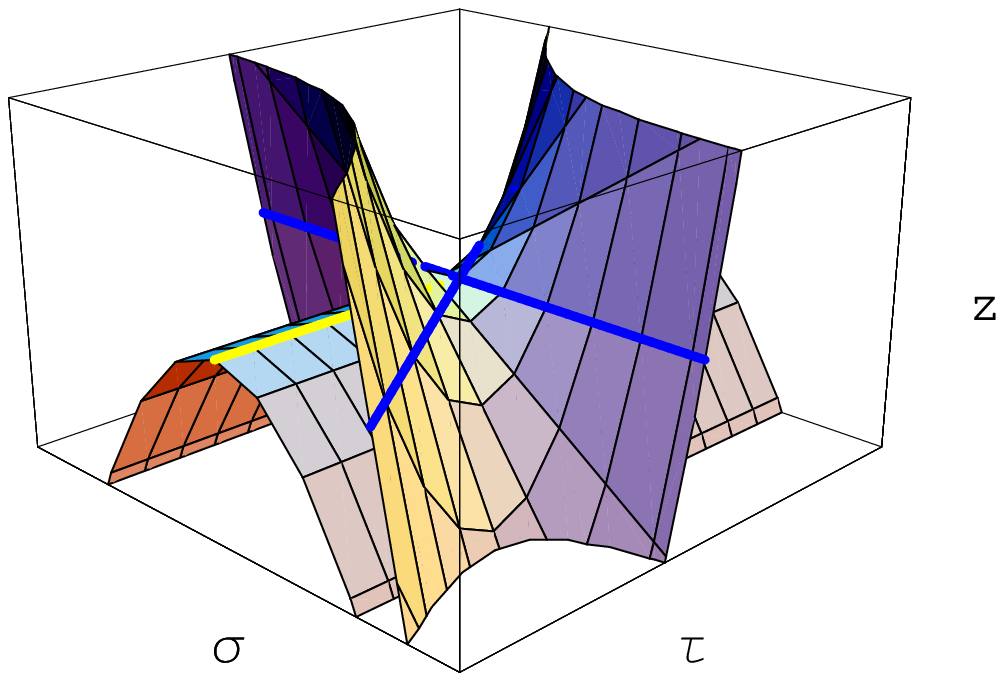}   \\
 (a) & & (b)
 \end{tabular}
\caption{\label{fig:discr-DD-DA} Discriminant locus $D'$ near codimension-3 
singularity points, where $D_{7}$ singularity is deformed to $D_5$ 
in the example (a), and $D_6$ singularity is deformed to $A_4$ 
in (b). The $z=0$ component $S$ is also shown partially in (a). 
The local coordinates $(\tilde{a}_3, \tilde{D})$ of (a) are the same as 
those of Figure~\ref{fig:C2-D7}~(b), and $(\sigma,\tau)$ of (b) are 
the same as those of Figures~\ref{fig:C-D6} and \ref{fig:C2-D6}.}
\end{center}
\end{figure}

We have also studied a codimension-3 singularity where $D_n$ type 
surface singularity is deformed to $D_m$; the type (c) codimension-3 
singularity points of SO(10) models is an example of this type of 
deformation, and we studied this case in section \ref{sssec:SO(10)-D7}.
Around the codimension-3 singularity of $D_7 \rightarrow D_5$
deformation, the non-$z=0$ component of the discriminant locus $D'$ 
is defined locally by 
\begin{equation}
 4 a_4^5 \left( y_*^2 \tilde{z}^2 - \tilde{D} \tilde{z} +
	  (\tilde{a}_3/2)^2 \right) \simeq 0,
\end{equation}
which was obtained from $\Delta/(z^7)$ of (\ref{eq:Det-E5})
by dropping the higher 
order terms under the scaling specified in (\ref{eq:D7-scaling}). 
$\tilde{z} \equiv z/a_4$.
Its behavior is shown in Figure~\ref{fig:discr-DD-DA}~(a).
The discriminant locus $D'$ in $B_3$ is irreducible around this 
codimension-3 singularity. At a generic point on the matter curve 
(codimension-2 singularity), the $z$ coordinate of $D'$ depends
quadratically on the normal coordinate of the matter curve 
(see Figure~\ref{fig:discr-DD-DA}~(a)). This behavior is due 
to orientifold projection, and is known since 1990's.

Just like in the case of deformation 
from $A_{N-1}$ to $A_{M-1}$, there is only one irreducible 
off-diagonal component---${\bf vect.}$ representation---in 
the deformation of $D_n$ to $D_m$. The defining equation of 
the spectral surface of this ${\bf vect.}$ representation was 
(\ref{eq:def-CV2}), and the defining equation of the discriminant 
locus $D'$ in $B_3$ and that of the spectral surface $C_{\bf vect.}$ 
in $\mathbb{K}_S$ are exactly the same, when $\tilde{z}$ is 
identified with $\xi^2$. Although there is no ``picture before
orientifold projection'' in generic F-theory compactification, 
the behavior of the spectral surface in Figure~\ref{fig:C2-D7}
gives an impression that the spectral surface can be regarded 
as such a picture (intersecting D7-branes) in generic F-theory 
compactification; the identification $\tilde{z} \leftrightarrow 
\xi^2$ now looks like the orientifold projection.

One begins to see more interesting phenomena when studying deformation 
of $D_n$ surface singularity to $A_{m-1}$. We already encountered this 
situation in section~\ref{sssec:SU(5)-D6}, where $D_6$ singularity is 
deformed to $A_4$ around the type (d) codimension-3 singularity of 
$\SU(5)$ models. In order to find out the defining equation of the 
$D'$ component of the discriminant locus around this codimension-3 
singularity, we drop higher-order terms of $\Delta/(z^5)$ 
of (\ref{eq:Det-E4}) under the scaling (\ref{eq:scale-E6-a-para}).
It is  
\begin{equation}
 - 4 a_4^7 \tilde{D} (\tilde{z}+\sigma^2)^2 (\tilde{z} - (2\sigma+\tau)\tau)
 = - 4 a_4^7 \tilde{D} \left(\tilde{z} + (\tilde{a}_5/2)^2 \right)^2
  (\tilde{z} - P^{(5)}/(4 a_4^3 \tilde{D})) = 0,
\label{eq:discr4D6}
\end{equation}
and factorizes. In the limit toward this codimension-3 singularity, 
the discriminant locus $D'$ splits into two irreducible pieces, 
$D'_{\rm asym}$ given by $(\tilde{z} + \sigma^2) = 0$ and 
$D'_{\rm fund}$ by $(\tilde{z} - (2\sigma + \tau)\tau) = 0$
Their behavior is shown in Figure~\ref{fig:discr-DD-DA}~(b). 
As a divisor, $D' = 2D'_{\rm asym} + D'_{\rm fund}$.
The matter curve $\tilde{a}_5 = 0$ for the anti-symmetric representation 
of $A_4 = \SU(5)$ gauge group is regarded as the intersection 
$S \cdot D'_{\rm asym}$, and the matter curve $P^{(5)} = 0$ for the 
fundamental representation as the intersection $S \cdot D'_{\rm fund}$.

There are three things to point out. 
First, there are three Hermitian conjugate 
pairs of off-diagonal irreducible components in the decomposition of 
$D_n$-{\bf adj.} representation. To each one of the pairs (and to each 
one of the irreducible pieces of matter curves) corresponds a spectral 
surface. The deformation of $D_n$ to $A_{m-1}$ has a Type IIB 
interpretation, as we already discussed in section~\ref{sssec:SU(5)-D6}, 
and the spectral surfaces carry information of the intersection of 
the two relevant D7-branes.

Secondly, we notice that the discriminant locus $D'_{\rm fund}$ 
is irreducible, whereas the corresponding matter curve $P^{(5)} = 0$ 
is split up into two irreducible pieces. Multiple irreducible 
components in the matter curve in $S$ does not imply at all 
that the discriminant locus $D'$ has multiple irreducible pieces.
On the other hand, there are corresponding irreducible pieces among 
the (desingularized) spectral surfaces.

Finally, the factorized form of the discriminant locus $D'$ 
in (\ref{eq:discr4D6}) was obtained only after the higher order 
terms in the scaling was dropped. The two irreducible pieces 
$D'_{\rm asym}$ and $D'_{\rm fund}$ touch each other along a curve 
(see Figure~\ref{fig:discr-DD-DA}~(b)), and they will reconnect to 
form a single irreducible piece $D'$ under even a small perturbation 
coming from higher order terms in the scaling. Thus, the discriminant 
locus $D'$ is a single irreducible piece for generic complex structure, 
even through the intersection $S \cdot D'$ sometimes have more than one 
irreducible components corresponding to matter curves of multiple 
representations.

At the end of the analysis in this section \ref{ssec:discr-spec}, 
let us look at the discriminant at the type (a) codimension-3 
singularity of SU(5) models, where $E_6$ singularity is deformed 
to $A_4$. The local defining equation is obtained just as we did 
above; we use the scaling (\ref{eq:scale-E6-a-para}) to drop higher 
order terms from (\ref{eq:Det-E4}). 
\begin{equation}
a_3^7 \left( \frac{27}{16}\tilde{z}^3 
    + \frac{\tilde{a}_4}{4} (4 \tilde{a}_4^2 - 9 \tilde{a}_5) \tilde{z}^2
    + \frac{1}{16}\tilde{a}_5^2 (8 \tilde{a}_4^2 - \tilde{a}_5)
    \tilde{z}
    + \frac{1}{16}\tilde{a}_4 \tilde{a}_5^4
      \right) = 0.
\end{equation}
We found that this does not factorize further into polynomials 
of local coordinates of $B_3$, that is 
$\tilde{z}$, $\tilde{a}_4$ and $\tilde{a}_5$. Thus, $D'$ remains
irreducible even when zooming in to the codimension-3 singularity point.
$D'$ does not split to irreducible pieces 
$D'_{(U_i,R_i) + (\bar{U}_i, \bar{R}_i)}$ for the two pairs of 
off-diagonal irreducible representations in (\ref{eq:e6-decomp}).
An intuition in Type IIB string theory that any charged matter
multiplets are in one-to-one correspondence with systems of
intersecting irreducible 7-branes breaks down here,\footnote{
Although we sometimes find in recent literature on F-theory that 
irreducible pieces of matter curves are identified with intersection 
of discriminant loci, this is not always true in F-theory.} 
if ``7-branes'' here mean $(p,q)$ 7-branes (discriminant locus) in $B_3$.
The spectral surfaces in $\mathbb{K}_S$, however, are defined separately 
for the pairs of off-diagonal irreducible components, and there remains 
a chance to extend the intuition in Type IIB string theory into generic 
F-theory compactification. This is done in the next section.

The irreducible nature of the $D'$ component of the discriminant locus 
is not a new discovery, in fact. Geometry of F-theory compactification 
was studied in 1990's, and $D'$ was often treated as an irreducible 
piece as a whole. There was no reason to think that it is reducible, 
to begin with, unless phenomenological constraints are introduced.
We have presented this detailed analysis of the behavior of discriminant 
locus around codimension-3 singularities, as a pedagogical introduction 
to the next section, where the spectral surfaces take center stage, 
not the discriminant locus.

\section{Higgs Bundle in F-Theory Compactification}
\label{sec:Higgs}

Massless charged chiral multiplets in low-energy effective 
theory are identified with {\it global} holomorphic sections 
of certain line bundles on (covering) matter 
{\it curves} \cite{Curio, DI, DW-1, BHV-1, HayashiEtAl}.
Using the standard techniques in algebraic geometry, one can 
determine the sections over the (covering) matter curves 
completely. It is impossible, however, to determine even 
the number of independent massless modes without looking 
at the entire compact curves. 
Field theory local models discussed in \cite{DW-1, BHV-1} and 
in earlier sections in this article may well be able to determine 
local behavior of zero-mode wavefunctions and calculate Yukawa 
couplings localized at codimension-3 singularities.  
But it is formulated on a non-compact base space $S$, 
and covers only a {\it local} region in a discriminant for 
the low-energy gauge group. Wavefunctions are defined on 
the {\it surface} $S$.
In order to combine both techniques and calculate Yukawa couplings 
of zero modes in the effective theory, therefore, one needs to 
figure out the relation between the two descriptions of matters. 
One of them provides a global description based on curves, and 
the other is for a local description on surfaces!

An important hint comes from our observation in 
section \ref{sec:warm-up}. Although zero-mode wavefunctions 
are often multi-component fields on a {\it surface} $S$, we found that they 
are better understood as single-component fields on (desingularized)
spectral cover surfaces $\widetilde{C}_{U_i}$. We also found that the 
zero-mode wavefunctions are ultimately determined by holomoprhic 
functions (sections) on the spectral surfaces modulo those 
that vanish on the (covering) matter curves. After taking 
the quotient, the remaining degrees of freedom are only the 
sections on the (covering) matter {\it curves}! This must be 
the way the two apparently different descriptions are compatible.

With this motivation in mind, we lay down in section 
\ref{ssec:Higgs-F} a firm foundation for describing charged matter 
multiplets in F-theory using sheaves on spectral covers, not 
on matter curves. We do find indeed that quotients of holomorphic 
sections of sheaves on spectral cover is the natural way to 
describe charged matters in F-theory [that is, (\ref{eq:matter-F-pre})], 
although in the end, this new description is equivalent to the 
one in \cite{HayashiEtAl}, where charged matters are sections on curves. 
Spectral surface has never been used in study of F-theory
compactification, but it turns out to be a key notion in 
generalizing such objects as D-branes and gauge bundles 
on them (or sheaves) in Calabi--Yau 3-fold in Type IIB 
compactification.

It is well-known that spectral surface is used in describing 
vector bundles in elliptic fibered compactification of Heterotic 
string theory \cite{FMW, Donagi-spec-ellip}. With the discovery 
of description of F-theory compactification using spectral covers, 
we have a new way to understand the duality 
between the Heterotic string and F-theory. In section \ref{ssec:Het}
we find that charged matters can be identified as quotients 
of holomorphic sections of sheaves on spectral cover, also 
in Heterotic string compactification. 
Moduli of spectral surfaces and sheaves on them in both theories 
are directly identified in (\ref{eq:spec-data-duality}), a totally 
new way to describe the duality map. The new duality map improves 
the one developed in \cite{CD, DW-1, HayashiEtAl}, and determines 
how to describe codimension-3 singularities in F-theory 
without a room of ambiguity.

\subsection{Higgs Bundle, Abelianization and ${\rm Ext}^1$ Group}
\label{ssec:Higgs-F}

\subsubsection{$K_S$-valued Higgs Bundle and Spectral Cover}

Physics of gauge fields, charged matter and some of complex 
structure moduli of F-theory compactification can be described 
by the field theory formulation of \cite{DW-1, BHV-1}, once 
the field vev's of a local model are determined properly from 
Calabi--Yau 4-fold and four-form fluxes. 
BPS conditions for background field configuration follows from 
vanishing vevs of the auxiliary fields ${\cal D}$, ${\cal H}_{mn}$ 
and ${\cal G}_m$ \cite{BHV-1}:
\begin{eqnarray}
\omega \wedge F^{(1,1)} - \frac{|\alpha|^2}{2} [\varphi,
 \overline{\varphi}]
 & = & 0, \label{eq:BPS-D}\\
F^{(0,2)} & = & 0, \qquad F^{(2,0)} = 0, \label{eq:BPS-F}\\
\bar{\partial}_{\bar{m}} \varphi & = & 0, \qquad 
\partial_m \overline{\varphi} = 0. \label{eq:BPS-phi}
\end{eqnarray}
Thus, the gauge-field background defines a holomorphic vector bundle 
$V$ on $S$ with structure group $G'$. The (0,1)-part of the gauge field
background can be ``gauged-away'' by complexified gauge transformation 
of $G'$. Thus, from\footnote{Remember that $S$ is still non-compact 
in this section.} (\ref{eq:BPS-phi}), 
\begin{equation}
 \vev{\varphi} \in H^0(S; \mathfrak{g}' \otimes K_S). 
\label{eq:phi-as-H0}
\end{equation}
Although F-theory compactifications down to 5+1 dimensions are 
described by a 1-form field $\varphi$, instead of a 2-form $\varphi$, 
(\ref{eq:phi-as-H0}) remains true.
The $\varphi$ field background induces a map 
\begin{equation}
 \varphi: V \rightarrow V \otimes K_S.
\label{eq:phi-multiply}
\end{equation}
Such a combination of $(V,\varphi)$ is called $K_S$-valued 
(canonical bundle valued) Higgs bundle in the literature.
For example, see \cite{Donagi-spec}.
Supersymmetric compactification of F-theory, therefore, 
is described locally by the 8-dimensional field theory of \cite{DW-1,
BHV-1} with a canonical bundle valued Higgs bundle as a background.

For compactification down to 5+1 dimensions, 
where $S$ is a complex curve, the condition (\ref{eq:BPS-F}) 
becomes trivial. $\varphi$ is a 1-form field, as mentioned, 
and the condition (\ref{eq:BPS-D}) becomes 
\begin{equation}
 F - i [\varphi, \overline{\varphi}] = 0 \qquad 
({\rm equivalently} \quad (iF) - [i\varphi, i\overline{\varphi}] = 0).
\label{eq:Hitchin-D}
\end{equation}
A combination of (\ref{eq:BPS-phi}) and (\ref{eq:Hitchin-D}) is called 
Hitchin equation.\footnote{Note that we adopt a convention throughout 
this article that gauge fields and $\varphi+\overline{\varphi}$ 
are Hermitian, not anti-Hermitian.} Historically, the Hitchin equation arose from 
dimensional reduction of anti-self-dual equation $\omega \wedge F = 0$ 
on a complex surface to a complex curve, which is also regarded as 
a generalization of the flat-connection condition on a complex 
curve \cite{Hitchin-1}. 
Correspondence between Hermitian gauge field and 
$\varphi+\overline{\varphi}$ satisfying 
(\ref{eq:BPS-D}--\ref{eq:BPS-phi}) and an object of algebraic geometry 
$(V, \varphi)$, therefore, is regarded as a generalization (or
dimensional reduction) of the works of Narasimhan--Seshadri \cite{NS}, 
Donaldson \cite{Donaldson} and Uhlenbeck--Yau \cite{UY}. 
Reference \cite{Hitchin-1} proves the correspondence for the case $S$ 
is a curve.

For a given $K_S$-valued Higgs bundle $(V, \varphi)$ with 
a rank-$r$ structure group $G'$, $G'$-invariant homogeneous functions 
of $\varphi$ of degree $d_j$ ($j = 1,\cdots, r$) can be constructed
out of $\varphi$; to be more precise, they are holomorphic 
sections of $K_S^{\otimes d_j}$ on $S$:
\begin{equation}
 p_i: H^0(S; \mathfrak{g}' \otimes K_S) \ni \varphi \mapsto 
  p_i(\varphi) \in H^0(S; K_S^{\otimes d_i}).
\end{equation} 
When the structure group is $\U(N)$, $p_i$'s ($i=1,\cdots,N$) 
are symmetric homogeneous functions of eigenvalues of 
$N \times N$ matrix valued $\varphi$ of degree $i$.
For the case $G' = \U(2)$, as we encountered in  
sections \ref{sec:warm-up}, \ref{sssec:SU(5)-E6}, \ref{sssec:SU(5)-A6} 
and \ref{sssec:SO(10)-E7}, $p_1$ is the trace and $p_2$ determinant.
Thus, $(-p_1)$ and $p_2$ are $(s_1,s_2)$, 
$(\tilde{a}_4, \tilde{a}_5)$, $(R^{(5)}, P^{(5)})$ and 
$(\tilde{a}_3, \tilde{a}_4)$ in these sections, respectively.
If we were to consider full deformation of $E_6$ or $E_7$ singularity, 
$\epsilon_{d_j}$ with $d_j = 2,5,8,6,9,12$ in (\ref{eq:E6-def-A}) 
and 
$\epsilon_{d_j}$ with $d_j = 2,6,8,12,10,14,18$ in (\ref{eq:def-E7-A})
would have been the $p_j$'s for $j=1,\cdots, r$.
The structure group $G'$ is $\SO(4)$ in the example of 
section \ref{sssec:SO(10)-D7}, and $\tilde{a}_3$ and $\tilde{D}$
are the two sections of $K_S^{\otimes 2}$.
These maps $p_i$ above are called Hitchin map and was introduced 
in \cite{Hitchin-2}.
Studies in the precedent sections correspond to the inverse process 
of the Hitchin map; it was to find $\varphi$ whose $p_i(\varphi)$ 
reproduces coefficients of the defining equation of local Calabi--Yau 
geometry. One could also say that Calabi--Yau 4-fold provides 
a description of the same physics in terms of gauge invariants.

It is actually $K_S$-valued $G'$-principal Higgs bundle, 
rather than $K_S$-valued Higgs bundle in certain representation, 
that the geometry of Calabi--Yau 4-fold (and 4-form flux) determines. 
Certainly the irreducible decomposition of adjoint representation of 
$\mathfrak{g}$ under the subgroup $G' \times G''$ 
in (\ref{eq:irr-decmp}) yields only one off-diagonal component 
up to Hermitian conjugation in the case both $G$ and $G''$ are of 
$A_n$ type, or both are of $D_n$ type. 
But, there are more than one if $G$ is $D_n$ type and $G''$ is
$A_{m-1}$ type, or $G = E_{6,7,8}$ in general. 
Geometry determines $\varphi$ in $\mathfrak{g}'$, from which 
we can read the collection of eigenvalues of $\varphi$ acting on 
individual irreducible components $(U_i, R_i)$. 
From a $K_S$-valued $G'$-principal Higgs bundle, all the 
$K_S$-valued Higgs bundle in representation $U_i$'s are constructed. 

For a given representation of $U_i$, the collection of
eigenvalues---spectrum---of $\rho_{U_i}(\varphi)$ can be extracted 
by defining a characteristic polynomial 
\begin{equation}
 \chi_{\varphi; U_i} \equiv {\rm det} 
  \left(\xi {\bf 1} - \rho_{U_i}(\varphi) \right). 
\end{equation}
The zero locus of this characteristic polynomial $\chi_{\varphi; U_i}$
defines a surface $C_{U_i}$, which is called spectral surface (or
spectral cover in more general). The spectral surfaces are regarded 
as divisors of $\mathbb{K}_S$, the total space of canonical bundle 
$K_S$, as we saw in sections \ref{sec:warm-up} and \ref{sec:GUT} and 
as also known in the literature. Although it is possible to introduce 
a more universal object, like principal bundles that does not rely 
upon specific choice of representation $U_i$, we will not do this 
in this article. There is such a need only for the cases with 
$G = E_{6,7,8}$, where there are multiple irreducible components 
appearing in the irreducible decomposition, yet practically all 
we need is to deal with doublet ${\bf 2}$ and singlet 
$\wedge^2 {\bf 2}$ representations of a Higgs bundle 
with $\U(2)$ structure group. Thus, there is no high demand for 
more abstract object; interested readers, however, are referred to 
\cite{Donagi-spec}.  

We have already seen many examples of spectral surfaces of Higgs
bundles in the preceding sections. 
The equations (\ref{eq:C2-An}), (\ref{eq:C-E6-A4}) and 
(\ref{eq:C2-E7-D5}) determine the spectral surface for the doublet 
representation $U_i = {\bf 2}$ of Higgs bundles with a structure 
group $\U(2)$. The spectral surface given by (\ref{eq:C1-E7-D5}) 
is for the representation $U_i = \wedge^2 {\bf 2}$. 
The $G=\SO(14)$ field theory in section \ref{sssec:SO(10)-D7} 
has a Higgs bundle with the structure group $\SO(4)$, and 
the spectral surface $C_{\rm vect.}$ for the vector representation 
of $\SO(4)$ was given by (\ref{eq:def-CV2}).

\subsubsection{Abelianization}

It is known in the literature that a $K_S$-valued Higgs bundle 
$(V,\varphi)$ with a structure group $\U(N)$ or $\SU(N)$ 
(in the fundamental representation $N$) has an equivalent 
description in terms of spectral data as long as $\varphi$ 
satisfies certain conditions.\footnote{$\varphi$ being ``regular'' 
is one of those conditions; see e.g. \cite{DG} for its precise 
definition. Precise conditions for $\varphi$ or spectral cover 
$C_V$ seem to vary over the literature, and we are not sure 
of the precise set of conditions.} Spectral data consists of a 
spectral surface $C_V$, which is a divisor of $\mathbb{K}_S$, and 
a line bundle ${\cal N}_V$ on it \cite{Hitchin-2, BNR, 
Donagi-spec, Donagi-Tani}. From spectral data, 
$(C_V, {\cal N}_V)$, the holomorphic vector bundle $V$ on $S$ 
is recovered by 
\begin{equation}
 V = \pi_{C_V *} {\cal N}_V, 
\label{eq:spec2Higgs}
\end{equation}
where $\pi_{C_V} = \pi_{\mathbb{K}_S} \circ i_{C_V}: C_V \rightarrow S$ is 
a degree-$N$ cover. 
Although the structure group of the theory on $S$ is non-Abelian, 
there is an equivalent description on the spectral cover $C_V$ that 
only involves a line bundle. 
This Abelianization makes our lives much easier.

The structure group of the $K_S$-valued Higgs bundle was $\U(2)$ 
for the field theory local model of codimension-3 singularities 
of type (a) and (c1) in SU(5) GUT models and those of type (a) 
in SO(10) GUT models.
For generic choice of complex structure moduli, the spectral 
surface $C_V$ is smooth and reduced, as in Figures~\ref{fig:C-An} and 
\ref{fig:C-E67-a}~(ib), and extra care does not have to be paid. 
Although there is a subtlety of whether $\varphi$ being regular or not, 
and uncertainty of how to determine $\varphi$ from geometry of a
Calabi--Yau 4-fold, not just their eigenvalues, we will rather take 
the spectral data $(C_V, {\cal N}_V)$ as the starting point, and 
consider only Higgs bundles that are given by such spectral data.
We need to deal with $K_S$-valued Higgs bundle with the structure 
group $\SO(4)$ for the field theory local model of the type (c) 
codimension-3 singularities, and the spectral surface $C_{\bf vect.}$ 
is necessarily singular as we saw in Figure~\ref{fig:C2-D7}. We will come back 
to this issue later, and for now, we will discuss the simplest case 
where the structure group is $\U(N)$.

References~\cite{Pantev-SB, DKS} also introduce a notion of Higgs 
sheaf, which is a sheaf of ${\cal O}_{\mathbb{K}_S}$-module 
on $\mathbb{K}_S$.
A Higgs sheaf ${\cal V}$ is constructed out of a Higgs bundle 
$(V,\varphi)$ on $S$ in a procedure described there, and 
the spectral surface $C_V$ is regarded as the support of this sheaf
${\cal V}$. That is,  
\begin{equation}
{\cal V} = i_{C_V*} {\cal N}_V, 
\end{equation}
and $V = \pi_{\mathbb{K}_S *} {\cal V}$. 
All the information of spectral data 
$(C_V, {\cal N}_V)$ is simply described by the Higgs sheaf ${\cal V}$.
Although we just refer 
to the literature \cite{Pantev-SB, DKS} and do not explain 
the procedure here, it is regarded as the inverse process of 
(\ref{eq:spec2Higgs}). 
Multiplying $2\alpha \varphi_{12}$ in (\ref{eq:phi-multiply}) 
is equivalent to multiplying $\xi$ on the spectral surface, 
where $\xi$ is the coordinate of the fiber direction of 
the canonical bundle $\pi_{\mathbb{K}_S}: \mathbb{K}_S \rightarrow S$.

\subsubsection{Zero-Modes and ${\rm Ext}^1$ Group}

In section~\ref{sec:warm-up}, we had two observations. One is that 
zero-mode wavefunctions of chiral matter multiplets may be solved as 
a singled valued field configuration on the spectral surface $C_V$, 
and the other is that the wavefunctions are ultimately determined by 
holomorphic functions $f$ on $C_V$ modulo those that vanish on the 
matter curve $\bar{c}_V$. We have yet to justify the first 
statement rigorously, however. We were not sure, either, sections of 
which bundle the functions $f$ correspond to, because our treatment 
of section \ref{sec:warm-up} was only local, and furthermore, all the 
gauge-field background was ignored. Now, using the language of Higgs
bundles and Higgs sheaves, however, we can address these problems.

It is known in Type IIB string theory \cite{KS} that open string 
zero modes from D7--D7 intersection are identified with ${\rm Ext}^1$
groups. To be more precise, let us suppose that a D7-brane is 
wrapped on a divisor $i_S: S \hookrightarrow X'$ of a Calabi--Yau 3-fold 
$X'$ with a gauge bundle ${\cal E}$ on it, and another wrapped on 
a divisor $i_T: T \hookrightarrow X'$ with a bundle ${\cal F}$ on it. 
Chiral massless multiplets in the open string sector from $S$ to $T$ 
are identified with 
\begin{equation}
 {\rm Ext}^1 \left( i_{S*} {\cal E}, i_{T*} {\cal F} \right).
\label{eq:Ext1forIIB}
\end{equation} 
Those from $T$ to $S$ are identified with the ${\rm Ext}^2$ group.
Those extension groups are calculated by using spectral sequence, 
with 
\begin{equation}
 E_2^{p,q} = H^p(X'; {\cal E}xt^q (i_{S*} {\cal E}, i_{T*} {\cal F})),
\label{eq:E2forIIB}
\end{equation}
and ${\cal E}xt^q(i_{S*} {\cal E}, i_{T*} {\cal F})$ are obtained by 
tensoring $i_{S*} {\cal E}^\times$ to cohomology sheaves of this complex
\begin{equation}
 0 \rightarrow 
 {\cal H}om_{{\cal O}_{X'}} \left({\cal O}_{X'} , i_{T*} {\cal F} \right) 
 \rightarrow 
 {\cal H}om_{{\cal O}_{X'}} \left({\cal O}_{X'}(-S), i_{T*} {\cal F} \right)
 \rightarrow 0.
\end{equation}

Zero modes of matter multiplets in F-theory are characterized by 
(\ref{eq:EOM1}--\ref{eq:EOM4}) in the field-theory formulation.
For supersymmetric compactification, the wavefunctions of bosonic 
fields $(i A_{\bar{m}},\varphi_{12})$ should be the same as those of 
fermionic fields $(i \psi_{\bar{m}},\chi_{12})$. If one requires that 
an infinitesimal deformation from a BPS background still satisfies 
the BPS conditions (\ref{eq:BPS-D}--\ref{eq:BPS-phi}), then the 
conditions on an infinitesimal deformation 
$(i A_{\bar{m}},\varphi_{12})$ is exactly the same as the zero-mode 
equations. Thus, matter zero modes are regarded as infinitesimal 
deformation of Higgs bundle that still preserves the BPS conditions.
Is this deformation characterized as an extension group?
Is it possible to generalize the argument above, in a way 
suitable for F-theory compactification?

Inspired by the appendix of \cite{DKS}, we propose to
generalize\footnote{A similar attempt was made in \cite{DW-1}.} 
(\ref{eq:Ext1forIIB}, \ref{eq:E2forIIB}) as follows:
matter multiplets are identified with the extension groups 
\begin{equation}
 {\rm Ext}^1 \left(i_{\sigma*} {\cal O}_S, {\cal V}\right), 
\label{eq:Ext1forF}
\end{equation}
which is calculated by using spectral sequence from 
\begin{equation}
 E_2^{p,q} = H^p \left(\mathbb{K}_S; 
  {\cal E}xt^q (i_{\sigma * }{\cal O}_S, i_{C_V *}{\cal N}_V)\right).
\label{eq:E2forF}
\end{equation}
Here, $i_{\sigma}: S \hookrightarrow \mathbb{K}_S$ is the embedding of 
$S$ to $\mathbb{K}_S$ through the zero section of the canonical 
bundle $K_S$. Reference \cite{DKS} discussed matter fields arising 
from D-branes sharing the same support $S$ in a Calabi--Yau 3-fold $X'$
in Type IIB string theory, and an ${\rm Ext}^1$ group on the 
normal bundle of $S$, $N_{S|X'}$ was used. We replaced $N_{S|X'}$ 
by $\mathbb{K}_S$ for F-theory. Higgs sheaves for bi-fundamental 
representations of Type IIB string theory in \cite{DKS} will most 
naturally be generalized in F-theory to Higgs sheaves ${\cal V}$ for 
irreducible components of the decomposition like (\ref{eq:irr-decmp}).
Thanks to the Abelianization, Higgs bundles with non-Abelian structure 
group can be dealt with as if they were 7-branes with Abelian background 
on them.
If a line bundle is turned on on $S$ to break the GUT gauge group 
(such as SU(5) and SO(10)) to the Standard Model gauge group 
\cite{GUT-break-F-A, TW-GUT}\footnote{See also \cite{GUT-break-IIB-A, WY} 
for the preceding discussion in Type IIB language.}, 
then $i_{\sigma * } {\cal O}_S$ can be replaced by the line bundle.

To see whether this proposal is reasonable, let us obtain more 
explicit expressions for (\ref{eq:E2forF}).\footnote{We assume 
that the spectral surface $C_V$ is well-behaved, and is different 
from trivial $C_V = S$ corresponding to $\varphi = 0$.}
The extension sheaves 
${\cal E}xt^q \left(i_{\sigma * }{\cal O}_S, i_{C_V *}{\cal N}_V)\right)$
are cohomology sheaves of this complex, 
\begin{equation}
 0 \rightarrow {\cal H}om_{{\cal O}_{\mathbb{K}_S}} 
  \left( {\cal O}_{\mathbb{K}_S}, i_{C_V *} {\cal N}_V \right)
  \rightarrow {\cal H}om_{{\cal O}_{\mathbb{K}_S}} 
  \left( {\cal O}_{\mathbb{K}_S}(-S), i_{C_V *} {\cal N}_V \right)
  \rightarrow 0, 
\end{equation}
which is equivalent to 
\begin{equation}
\begin{array}{rcl}
 & \varphi & \\
0 \longrightarrow i_{C_V *} {\cal N}_V & \longrightarrow & 
i_{C_V *} ({\cal N}_V \otimes \pi_{C_V}^* K_S) \longrightarrow 0.
\end{array}
\end{equation}
Now the support of all these sheaves is the spectral surface $C_V$.
It thus follows that 
\begin{eqnarray}
 E_2^{p,0} & = & 0, \\
 E_2^{p,1} & = & H^p(C_V ; ({\cal N}_V \otimes \pi_{C_V}^* K_S) 
 {\rm ~mod~image~of~}\varphi), \label{eq:matter-F-pre} \\
  & = & H^p(\bar{c}_V; ({\cal N}_V \otimes \pi_{C_V}^* K_S)|_{\bar{c}_V}).
\end{eqnarray}
Here, $\bar{c}_V$ is the matter curve, locus on $C_V$ where 
$\xi = 2\alpha \varphi_{12}$ vanishes.

The spectral sequence already converges, because $E_2^{p,q}$ does not 
vanish only for $q = 1$. Thus, $E_2^{n-1,1}$ becomes the extension 
groups ${\rm Ext}^n$. Matter multiplets correspond to $n=1,2$.
Therefore, we obtain 
\begin{eqnarray}
 {\rm Ext}^1(i_{\sigma *}{\cal O}_S; {\cal V}) & = & 
  H^0(\bar{c}_V; ({\cal N}_V \otimes \pi_{C_V}^* K_S)|_{\bar{c}_V}), 
  \label{eq:matter-F} \\
 {\rm Ext}^2(i_{\sigma *}{\cal O}_S; {\cal V}) & = & 
  H^1(\bar{c}_V; ({\cal N}_V \otimes \pi_{C_V}^* K_S)|_{\bar{c}_V}).
\end{eqnarray}
The holomorphic functions ``$f$'' we encountered in 
section~\ref{sec:warm-up} exactly have these properties. 
They are locally holomorphic functions on the spectral surface $C_V$, 
transform like $\chi_{12}$ or $\varphi_{12}$, which accounts for 
the tensor factor $\pi_{C_V}^* K_S$. What is really in one-to-one 
correspondence with the zero modes is the ``functions'' 
(in a local description) modulo those that vanish along $\varphi = 0$, 
and (\ref{eq:matter-F-pre}) literally provides such a characterization 
of zero-mode matter fields. We will further see in section
\ref{ssec:Het} that matter multiplets can also be characterized exactly 
in the same way in Heterotic string theory, and all these observations 
combined gives us a confidence in the proposal 
(\ref{eq:Ext1forF}, \ref{eq:E2forF}).

We also find at the same time that the ``$f$'' should be regarded 
as sections of ${\cal N}_V \otimes \pi_{C_V}^* K_S$. 
The ``${\cal N}_V$'' part was not clear from the argument 
in sections~\ref{sec:warm-up} and \ref{sec:GUT}, partly because 
we ignored gauge-field background coming from 4-form fluxes and 
partly because it was not clear how to treat the branch locus 
of the field theory local models of F-theory.
By now, however, we can start from the spectral data 
$(C_V, {\cal N}_V)$ in F-theory compactification, instead of 
Calabi--Yau 4-fold and 4-form flux on it. In the spectral data, 
there is nothing ambiguous in how to deal with the ramification locus.

\subsection{Heterotic--F Duality Revisited}
\label{ssec:Het}

We have not used duality between the Heterotic string theory 
and F-theory so far. We have not even assume that Heterotic dual 
exists for a Calabi--Yau 4-fold compactification of F-theory.
It is not difficult, however, to see how the duality relates 
both theories. 

We have determined spectral surfaces of Higgs bundles from defining 
equations of Calabi--Yau 4-fold for F-theory compactifications. 
On the Heterotic theory side, where the Heterotic $E_8 \times E_8$
string theory is compactified on an elliptic fibered Calabi--Yau 3-fold 
$Z$ on a common base 2-fold $S$, 
\begin{equation}
 \pi_Z : Z \rightarrow S, 
\label{eq:pi-Z}
\end{equation}
a vector bundle within one of the $E_8$ factors is described by 
yet another spectral surface \cite{FMW, Donagi-spec-ellip}. 
When the elliptic fibration $Z$ is given by a Weierstrass equation 
\begin{equation}
 y^2 = x^3 + f_0 x + g_0, 
\label{eq:Weierstrass-Het}
\end{equation}
with $(x,y)$ as the coordinates of the elliptic fiber, then 
the spectral surface of an $\SU(5)$ bundle is given by 
\begin{equation}
 5\sigma + {\rm div} (a_0 + a_2 x + a_3 y + a_4 x^2 + a_5 x y),
\label{eq:spectral-surface}
\end{equation}
where $a_r$ ($r = 0,2,3,4,5$) are sections of line bundles on 
$S$, ${\cal O}(r K_S + \eta)$. The divisor $\eta$ on $S$ should be 
chosen so that the normal bundle $N_{S|B_3}$ of the F-theory 
compactification corresponds to ${\cal O}(6K_S + \eta)$ 
\cite{MV1,MV2,FMW,Het-F-4D,Rajesh}.
The same notation $a_{0,2,3,4,5}$ are used in (\ref{eq:defeq}) 
and (\ref{eq:spectral-surface}), because this is known to be 
the right duality map \cite{KMV-BM, CD, DW-1, HayashiEtAl}. 
Let us see in explicit examples in the following that the defining 
equations of the spectral surfaces of Higgs bundles in F-theory 
perfectly agree with those of spectral surfaces of vector bundles 
in elliptic fibered compactification of Heterotic string theory.

In the elliptic fibration given by (\ref{eq:Weierstrass-Het}), 
the zero section $\sigma$ corresponds to $(x,y)=(\infty,\infty)$.
One can choose a local coordinate $\xi$ of the elliptic fiber 
around the zero section, so that $x \simeq 1/\xi^2$ and 
$y \simeq 1/\xi^3$ around the zero section. 
Around a type (a) codimension-3 singularity point of a theory 
with unbroken $\SU(5)$ symmetry, where both $a_4$ and $a_5$ 
become small, the equation of the spectral surface 
(\ref{eq:spectral-surface}) becomes 
\begin{equation}
 5 \sigma - {\rm div} \xi^5 + 
{\rm div} (a_0 \xi^5 + a_2 \xi^3 + a_3 \xi^2 + a_4 \xi + a_5)
 \simeq  {\rm div} (a_3 \xi^2 + a_4 \xi + a_5),
\end{equation}
where now we only pay attention to a region near the zero section.
This behavior of the spectral surface of an $\SU(5)$ vector bundle 
in the Heterotic string theory is exactly the same as 
the spectral surface of Higgs bundle in (\ref{eq:C-E6-A4}) in F-theory.
Similarly, the spectral surface of $\SU(4)$ bundle in Heterotic 
string theory, $a_2 \xi^2 + a_3 \xi + a_4 \simeq 0$, agrees perfectly 
with the spectral surface of $\U(2)$ Higgs bundle (\ref{eq:C2-E7-D5}) 
of F-theory.

Reference \cite{HayashiEtAl} determined a defining equation of 
spectral surface of $\wedge^2 V$ bundle for an $\SU(4)$ vector 
bundle $V$ in Heterotic string theory, 
in a neighborhood of a type (c) codimension-3 singularity,
$(\tilde{a}_3, \tilde{D}) \sim (0,0)$. The spectral surface 
(\ref{eq:def-CV2}) of a Higgs bundle of F-theory perfectly 
agrees with the defining equation of $C_{\wedge^2 V}$ in the 
appendix of \cite{HayashiEtAl} (after a few typos there are corrected).

The duality map between the vector bundle moduli of Heterotic string 
theory and a part of complex structure moduli of F-theory has been 
constructed by studying how the complex structure moduli of 
$dP_8$ (del Pezzo 8) surface changes a flat bundle on $T^2$. 
Now that we are familiar with the notion of Higgs bundle and its 
spectral data for F-theory compactification, 
we have a new channel (though essentially equivalent) of 
establishing a duality map: map the moduli parameters of both sides 
so that the spectral surfaces of vector bundles on elliptic fibration 
correspond to the spectral surfaces of Higgs bundle. 
The spectral surface in Heterotic side describes the behavior of 
Wilson line in the elliptic fiber direction. That is, degree-$N$ 
cover spectral surface over $S$ determines $N$ different values 
of Wilson lines $A_{\bar{3}}$ (and $A_{3}$) for a given point in $S$, 
where $A_{\bar{3}} \equiv (A_8 + i A_9)/2$. The spectral surface 
in F-theory side describes $N$ different eigenvalues of $\varphi_{12}$
at a given point in $S$. The Heterotic--F theory duality simply 
replaces $A_{\bar{3}}$ of Heterotic string theory by $\varphi_{12}$ 
in F-theory and vice versa at each point on $S$. The Calabi--Yau 
description of F-theory compactification extracts gauge invariants 
of the field-theory formulation through the Hitchin/Katz--Vafa map.

When a couple of irreducible components are involved 
in (\ref{eq:irr-decmp}), which is usually the case, duality should 
be better stated as the correspondence between cameral covers of 
both sides. But this is still equivalent to equating the 
spectral surfaces for all the irreducible representations 
of those bundles that appear in (\ref{eq:irr-decmp}), and we 
are not pursing in phrasing the duality in terms of cameral cover 
in this article. 

The duality correspondence of zero-modes of both sides is 
also easier to see in the aid of spectral surfaces of Higgs 
bundle in F-theory side.\footnote{Since the zero-mode charged 
matters correspond to deformation of vector and Higgs bundles 
of both theory, the Heterotic--F duality (vector and Higgs bundle
duality) for arbitrary structure group already implies the duality 
between charged massless matters, though.}  To begin with, let us remind ourselves how the zero-mode 
matter multiplets are identified in elliptic fibered Calabi--Yau 3-fold 
compactification of Heterotic $E_8 \times E_8$ string theory.
Suppose that a vector bundle $V$ with a structure group 
$\SU(N) \subset E_8$ is turned on $Z$, and that $V$ is given 
by a Fourier--Mukai transform of a line bundle ${\cal N}_V$ on 
a spectral surface $C_V$:
\begin{equation}
\vcenter{\xymatrix{
 & C_V \times_{S} Z \ar[dl]_{p_1} \ar[dr]^{p_2} & \\
   C_V \ar[dr]_{\pi_{C_V}}& & Z \ar[dl]^{\pi_Z}\\
 & S & 
}}, \qquad 
V = p_{2*} \left({\cal P}_{S} \otimes p_1^* ({\cal N}_V)\right).
\end{equation}
%
For now, we assume that the structure group of $V$ is not a proper
subgroup of $\SU(N)$, but $\SU(N)$ itself, and $C_V$ is irreducible 
and smooth. Fourier--Mukai transform of the bundle $V$ is represented 
by the original line bundle ${\cal N}_V$.
In the irreducible decomposition (\ref{eq:irr-decmp}) with $G = E_8$ 
and $G' = \SU(N)$, there is an irreducible component $(U_i, R_i)$ where 
$U_i$ is the fundamental representation of $G' = \SU(N)$. 
The zero modes chiral multiplets from this component is identified with 
a vector space
\begin{equation}
 H^1(Z ; \rho_{U_i}(V)) = H^1(Z; V).
\end{equation}
It has been customary, when we are to compare this expression with 
its F-theory dual counter part, to use the spectral sequence associated 
with the elliptic fibration (\ref{eq:pi-Z}):
\begin{equation}
 H^1(Z; V) \simeq H^0(S; R^1\pi_{Z*} V),
\end{equation}
where $R^1\pi_{Z*} V$ is a sheaf on $S$, and we assumed that the bundle 
is non-trivial when restricted on a generic fiber of (\ref{eq:pi-Z}).
The sheaf $R^1\pi_{Z*} V$ has a support only on the matter curve of
bundle $\rho_{U_i}(V) = V$, and the localized sheaf on the matter curve 
$\bar{c}_V$ is compared with (or translated into) line bundles localized 
on the matter curves in F-theory compactification \cite{Curio, DI}. 
This is the traditional story. 

Now that we also have spectral cover $C_V$ in F-theory side, 
however, there is another channel to compare the zero modes 
on both sides of the duality. Because the vector bundle $V$ 
on $Z$ is given by a push-forward $p_{2*} = R^0 p_{2*}$ of 
a line bundle on $C_V \times_S Z$, we can rewrite the sheaf 
$R^1 \pi_{Z*} V$, by following the argument of section 7.4 
in \cite{Penn5}:
\begin{eqnarray}
 R^1 \pi_{Z*} V & = &  
 R^1 \pi_{Z*} \left( R^0 p_{2*} \left( 
       {\cal P}_S \otimes p^*_1({\cal N}_V)
                                \right)
              \right), \nonumber \\  
 & = & R^0 \pi_{C_V *} \left( R^1 p_{1*} \left(
       {\cal P}_S \otimes p^*_1({\cal N}_V)
                                  \right)
              \right)
 = \pi_{C_V *} \left({\cal N}_V \otimes R^1 p_{1*} {\cal P}_S \right).
\end{eqnarray}
Therefore, on the side of Heterotic string compactification, 
the zero mode matter multiplets from $\rho_{U_i}(V) = V$ 
can be regarded as 
\begin{equation}
 H^1(Z; V) \simeq H^0 \left( S; \pi_{C_V *}\left( 
  {\cal N}_V \otimes R^1 p_{1*} {\cal P}_S    \right) \right)
 = H^0 \left( C_V; {\cal N}_V \otimes R^1 p_{1*} {\cal P}_S \right),
\end{equation}
holomorphic sections of a sheaf on the spectral {\it surface} $C_V$, 
not on the matter {\it curve} $\bar{c}_V$. 
Zero modes are now characterized as holomorphic objects on the 
spectral surface (modulo redundancy as we see below) also 
in Heterotic string compactification.

The appendix A of \cite{HayashiEtAl} looked into the details 
of the sheaf $R^1 p_{1*} {\cal P}_S$. It is a sheaf on $C_V$, 
although its support is the matter curve $\bar{c}_V$.
It is characterized by this short exact sequence, 
\begin{equation}
 0 \rightarrow {\cal I}_{\bar{c}_V}(\pi_{C_V}^* K_S)
 \rightarrow {\cal O}_{C_V}(\pi_{C_V}^* K_S) \rightarrow 
 R^1 p_{1*} {\cal P}_S \rightarrow 0.
\label{eq:ex-sequence-Het}
\end{equation}
Here, ${\cal I}_{\bar{c}_V}$ is the ideal sheaf of the matter 
curve $\bar{c}_V$ on the spectral surface $C_V$, which is roughly 
speaking, holomorphic functions that vanish on $\bar{c}_V$.  
${\cal I}_{\bar{c}_V}(D) \equiv {\cal I}_{\bar{c}_V} \otimes {\cal O}(D)$.
All the three sheaves in the short exact sequence above are 
on the spectral surface $C_V$.
Local generators of the sheaf $R^1 p_{1*} {\cal P}_S$ 
can be expressed as those of ${\cal O}(\pi_{C_V}^* K_S)$ modulo 
image from ${\cal I}_{\bar{c}_V}(\pi^*_{C_V} K_S)$. 
To be more explicit, the generators of $R^1 p_{1*} {\cal P}_S$ 
in language of \v{C}ech cohomology are of the form \cite{HayashiEtAl}  
\begin{equation}
 \frac{y}{x-x_i} f, 
\end{equation}
where $f$ is a section of\footnote{Note that 
$x$, $x_i$ and $y$ transform like sections of line bundles 
$K_S^{\otimes (-2)}$, $K_S^{\otimes (-2)}$ and $K_S^{\otimes (-3)}$, 
respectively, and hence $f$ transforms as a section of 
$\pi_{C_V}^* K_S$. } ${\cal O}(\pi^*_{C_V} K_S)$. The image from 
${\cal I}_{\bar{c}_V}(\pi_{C_V}^*(K_S))$ corresponds to 
``$f$'' that at least involves single power of the normal coordinate of 
the curve $\bar{c}_V$ in $C_V$. 
Explicit study of the structure of $R^1 p_{1 *} {\cal P}_S$ 
in \cite{HayashiEtAl} using \v{C}ech cohomology shows that 
$R^1 p_{1 *} {\cal P}_S$ is generated by $f$'s modulo the image 
from ${\cal I}_{\bar{c}_V}(\pi_{C_V}^* K_S)$, and hence the 
characterization by (\ref{eq:ex-sequence-Het}) follows.

Simply tensoring ${\cal N}_V$ to the argument above, we find that the 
zero-mode matter multiplets in representation $V$ are identified with 
global holomorphic sections $f$ of 
${\cal N}_V \otimes {\cal O}(\pi^*_{C_V} K_S)$ on the spectral 
surface $C_V$, modulo those involving at least one power of normal 
coordinate of the matter curve. 
Now we will not need even a single word to see the dual correspondence 
with the characterization of zero mode matter multiplets in F-theory 
description, (\ref{eq:matter-F-pre}, \ref{eq:matter-F}).
Alternatively, this agreement can be regarded as a justification 
of the proposal (\ref{eq:Ext1forF}, \ref{eq:E2forF}).

\subsection{Ramification and Four-form Flux}
\label{ssec:rami_flux}

Spectral data $(C_V, {\cal N}_V)$ are used on both sides of the 
duality between the Heterotic string and F-theory. The duality 
map is simply stated as 
\begin{equation}
 (C_V, {\cal N}_V)^{\rm Het} = (C_V, {\cal N}_V)^{\rm F}.
\label{eq:spec-data-duality}
\end{equation}
Not only the moduli of the spectral surfaces but also discrete 
as well as continuous data of the line bundles ${\cal N}_V$ on 
both sides are identified. This novel way to see the duality 
allows us to bypass the complicated discussion involving 
$dP_8$-fibration and its blow up to $dP_9$-fibration, and 
(projected) cylinder map in \cite{CD,DW-1,HayashiEtAl} 
in mapping the discrete data of ${\cal N}^{\rm Het}_V$ into F-theory.

This observation goes beyond a discovery of academic interest, 
and brings a practical benefit. 
The spectral cover $\pi_{C_V}: C_V \rightarrow S$ is generically 
ramified in Heterotic and F-theory compactifications alike.
References \cite{DW-1, HayashiEtAl} used the Heteroic--F theory
duality to study how the four-form flux determines the sheaves/bundles 
on the (covering) matter curves for F-theory on a Calabi--Yau with 
generic complex structure. 
But, there was one weakness in the argument. The duality map of 
the discrete data of ${\cal N}_V$ through del Pezzo fibration 
assumes that there is a well-defined choice of independent basis 
of 2-cycles of del Pezzo fiber. But, two 2-cycles of del Pezzo fiber 
become degenerate where the spectral surface ramifies, and it is not 
obvious how to choose independent 2-cycles at the branch locus. 
This weakness of the argument in \cite{DW-1, HayashiEtAl}, however, 
is overcome under the new duality map.
Here, we do not have to go through del Pezzo fibration, and no ambiguity 
arises in translating (copying) the line bundle ${\cal N}_V$ from 
Heterotic string to F-theory, even along the ramification locus of 
the spectral cover.

The line bundles ${\cal N}_V$ corresponding to a bundle 
with structure group $\SU(N)$ has a description 
\begin{equation}
 {\cal N}_V = {\cal O}\left(\frac{1}{2} r + \gamma \right),
\label{eq:r+gamma}
\end{equation}
where $r$ is the ramification divisor of $\pi_{C_V}: C_V \rightarrow S$, 
and $\gamma$ another divisor (allowing half integral coefficients) of 
$C_V$. This result is known to most physicists through \cite{FMW}, 
where ${\cal N}_V$ arises as a part of spectral data for Heterotic 
compactification. The $\gamma$ part in Heterotic string corresponds to 
four-form flux in F-theory
\cite{CD, DW-1, HayashiEtAl}. 
The separation between the $r/2$ piece and the rest also appeared 
in mathematical literature describing abelianization of Higgs bundles; 
see \cite{Donagi-spec} and references therein. Thus, the separation 
between the $r/2$ part and the rest goes hand in hand on the both sides 
of the duality.

In order to understand physics associated with codimension-3 
singularity of F-theory compactification, it turns out that 
it is best to study the effects of $r/2$ and $\gamma$ separately. 
In the following, we will work on issues associated with 
$r/2$ piece in section \ref{sssec:ram} and those with $\gamma$ 
(equivalently the four-form flux) in section \ref{sssec:4-form}.

\subsubsection{Ramification}
\label{sssec:ram}

Using the expression (\ref{eq:r+gamma}), the zero-modes from 
the fundamental representation of (a relevant local part of) $\SU(N)$ 
vector/Higgs bundle is rewritten as \cite{Curio, DI, HayashiEtAl} 
\begin{equation}
 H^0(\bar{c}_V; {\cal O} (i_V^* K_S + j_V^* r/2 + j_V^* \gamma ) );
\end{equation}
maps $i_V$ and $j_V$ are those in the diagram below:
\begin{equation}
 \vcenter{\xymatrix{
 \tilde{\bar{c}}_{\rho_U(V)} \ar[r]^{\tilde{\jmath}_{\rho_U(V)}} \ar[d]_{\nu_{\bar{c}_{\rho_U(V)}}} &
 \widetilde{C}_{\rho_U(V)} \ar[d]_{\nu_{C_{\rho_U(V)}}} & \\
 \bar{c}_{\rho_U(V)} \ar[r]^{j_{\rho_U(V)}} \ar[dr]_{i_{\rho_U(V)}} &
  C_{\rho_U(V)} \ar[r]^{i_{C_{\rho_U(V)}}} \ar[d]_{\pi_{C_{\rho_U(V)}}} & Z [\mathbb{K}_S]
  \ar[dl]^{\pi_Z [\pi_{\mathbb{K}_S}]} \\
 & S & 
%
}}. \label{eq:all-maps}
\end{equation}
Some of the objects and maps in this diagram have not been used so far, 
but they will appear later on.
Reference \cite{Curio} showed by using the adjunction formula that 
%
\begin{equation}
 i_V^* K_S + \frac{1}{2} j_{V}^* r = K_{\bar{c}_V}^{\frac{1}{2}},
\label{eq:half-Kc-V}
\end{equation}
and hence the Serre duality accounts for the anticipated relation 
\begin{equation}
 h^1(\bar{c}_V; {\cal O}((i_V^* K_S + j_V^* r/2) + j_V^* \gamma) 
 =
 h^0(\bar{c}_V; {\cal O}((i_V^* K_S + j_V^* r/2) - j_V^* \gamma). 
\label{eq:Serre-on-curve-V}
\end{equation}
We have nothing more to add for the cases with $\SU(N)$ structure group, 
except that the spectral surface $C_V$ can be defined intrinsically 
from Calabi--Yau 4-fold for F-theory compactification, and that 
the same logic as in \cite{Curio} can be used to 
show (\ref{eq:half-Kc-V}) even for F-theory compactification without 
Heterotic dual.\footnote{A proof for (\ref{eq:half-Kc-V}) in
F-theory compactification without Heterotic dual was obtained 
in \cite{HayashiEtAl}, by counting the number of type (a) codimension-3 
singularity points. We find the argument here more elegant, however.}

Spectral surfaces for the matter of SU(5) 
${\bf 5}+\bar{\bf 5}$ representations and for those of 
SO(10) {\bf vect.} representation are not smooth everywhere along 
the matter curve, even for the most generic choice of complex 
structure moduli of F-theory compactifications.
$C_{6+12}$ (\ref{eq:def-CV2-d}) has a double curve singularity around 
type (d) codimension-3 singularity points, and $C_{\bf vect.}$
in (\ref{eq:def-CV2}) has a double curve along $\tilde{a}_3 = 0$ 
and a pinch point singularity at $(\tilde{a}_3, \tilde{D}) = (0,0)$ 
around type (c) codimension-3 singularity points. 
Those matter multiplets are both from $\wedge^2 V$ bundles in 
Heterotic string compactification, and Ref. \cite{HayashiEtAl} 
studied the structure of the sheaf ${\cal N}_{\wedge^2 V}$. 
The sheaf ${\cal N}_{\wedge^2 V}$ is represented by a pushforward 
of a line bundle $\widetilde{\cal N}_{\wedge^2 V}$ on the 
desingularization $\widetilde{C}_{\wedge^2 V}$, 
\begin{equation}
 \nu_{C_{\rho=U(V)}}: 
\widetilde{C}_{\rho_U(V)} \rightarrow C_{\rho_U(V)}, \qquad 
{\cal N}_{\wedge^2 V} = \nu_{C_{\wedge^2 V}*} 
\widetilde{\cal N}_{\wedge^2 V}.
\end{equation}
In the context of F-theory 
compactifications / Abelianization of Higgs bundles, it is just 
natural from D7-brane interpretation to consider desingularized 
spectral cover $\widetilde{C}_{6+12}$ in (\ref{eq:def-CV2-d-resolve})
and gauge bundles on individual irreducible components.
Reference \cite{Hitchin-2} already points out the need of resolving 
double-{\it point} singularity on the spectral cover when the structure 
group is $\mathfrak{so}(2n)$ when $S$ is a curve; the structure group 
around the type (c) codimension-3 singularities in SO(10) GUT 
models was $\mathfrak{so(4)}$, and this is one of the cases Hitchin's 
observation is applied. Now, the double-{\it curve} singularity has to be 
resolved. The necessary resolution of the double curve also resolves 
the pinch-point singularity, and that is $\widetilde{C}_{\bf vect.}$.

For any representations $\rho_U(V)$ of vector/Higgs bundles, 
the line bundle $\widetilde{\cal N}_{\rho_U(V)}$ on 
$\widetilde{C}_{\rho_U(V)}$ should have a factor ${\cal O}(r/2)$, apart 
from the one coming from $\gamma$ (or from three-form potential in
F-theory). Here, $r$ is the ramification divisor associated with 
$\tilde{\pi}_{\rho_U(V)} \equiv \pi_{C_{\rho_U(V)}} \circ 
\nu_{C_{\rho_U(V)}}: \widetilde{C}_{\rho_U(V)} \rightarrow S$.
By repeating the same argument as for (\ref{eq:half-Kc-V}) and 
noting that the covering matter curve $\tilde{\bar{c}}_{\rho_U(V)}$ 
in $\widetilde{C}_{\rho_U(V)}$ is defined as the zero locus of the 
coordinate $\nu_{\rho_U(V)}^* \xi$ \cite{HayashiEtAl}, one finds that 
\begin{equation}
 (\tilde{\imath}_{\rho_{U}})^* K_S + 
 \frac{1}{2} \tilde{\jmath}_{\rho_U(V)}^* r = 
 K_{\tilde{\bar{c}}_{\rho_U(V)}}^{\frac{1}{2}}.  
\end{equation}
Here, $\tilde{\imath}_{\rho_U(V)}$ is a map from the covering matter 
curve $\tilde{\bar{c}}_{\rho_U(V)}$ to $S$ passing any routes 
in (\ref{eq:all-maps}). Using this relation and Serre duality 
on the covering matter curve $\tilde{\bar{c}}_{\rho_U(V)}$, 
it is now easy to see the relation analogous to (\ref{eq:Serre-on-curve-V})
for the number of multiplets in $\SU(5)$-{\bf 5} representation.
This observation greatly simplifies the argument\footnote{References 
\cite{Penn5, BMRW, HayashiEtAl} eventually arrived at a divisor 
$\tilde{\pi}_{D*} (r_V|_D - R)$ on $\tilde{\bar{c}}_{\wedge^2 V}$; 
here, $r_V$ is the ramification divisor of 
$\pi_{C_V}: C_V \rightarrow S$. See \cite{HayashiEtAl} for notations. 
It is possible to understand intuitively that this divisor corresponds 
to $r_{\wedge^2 V}|_{\tilde{\bar{c}}_{\wedge^2 V}}$ as follows. The
desingularized spectral surface $\widetilde{C}_{\wedge^2 V}$ consists of
points $p_i \boxplus p_j$, where both $p_i$ and $p_j$ are points in 
$C_V$ that are mapped to the same point in $S$, and $\boxplus$ stands
for the group-law summation. When a layer of $C_V$ containing either
$p_i$ or $p_j$ is ramified over $S$, so is $\widetilde{C}_{\wedge^2 V}$.
The only exception is the case when the layers of $C_V$ containing 
$p_i$ and $p_j$ locally form a degree-2 ramified cover over $S$. 
The spectral surface $\widetilde{C}_{\wedge^2 V}$ is not ramified 
in this case. This is why the component $R$ has to be subtracted to 
obtain $r_{\wedge^2 V}$ from $r_V$.} 
in \cite{Penn5, BMRW, HayashiEtAl}.

Reference \cite{HayashiEtAl} found that the charged matter zero modes 
of F-theory are global holomorphic sections of line bundles on covering 
matter curves, $\tilde{\bar{c}}_{\rho_U(V)}$; the line bundles on 
$\tilde{\bar{c}}_{\rho_U(V)}$ are described by their divisors, and 
the divisors were determined by using the Heterotic--F theory duality. 
The divisors have support on codimension-3 singularity points of 
F-theory. The coefficients of the divisors at these singularities 
were determined by the duality, but an explanation intrinsic to F-theory 
was not found, and remained an open problem. Now a complete answer 
to this question is available. Codimension-3 singularity is formed 
whenever the ramification divisor of (desingularized) spectral surface 
intersect the (covering) matter curve, and the $r/2$ piece of 
$\widetilde{\cal N}_{\rho_U(V)}$ always leaves a contribution 
to the divisor on the (covering) matter curve. The coefficients 
at the codimension-3 singularities are determined purely from the 
ramification behavior of the (desingularized) spectral surfaces 
of the Higgs bundle of F-theory compactification.

Now that all the subtleties associated with ramification / branch 
locus have been clarified, we can fill-in the missing details of our 
observation in section \ref{sec:warm-up}. 
The zero-mode equations (\ref{eq:EOM1}--\ref{eq:EOM4}) on $S$ for 
an irreducible component $(U_i, R_i)$ are for a set of fields 
$\Psi \equiv (i \psi_{\bar{m}}, \chi_{12})$ with ${\rm dim}. \; U_i$ 
components. We found in section \ref{sec:warm-up}, however, that 
it is possible to write down an equivalent set of equations on 
the desingularized spectral surface $\widetilde{C}_{\rho_{U_i}(V)}$, 
where we only need a single component of $\Psi$. This idea 
now seems very convincing, because charged matter zero modes 
are characterized as holomorphic sections $f$ of sheaves 
$\widetilde{\cal F}_{\rho_{U_i}(V)} \equiv 
 \widetilde{\cal N}_{\rho_{U_i}(V)} \otimes 
 \tilde{\pi}_{C_{\rho_{U_i}(V)}}^* K_S$
on the (desingularized) spectral surfaces, modulo redundancy, 
both in F-theory and Heterotic string theory. 
Covariant derivatives in the set of equations should involve gauge 
fields, and the gauge field backgrounds are determined from the line bundle 
$\widetilde{\cal N}_{\rho_{U_i}(V)}$. This line bundle is 
determined by half-the-ramification-divisor twist of 
$\widetilde{C}_{\rho_{U_i}(V)} \rightarrow S$ and 
by three-form potential background. 

\subsubsection{Four-form Flux}
\label{sssec:4-form}

In this section \ref{sssec:4-form}, we address two issues 
associated with four-form flux background, or almost 
equivalently $\gamma$ in (\ref{eq:r+gamma}). This part is 
digression in nature, and readers may skip and proceed to 
section \ref{sec:Yukawa}.

The first goal we achieve in this section \ref{sssec:4-form} 
is to obtain better understanding of what is really going on 
in the presence of both the ramification divisor $r$ and 4-form 
flux $\gamma$ in (\ref{eq:r+gamma}). The second objective is 
to illustrate how one and the same four-form background 
determines gauge field backgrounds for fields in different 
representations.

${\cal N} = 1$ supersymmetry is preserved when 
a 4-form flux background in a Calabi--Yau 4-fold $X$ 
is primitive and is in the (2,2) part. One needs to know 
global geometry of $X$ in order to determine the whole 
variety available in $H^{2,2}(X)$. For the purpose above, 
however, we only need illustrative examples, and hence 
we will use a (topological) form of $\gamma$ that has been 
known in elliptic-fibered Calabi--Yau 3-fold compactification 
of Heterotic string theory.

In Heterotic string compactification on an elliptic fibration, 
a ``divisor'' $\gamma$ on a spectral surface $C_V$ of an $\SU(N)$ 
vector bundle in (\ref{eq:r+gamma}) has to satisfy 
\begin{equation}
 \pi_{C_V *} \gamma = 0.
\label{eq:1stCHN}
\end{equation}
We can choose 
\begin{equation}
 \gamma = \lambda N \bar{c}_V - \lambda \pi_{C_V}^{-1}(a_N = 0)
\label{eq:expl-gamma}
\end{equation}
as a solution to (\ref{eq:1stCHN}) \cite{FMW}. Here, $a_N$ are those 
appearing in the defining equation of the spectral surface $C_V$
(\ref{eq:spectral-surface}), and $N = 5$ for SU(5) GUT models and 
$N=4$ for SO(10) GUT models. Although we use a language in 
Heterotic string compactification and deal with the (spectral 
surface of) full rank-$N$ vector bundle on, it will not be difficult 
to extract the rank-1 or rank-2 part relevant to the matter 
curve and codimension-3 singularities. Because of the duality 
relation (\ref{eq:spec-data-duality}) the whole story below is 
for F-theory compactifications as well.

Let us first focus on a region around a type (a) codimension-3 
singularity point of SU(5) / SO(10) GUT models. There, only 
rank-2 part is relevant, and the relevant part of the spectral 
surface is defined by (\ref{eq:C-E6-A4}, \ref{eq:C2-E7-D5}), which 
are reproduced here as 
\begin{equation}
 \xi^2 + \tilde{a}_{N-1} \xi + \tilde{a}_{N} \simeq 0,
\end{equation}
where $\tilde{a}_{N-1} = a_{N-1}/a_{N-2}$ and 
$\tilde{a}_N = a_N/a_{N-2}$, and $(\tilde{a}_{N-1}, \xi)$ can 
be chosen as a set of local coordinates on the spectral surface.
The divisor $\gamma$ in (\ref{eq:expl-gamma}) can be expressed 
locally as 
\begin{equation}
 \gamma = {\rm div} \; \left( \xi^{\lambda N} \cdot \tilde{a}_N^{- \lambda}
		       \right).
\label{eq:div-form-gamma}
\end{equation}
When a divisor $D$ is locally given by ${\rm div} f_\alpha$, where 
$f_\alpha$ is a rational function, a line bundle ${\cal O}(D)$ 
has a description in terms of gauge field, with its singular component 
given by\footnote{This singular component does not depend on the choice 
of $f_\alpha$, because for another choice $f_\beta$, 
$f_\alpha/f_\beta$ should be holomorphic, and neither zero or infinite.} 
\begin{equation}
 i A \sim f_\alpha^{-1} \partial f_\alpha.
\end{equation}
Thus, for $\gamma$ in (\ref{eq:expl-gamma}, \ref{eq:div-form-gamma}), 
the singular component of the gauge field is 
\begin{equation}
 i A \sim \lambda \left[ (N-1) \frac{1}{\xi}d\xi - 
 \frac{1}{\xi+\tilde{a}_{N-1}}(d\xi + d \tilde{a}_{N-1})\right].
\end{equation}
We will be satisfied in this section \ref{sssec:4-form} only with 
following this singular component\footnote{This singular component
disappears in local trivialization frame of a line bundle, and smooth 
component remains. We still use only the singular component, because 
it is easily determined, and because just the singular component is
sufficient in illustrating a topological configuration of the
corresponding four-form flux around type (a) codimension-3 singularity 
points, as well as illustrating how the gauge field backgrounds for 
{\bf 2} and $\wedge^2 {\bf 2}$ representations are related.} 
around the type (a) codimension-3 singularities. 

It is also possible to obtain $2 \times 2$ matrix representation 
of the gauge field background for the field-theory formulation 
on $S$. 
The $\xi$ coordinate values of the degree-2 cover $C_V$ is 
\begin{equation}
 \xi_{\pm} = - (\tilde{a}_{N-1}/2) \pm 
  \sqrt{(\tilde{a}_{N-1}/2)^2 - \tilde{a}_N} 
\end{equation}
as functions of local coordinates $(\tilde{a}_{N-1}, \tilde{a}_N)$ 
of $S$. Using this expression for $\xi$, $iA$ above can be rewritten. 
If we denote $iA$ for $\xi_{\pm}$ branch as 
\begin{equation}
i A = \left( \begin{array}{cc}
       i A_+ & \\ & i A_-
	     \end{array}\right) =
 \left( \begin{array}{cc}
  i A^{\rm ave.} + i A^{\rm diff} & \\ & i A^{\rm ave.} - i A^{\rm diff}
	\end{array}\right),
\end{equation}
then 
\begin{eqnarray}
 i A^{\rm diff} & \sim & \frac{\lambda N}
    {2\sqrt{(\tilde{a}_{N-1}/2)^2 - \tilde{a}_N}  }
  \left(-d\tilde{a}_{N-1}+\frac{\tilde{a}_{N-1}}{2}d\tilde{a}_{N} \right), \\
 i A^{\rm ave.} & \sim & \lambda \frac{N-2}{2 \tilde{a}_N} d\tilde{a}_{N}.
\end{eqnarray}
$A^{\rm diff}$ becomes negative of what it is after going around 
the branch locus $\tilde{a}_{N-1}^2 - 4 \tilde{a}_N = 0$. 
Noting that the topological 2-cycle also becomes negative of 
what it is around the branch locus (as we saw in section \ref{sec:warm-up}), 
we see that the 3-form potential---product of $A^{\rm diff}$ and 
the 2-form to be integrated over the 2-cycle---is invariant under 
the monodromy.\footnote{Although a massless vector field for this 
$\mathfrak{su}(2)$ Cartan part is absent in the effective theory 
below the Kaluza--Klein scale because of the monodromy, topological 
background flux can be introduced in this Cartan part. These two 
things should not be confused.} 

The gauge field background for the $\bar{\bf 5}$ [resp. ${\bf vect.}$] 
representation is given by $\wedge^2 {\bf 2}$ representation 
of what we found above. Therefore, 
\begin{equation}
 i A^{\wedge^2 {\bf 2}} \sim \lambda (N-2) \frac{d \tilde{a}_N}{\tilde{a}_N}.
\end{equation}
It is needless to say that $- i A^{\wedge^2 {\bf 2}} + iA_+ + i A_- =
0$, corresponding to the fact that 
\begin{equation}
 \wedge^2 \bar{\bf 2} \otimes \wedge^2 {\bf 2} \subset 
 \wedge^2 \bar{\bf 2} \otimes {\bf 2} \otimes {\bf 2}
\end{equation}
contains a singlet of the structure group $\U(2)$.

\section{Yukawa Couplings}
\label{sec:Yukawa}

For Heterotic string compactification on a Calabi--Yau 3-fold $Z$, 
the chiral matter multiplets $\Phi_I$ correspond to
independent basis $\{ A_I \}$ of a vector space 
\begin{equation}
 H^1(Z; U)
\end{equation}
where the vector bundle $U$ is a part of ${\rm ad} E_8$ and 
is not necessarily assumed to be irreducible. The Heterotic string 
theory has a superpotential 
\begin{equation}
 \Delta W = \int_Z \Omega \wedge \tr {}_{E_8\; {\rm adj.}}
\left( A dA + i \frac{2}{3}A A A\right),
\label{eq:Het-super}
\end{equation}
and Yukawa interactions $\Delta W = \lambda_{IJK} \Phi_I \Phi_J \Phi_K$ 
are calculated by substituting $A = \sum_I A_I \Phi_I$ into 
the above superpotential. In Type IIB Calabi--Yau orientifold
compactification with D7-branes wrapped on holomorphic 4-cycles, 
the chiral multiplets are identified with certain global holomorphic sections 
on vector bundles (often line bundles) on complex curves of D7--D7 
intersection, and Yukawa couplings of three chiral multiplets 
are generated at points where three stacks of D7-branes intersect. 
Since the Type IIB string theory is formulated so that it can compute 
any short-distance physics processes, Yukawa couplings can be calculated 
although D7-branes and their intersection curves form a codimension-3
singularity there. 

References \cite{DW-1, BHV-1, HayashiEtAl} clarified how chiral 
matter multiplets are characterized in terms of geometry in F-theory, 
and the preceding sections of this article elaborate on it.
One of the most important motivations of studying F-theory 
compactification, as opposed to limiting to Type IIB Calabi--Yau orientifold 
compactification, is that Yukawa couplings of the form 
(\ref{eq:10105}) can be generated. However, it has been 
a long time puzzle how to calculate Yukawa couplings 
of low-energy effective theory for a given geometry for compactification.

It is not that we do not know anything about Yukawa interactions 
in F-theory. F-theory is regarded as the small fiber limit of 
M-theory compactification on an elliptic fibered manifold. The algebra 
of vanishing cycle suggests what kind of Yukawa couplings can be
generated \cite{TW1}. We have also learnt that such Yukawa couplings 
can be attributed to codimension-3 singularities \cite{HayashiEtAl}. 
In the absence of a microscopic quantum formulation of F-theory, however, 
it appears to be difficult to find a way to calculated Yukawa 
interactions that are associated with singularity points.

The field-theory formulation of ``particle physics sector'' of 
F-theory initiated by \cite{DW-1, BHV-1}, however, offers a solution.
The singular geometry is encoded as a background of a 8-dimensional gauge theory 
the and microscopic aspects (quantum UV completion 
of F-theory) may well be dealt with by 
$\alpha'$-(or $\kappa_{11}$-)expansion. It has a
superpotential \cite{DW-1, BHV-1} \footnote{
The F-term part of the BPS conditions 
(\ref{eq:BPS-F}, \ref{eq:BPS-phi}) follows from the superpotential 
(\ref{eq:F-super}). The F-term scalar potential from (\ref{eq:F-super})
gives rise to quartic interactions among zero mode scalar fields in 
effective field theory below the Kaluza--Klein scale. This interaction 
corresponds to the fact that not all the infinitesimal BPS 
deformation (massless modes) can simultaneously turned on together 
to obtain finite BPS deformation of the Higgs bundle. Yukawa
interactions are regarded as the supersymmetrization of this quartic 
interactions. } 
\begin{equation}
 \Delta W = \int_S \tr \left(\Phi^{(2,0)} \wedge F^{(0,2)} \right),
\label{eq:F-super}
\end{equation}
which is the F-theory counter part of the Heterotic result (\ref{eq:Het-super})
and of type IIB result

\begin{equation}
 \Delta W = \int_S \Omega_{lmn} 
 \tr \left(\zeta^l F^{(0,2)}_{\bar{m}\bar{n}}\right) 
  du_1 \wedge du_2 \wedge d\bar{u}_1 \wedge d\bar{u}_2
\end{equation}
in Type IIB string theory. In this section, we will study 
how (\ref{eq:F-super}) can be used to calculate Yukawa couplings 
of charged matter multiplets in most generic compactification 
of F-theory.

\subsection{Combining Local Models with Different Gauge Groups}

The first step of the process of calculating the Yukawa couplings 
is to identify the zero mode chiral multiplets. This cannot be done 
without looking at the entire compact discriminant locus $S$. 
The zero modes are 
\begin{equation}
 H^0 \left(\tilde{\bar{c}}; \tilde{\imath}^* K_S \otimes 
  {\cal O}\left(\tilde{\jmath}^* r / 2 \right) \otimes {\cal L}_{G} \right),
\label{eq:F-matter-abstract}
\end{equation}
global holomorphic sections of line bundles along the covering matter
curves, which are $\nu^*(\xi) = 0$ loci of desingularization of the 
spectral surfaces. The divisor $r$ is the ramification divisor of 
$\widetilde{C}$ over $S$, but only a geometry of $\widetilde{C}$ 
along $\tilde{\bar{c}}$ is needed to determine this divisor restricted 
on $\tilde{\bar{c}}$. See \cite{HayashiEtAl} for more details.\footnote{ 
All topological aspects of elliptic fibered Calabi--Yau 4-fold 
$X$, $B_3$, $S$ and four-form flux on $X$ have to be arranged properly, 
so that there are three independent global holomoprhic sections 
corresponding to left-handed quark doublets etc., and there is 
just one corresponding to up-type Higgs doublet, for example.
Finding an explicit example of topological choice of $X$, $B_3$, $S$ 
and four-form flux right for the minimal supersymmetric standard model 
is not the subject of this article. We just assume that there are 
the right number of independent global holomorphic sections for certain 
choice of geometry, and we just move on.}

By definition, line bundles have descriptions of local trivialization, 
where holomorphic sections are expressed as holomorphic functions in 
any local patches covering the curve.\footnote{At the same time, 
the singular component of the gauge-field background that we discussed 
in section \ref{sssec:4-form} also goes away when we take the trivialization 
frame.} Zero modes, and hence the global 
holomorphic sections are cast into holomorphic functions in any 
given local patches within $S$. Let us take a basis of the vector space 
(\ref{eq:F-matter-abstract}) for representation $R$, 
$\{\tilde{f}_{R;i}\}_{i \in I_R}$ and
denote the holomorphic function expression of $\tilde{f}_{R;i}$ 
in a trivialization patch $U_\alpha$ as $f_{R;i \alpha}$.

The Yukawa couplings of charged matter multiplets are expected to 
be generated around points of codimension-3 singularities. 
Thus, one chooses a local patch $S_\alpha \subset S$ around a given 
point $p_\alpha$ of codimension-3 singularity in $S$, so that 
the intersection of $\tilde{\pi}^*_{C}(S_\alpha)$ and (covering) 
matter curves $\tilde{\bar{c}}$ is contained in a trivialization 
patch $U_\alpha$ on the (covering) matter curves. For such a
given local patch $S_\alpha$ an appropriate field-theory local model 
with an appropriate gauge group and background is chosen from those 
in sections~\ref{sec:warm-up} and \ref{sec:GUT}. 
The zero-mode wavefunctions 
$(A_{\bar{m};R;i\alpha},\varphi_{12; R;i\alpha})$ are determined 
on a patch of desingularized spectral surface 
$\tilde{\pi}_{C_U}^{-1}(S_\alpha)$ from the holomorphic functions 
$f_{R;i\alpha}$ $(i \in I_R)$. The holomorphic functions $f_{R;i \alpha}$ 
on the curve $\tilde{\bar{c}}_U$ are regarded as holomorphic functions 
on (desingularized) spectral surface 
$\tilde{\pi}^*_{C_U}(S_\alpha) \subset \widetilde{C}_U$, modulo those
that vanish along the curve $\tilde{\bar{c}}_U$.
We found in sections \ref{sec:warm-up}, \ref{sec:GUT} and the 
appendix \ref{sec:0-mode} that the zero-mode wavefunctions are 
smooth on the desingularized spectral surfaces. 
From knowing a wavefunction $(A_{\bar{m};R;i\alpha},\varphi_{12;R;i\alpha})$ 
on the desingularized spectral surface, a wavefunction on $S_\alpha$ is 
constructed. This wavefunction becomes multi-component, if 
$\tilde{\pi}_{C_U}: \tilde{\pi}^{-1}_{C_U}(S_\alpha) \rightarrow S_\alpha$ 
is a multi-degree cover. Note that the same gauge field configuration 
has to be used for different representations in determining the 
wavefunctions in a given patch $S_\alpha$.
We have already studied in sections \ref{sec:GUT} all types of 
codimension-3 singularities that appear in F-theory compactifications 
for SU(5) or SO(10) models, and found that zero-modes are either 
in the Gaussian form \cite{KV, BHV-1} (when the fields are in 
linear Abelian background) or in a form studied 
in section \ref{ssec:AN-gen} and the appendix \ref{sec:0-mode} when fields are in the doublet background.
This is how the field-theory {\it local} models on complex {\it surfaces} 
with specific choices of their own (often non-Abelian) structure groups 
are combined by {\it global} descriptions of zero-modes on compact complex 
{\it curves} given in terms of (Abelian) line bundles.

Such wavefunctions are inserted into 
(\ref{eq:F-super}), and integration should be carried out over the 
patch $S_\alpha$. The wavefunctions are expected to damp very quickly 
in a direction normal to matter curves even in the doublet background, 
as we saw in sections \ref{sec:warm-up}, \ref{sec:GUT} and the appendix 
\ref{sec:0-mode}. 
If the size of the discriminant locus $S$ is 
sufficiently large compared with the width of the exponentially 
decaying zero-mode wavefunctions,\footnote{We will discuss whether this 
assumption is reasonable late on.} the integration is likely to be 
concentrated around the points of codimension-3 singularity. 
Because of the $i A \wedge A$ term in $F = dA + i A \wedge A$, 
three wavefunctions can be plugged in. When the wavefunctions for 
$\tilde{f}_i$, $\tilde{f}_j$ and $\tilde{f}_k$ ($i,j,k \in I$) are 
plugged in, the commutator algebra of the structure group determines 
what kind of Yukawa couplings are generated, and the integration yields 
a contribution to the Yukawa couplings $\lambda_{ijk;\alpha}$ associated 
with the point of codimension-3 singularity $p_\alpha$.

\subsection{Yukawa Couplings from a Single Codimension-3 Singularity Point}

There is nothing more to add in this article about Yukawa couplings 
generated around a codimension-3 singularity point where the structure 
group of its local model is Abelian, and three matter curves pass through 
the point. 

What we call type (d) codimension-3 singularities of SU(5) models 
satisfy this criterion for generic complex structure of Calabi--Yau 
4-fold. That is where Yukawa couplings of the form 
(\ref{eq:5105})---down-type Yukawa and charged lepton Yukawa couplings 
in Georgi--Glashow SU(5)---are generated.  
Three matter curves run through any given point of type (d) 
codimension-3 singularity, and wavefunctions are Gaussian 
in the normal directions of the matter curves for all three matter 
curves.
One just needs to calculate the overlap of Gaussian wavefunctions.
Because of the $\mathfrak{so}(12)$ algebra, one wavefunction is picked 
up from one of the two branches of the (covering) matter curve of 
$\bar{\bf 5}$ representation, and another from the other branch.
If the width of the Gaussian profile is sufficiently small compared 
with typical radius of the compact discriminant locus $S$, then 
the Yukawa coupling $\lambda_{ijk;\alpha}$ from a given codimension-3 
singularity point is approximately given by \cite{BHV-1}
\begin{equation}
 \lambda_{ijk;\alpha} = c_\alpha f_{i;\alpha}(p_\alpha) \; f_{j;\alpha}(p_\alpha) 
  \; f_{k; \alpha}(p_\alpha),
\label{eq:dtype-rank1}
\end{equation}
where $f_{i, j,k;\alpha}(p_\alpha)$ are values of those holomorphic functions at 
the codimension-3 singularity point, and $c_\alpha$ is a numerical
constant that depends on the width parameter of the Gaussian profile 
and the angle of the intersection of the curves. 

The Yukawa coupling of the form (\ref{eq:10105})---up-type Yukawa couplings 
of Georgi--Glashow SU(5)---or (\ref{eq:1616vect}) in SO(10) models, 
on the other hand, are expected to be generated at type (a) 
codimension-3 singularity points of those compactifications. 
For generic choice of complex structure, only two matter curves 
run through the codimension-3 singularity points. The Yukawa couplings 
at such singularities or even the field theory local models 
for this type of codimension-3 singularities have not been studied 
so far. Since the discussion proceeds exactly the same way for 
the Yukawa couplings (\ref{eq:10105}) for SU(5) models and 
(\ref{eq:1616vect}) for SO(10) models, we will only discuss the 
SU(5) cases in the following.

The first thing to check is the algebra. A local geometry around 
a type (a) codimension-3 singularity point of SU(5) models is a family 
of deformed $E_6$ singularity.
The irreducible decomposition of $E_6$ 
algebra in terms of $\U(2) \times \SU(5)$ subgroup
is given in (\ref{eq:e6-decomp}). 
If we denote the generators of $E_6$ in the 
$({\bf 2}, {\bf 10})$ component as $t_{{\bf 10}{A; ab}}$, 
the ones in the $(\wedge^2 {\bf 2}, \bar{\bf 5})$ component as $t_{\bar{\bf 5}}^b$, 
and the ones in the $(\wedge^2 \bar{\bf 2}, {\bf 5})$ as $t_{{\bf 5};a}$, then 
\begin{equation}
 \left[ t_{{\bf 10}{A; ab}}, t_{{\bf 10}{B; cd}} \right] 
 \propto \epsilon_{AB} \epsilon_{abcde} t_{\bar{\bf 5}}^e,
\end{equation}
and hence 
\begin{equation}
 \tr {}_{E_6~{\rm adj.}} \left(t_{{\bf 5};e} 
   [ t_{{\bf 10}{A; ab}}, t_{{\bf 10}{B;cd}}]\right) 
  \propto \epsilon_{AB} \epsilon_{abcde}.
\end{equation}
Here, $A,B = 1,2$ are indices labelling $\U(2)$ doublets, and 
$a,b,c,d,e = 1, \cdots, 5$ are SU(5) indices. The totally
anti-symmetric tensor $\epsilon_{abcde}$ comes from the 
structure constant of $\mathfrak{e}_6$ Lie algebra. The doublet 
indices are contracted in order to extract singlet part out of 
\begin{equation}
 \wedge^2 \bar{\bf 2} \otimes {\bf 2} \otimes {\bf 2}.
\end{equation}

The $E_6$-valued gauge field on $S_\alpha$ is expanded as 
\begin{eqnarray}
 A_{\bar{m}}(x,y) & = & \vev{A_{\bar{m}}}(y) + 
 t_{{\bf 10}{A; ab}} 
 A_{\bar{m}; {\bf 10};i\alpha}^{A}(y)\phi_{{\bf 10};i}^{ab}(x) 
\nonumber \\ 
 & + &  
 t_{{\bf 5};e}
 A_{\bar{m}; {\bf 5};\alpha}(y) h^e(x) + 
 t_{\bar{\bf 5}}^a
 A_{\bar{m}; \bar{\bf 5}; k\alpha}(y) \phi_{\bar{\bf 5};a;k}(x) + \cdots.
\end{eqnarray}
Here, $A_{\bar{m}; {\bf 10};i\alpha}^{A}(y)$ is the $\U(2)$-doublet 
wavefunction in $S_\alpha$ for a zero mode $\tilde{f}_{{\bf 10};i}$, and 
$\phi_{{\bf 10};i}^{ab}(x)$ the corresponding complex scalar field 
in the SU(5)-{\bf 10} representation in the low-energy effective 
theory. $A_{\bar{m}; {\bf 5};\alpha}(y)$ and 
$A_{\bar{m}; \bar{\bf 5}; k\alpha}(y)$ are zero-mode wavefunctions 
in the $\wedge^2 \bar{\bf 2}$ and $\wedge^2 {\bf 2}$ representations 
of the structure group $\U(2)$, and $h^e(x)$ and 
$\phi_{\bar{\bf 5};a;k}(x)$ their corresponding scalar fields 
in the effective theory. Another field $\varphi_{12}$ is also 
expanded similarly. Because of the structure constant of
$\mathfrak{e}_6$, the up-type Yukawa couplings 
$\Delta W = \lambda_{ij} \; \phi^{ab}_{{\bf 10};i} \; \phi^{cd}_{{\bf 10};j} \; h^e
 \epsilon_{abcde}$ in Georgi--Glashow SU(5) models have a contribution 
\begin{equation}
 \lambda_{ij; \alpha} \sim \int_{S_\alpha} \epsilon_{AB} \left(
  \varphi_{{\bf 5};\alpha} \wedge A_{{\bf 10};i\alpha}^A 
  \wedge A_{{\bf 10}; j\alpha}^B 
- \left(\varphi_{{\bf 10};i\alpha}^A \wedge A_{{\bf 5};\alpha} 
        \wedge A_{{\bf 10};j\alpha}^B + (i\leftrightarrow j)  \right) \right)
\label{eq:up-Yukawa-overlap}
\end{equation}
from a local region $S_\alpha$ around a type (a) codimension-3 
singularity point $p_\alpha$. 

Let us take a set of local coordinates $(u_1,u_2)$ on $S_\alpha$, 
so that $\tilde{a}_4 \sim 2u_1$ and $\tilde{a}_5 \sim u_2$. 
Then, the $\U(2)$-singlet wavefunction of the Higgs boson 
$(A_{\bar{m}},\varphi)$ vanishes in the $A_{\bar{2}}$ component, 
and the other two components are localized along the matter curve 
$u_1 = 0$. The wavefunction in the remaining two components 
decay as in a Gaussian form $e^{-|u_1|^2}$.
Therefore, the behavior of the quark wavefunctions 
is relevant to the overlap integration (\ref{eq:up-Yukawa-overlap}) 
only in the large $|u_2|$ and small $|u_1|$ region, which is just the behavior 
already studied in sections \ref{sec:warm-up}, \ref{sec:GUT} and 
the appendix \ref{sec:0-mode}. 
The $\U(2)$-doublet wavefunctions of quarks vanish in the $A_{\bar{1}}$ 
component at the leading order, $\varphi^{A=1} = \varphi^{A=2}$ 
and $A_{\bar{2}}^{A=1} = - A_{\bar{2}}^{A=2}$, and these two components 
$(A_{\bar{2}}^A,\varphi^A)$ decay as $e^{-(4/3)|u_2|^{3/2}}$ for large 
$|u_2|$. Using this information, we see first that the first term 
in (\ref{eq:up-Yukawa-overlap}) vanishes at the leading order. 
The second (and the third) term does not vanish; neither  
$\varphi^{A=1}_{12;{\bf 10}i\alpha} \; A_{\bar{1}; {\bf 5}\alpha} \; 
 A^{B=2}_{\bar{2};{\bf 10};j\alpha}$ nor 
$\varphi^{A=2}_{12;{\bf 10}i\alpha} \; A_{\bar{1}; {\bf 5}\alpha} \; 
 A^{B=1}_{\bar{2};{\bf 10};j\alpha}$ vanish, and they are different only 
in sign. Being multiplied by the anti-symmetric structure constant 
$\epsilon_{AB}$, the last two terms of (\ref{eq:up-Yukawa-overlap}) 
give rise to non-vanishing unsuppressed contributions to the 
up-type Yukawa couplings. The overlap integration is concentrated 
around the type (a) codimension-3 singularity point at $(u_1,u_2)=(0,0)$,
because of the exponentially decaying product of wavefunctions 
$e^{-(8/3)|u_2|^{3/2} - |u_1|^2}$.

References \cite{BHV-1, BHV-2} considered that there are always 
three matter curves intersecting at a point of $E_6$ enhanced 
singularity, and the up-type Yukawa couplings were considered 
to be generated only at triple intersection of matter curves. 
This is why \cite{BHV-2, Ibanez, HV-2} were led to consider that complex 
structure has to be chosen so that the matter curve of 
$\SU(5)$-{\bf 10} representation has a double point right 
on the matter curve of $\SU(5)$-${\bf 5}$ representation, and  
diagonal entries of the up-type Yukawa matrix are generated. 
As we saw above, however, for generic choice of complex structure 
moduli in F-theory compactifications, the type (a) codimension-3 
singularity points of SU(5) models have only two intersecting 
matter curves, and yet the up-type Yukawa couplings (\ref{eq:10105}) 
are generated. 
In the evaluation of the overlap integration above, generation indices 
$i,j$ did not play an important role. The argument is valid for 
diagonal and off-diagonal entries alike.\footnote{
In Heterotic $E_8 \times E_8$ string compactification, the up-type 
Yukawa couplings comes from 
\begin{equation}
  H^1(Z; \wedge^2 V^\times) \times H^1(Z;V) \times H^1(Z; V) 
  \rightarrow H^3(Z; {\cal O}_Z),
\end{equation}
where $V$ is a rank-5 (rank-4 for SO(10) models) vector bundle 
turned on on a Calabi--Yau 3-fold $Z$. In F-theory compactification, 
we do the same thing; vector bundles on $Z$ are replaced by 
Higgs bundles on $S$, and that is basically it. In Heterotic 
string compactification, one is not usually worried whether diagonal 
entries are generated or not. It is dangerous, though, to rely too 
much on such an intuition coming from duality, because even in a 
compactification described by both theories, Heterotic string and F-theory 
provide good descriptions of different parts of the moduli space. 
} Thus, such a specific choice of 
complex structure with the double points of 
the $\SU(5)$-{\bf 10} representation matter curve on the $\SU(5)$-${\bf 5}$
representation matter curve is not required from phenomenology. 
Just a generic complex structure is fine in obtaining a large 
third generation Yukawa coupling of the up-type quark.

If one assumes that the wavefunction $f_{{\bf 10};i\alpha}$ and 
$f_{h;\alpha}$ vary only slowly on their matter curves around $p_\alpha$, 
then the same argument for the down-type Yukawa couplings is applied 
to the up-type Yukawa matrix. Just like in (\ref{eq:dtype-rank1}), 
the up-type Yukawa matrix is rank-1 at the leading order.\footnote{
If the matter curve of the $\SU(5)_{\rm GUT}$-${\bf 10}$ representation 
is irreducible (and hence all the three independent zero modes are 
localized on it) and has a double point at $p^{up}_{\alpha}$, so that 
a triple intersection of matter curves is realized by a self-pinching 
irreducible matter curve of {\bf 10} representation, then the 
up-type Yukawa matrix $\lambda_{ij}$ of 
$\Delta {\cal L} = - \lambda_{ij} u^c_i q_j h_u$ is given by 
\begin{equation}
\lambda_{ij} \sim f_{h;\alpha}(p^{up}) \; 
 \left( f_{q;j\alpha}(p^{up}_{3\tau;\alpha}) 
        f_{u^c;i\alpha'}(p^{up}_{-3(\tau+\sigma);\alpha})
      + f_{u^c;i\alpha}(p^{up}_{3\tau;\alpha}) 
        f_{q;j\alpha'}(p^{up}_{-3(\tau+\sigma);\alpha})
 \right)
\label{eq:rk-2-selfpinch}
\end{equation}
at the leading order. Note that $p^{up}_{3\tau;\alpha}$ and 
$p^{up}_{-3(\tau+\sigma);\alpha}$ are projected to the same point 
$p^{up}_{\alpha}$, but they are different points on the covering matter 
curve, where the double point of the matter curve is resolved. 
Holomorphic sections on the two irreducible branches of the 
matter curve of the {\bf 10} representation around the double point
are independent (see discussion in p. \pageref{page:covering4tripleintersect}
if necessary; $f(p^{up}_{3\tau})$ are values of $f_{{\bf 1}_+}(u_1)$ 
at $u_1=0$, and $f(p^{up}_{-3(\tau+\sigma)})$ those of $f_{{\bf 1}_-}(u')$). 
Certainly one can choose a basis of the three quark 
doublets so that $f_{q;j} \sim u_1^{3-j}$ around 
$p^{up}_{3\tau;\alpha}$ on the $3\tau = 3u_2=0$ branch of the 
covering matter curve, and the basis of the three anti-up-type 
quarks similarly on the $-3(\tau+\sigma)=0$ branch. 
The first term of (\ref{eq:rk-2-selfpinch}) does not vanish 
only in the (3,3) entry in this basis, and is clearly rank-1.
The second term is also another rank-1 matrix, and is generically 
different from the first one. Thus, the situation here predicts 
a rank-2 up-type Yukawa matrix at the leading order. 
}

The Yukawa matrix (\ref{eq:dtype-rank1}) and its up-type counter 
part are both rank-1, whereas the Yukawa matrices are supposed 
to be rank-3. Reference \cite{HV-2} proposed that there are only 
one pair of type (d) and type (a) codimension-3 singularity points, 
$p^{down}_{\alpha=1}$ and $p^{up}_{\alpha = 1}$ in the SU(5) discriminant 
locus, and showed that the derivative expansion of wavefunctions\footnote{
Reference \cite{Gaussian} also discussed the derivative expansion of 
wavefunctions as one of the origin of the hierarchical mass eigenvalues.} 
$f_{R;i\alpha}$ around $p^{down}_{\alpha=1}$ and distortion of 
wavefunctions due to the presence of fluxes give rise to 
deviation from the rank-1 form, suppressed typically by powers 
of $\sqrt{\alpha_{\rm GUT}}$. Although the discussion in \cite{HV-2}
explicitly relies on codimension-3 singularities at a triple 
intersection of matter curves, the idea of controlling 
the corrections from derivative expansion and flux distortion 
is more general, and Froggatt--Nielsen-like pattern will be 
expected also for the up-type Yukawa matrix generated around 
a type (a) codimension-3 singularity with only two matter curves 
intersecting normally. Details of the mass-eigenvalue pattern may 
or may not be different from the prediction of \cite{HV-2}. 
This is an interesting question, because the hierarchy between the 
up-quark and top-quark Yukawa couplings is larger than that between 
the down-type and bottom-quark Yukawa couplings, but we leave it 
as an open problem.   

There is another contribution to Yukawa matrices, even when 
there is only one codimension-3 singularity point of a given type.
To see this, we begin with the following observation.
As noted in \cite{BHV-2}, 
the GUT unification gauge coupling and the GUT scale are 
given by 
\begin{eqnarray}
 \frac{1}{\alpha_{\rm GUT}} & \sim & 
  \frac{1}{g_s}\left(\frac{L}{l_s}\right)^4, \\
 M_{\rm GUT} & \sim & \frac{1}{L},
\end{eqnarray}
where $L$ is the typical size of $S$, $l_s$ the string length and 
$g_s$ ``typical'' dilaton vev. The GUT symmetry breaking is assumed 
to be due to fluxes in the U(1)$_{\rm Y}$ direction, and hence 
the GUT scale is identified with the Kaluza--Klein scale.
Here, Type IIB language is used just to make a crude estimate 
of the parameters of compactification.
The definition of $l_s$ above is $2\pi \sqrt{\alpha'}$; 
factors such as $2$ are ignored in the equations above, but 
that would not significantly affect the estimate of $L/l_s$.
In F-theory, however, it is not actually clear what this $g_s$ means. 
Since we are interested in F-theory configurations that have enhanced 
singularity of exceptional type, some mutually non-local 7-branes are 
involved. The dilation vev $g_s$, therefore does not remain constant 
because of $\SL_2 \Z$ monodromy, and in particular, it cannot remain 
much smaller or much larger than unity. Setting $g_s \approx {\cal O}(1)$ 
as a crude approximation, and using experimentally inferred value 
$\alpha_{\rm GUT} \sim 1/24$, we conclude that 
\begin{equation}
 \frac{L}{l_s} \sim (24)^{\frac{1}{4}}  \sim 2\mbox{--}3.
\label{eq:ratio}
\end{equation}

We have so far assumed that the zero-mode wavefunctions are 
localized along the matter curves quite well. To what extent, 
however, is this assumption correct?
Wavefunctions are in the Gaussian profile in the normal 
directions of the matter curves (except at codimension-3 
singularities), and its width parameter $d$ is given by the 
$d \sim 1/\sqrt{F}$, where $F$ is the coefficient of 
$\varphi_{12} \sim F u$, and $u$ is the normal coordinate. 
If the typical value of $F$ is of order unity in unit of 
$(1/l_s)^2 = 1/[(2\pi)^2 \alpha']$, then 
$(L/d) \sim 2\mbox{--}3$. Gaussian tails of wavefunctions on 
near-by matter curves have significant overlap, and the picture 
we have assumed is not actually correct. 

It is not clear what the right choice of ``string unit'' is; 
$\sqrt{\alpha'}$ instead of $l_s = (2\pi) \sqrt{\alpha'}$ might be 
the right characteristic 
scale of ``stringy'' regime. If the coefficient $F$ is typically 
of order unity in unit of $(1/\alpha')$, then $(L/d) \sim 15$. 
Reference \cite{Gaussian} showed that the overlapping tails of Gaussian 
wavefunctions alone generate hierarchical Yukawa eigenvalues of 
the right amount of hierarchy for $(L/d) \sim 10$.
One can ignore this contributions from overlapping Gaussian tails, 
only when $(L/d)^2 \gg (10)^2$. Since the set-up in \cite{Gaussian} 
is not exactly the same as the situation here, it is a little 
dangerous to use the criterion $10^2$ literally, and conclude that 
$(L/d) \sim 15$ is safe. 
In principle, typical values of $F$ relatively to $1/\alpha'$ 
can be studied by flux compactification. Such study will tell 
us whether this contribution can be ignored, or 
rather accounts for the hierarchical Yukawa eigenvalues.

\section*{Acknowledgements} 

This work is supported by Global COE Program "the Physical 
Sciences Frontier", MEXT, Japan (HH), 
a Grant in-Aid No. 19540268 from the MEXT, Japan (TK), 
PPARC (RT), and by WPI Initiative, MEXT, Japan (TW).


 \appendix

\section{$D=8$ Field Theory Lagrangian for 7-Branes}
\label{sec:action}

Reference \cite{BHV-1} obtained a local field theory 
Lagrangian of 8-dimensions, which can be used to study 
localization of matter multiplets on 7-branes as well 
as how Yukawa couplings are generated. In this appendix, 
we quote the Lagrangian \cite{BHV-1} with corrected 
coefficients, along with detailed explanation of conventions.

{\bf Conventions}

The Lagrangian is described in terms of a vector multiplet 
$V^a = (A^a_\mu(x,y),\eta^a_\alpha(x,y),\bar{\eta}^a(x,y), 
D^a(x,y) )$ 
and chiral multiplets 
$A_{\bar{m}}(x,y) = (A_{\bar{m}},\psi_{\alpha, \bar{m}}, {\cal
G}_{\bar{m}})$ and 
$\Phi_{mn}(x,y) = (\varphi_{mn},\chi_{\alpha,mn}, {\cal H}_{mn})$. 
$V^a t^a$ takes its value in $\mathfrak{g}$; $V^a$ is real and 
generators $t^a$ of Lie algebra $\mathfrak{g}$ are in a Hermitian 
representation. $A_{\bar{m}}(x,y)$ and $\Phi_{mn}(x,y)$, on the other 
hand, take their values in $\mathfrak{g} \otimes \C$.

We stick to all the conventions adopted in \cite{WessBagger}, 
and in particular, the metric is $+$ in space directions.
$V^a$ is real as in \cite{WessBagger}, as opposed to the convention 
in \cite{BHV-1}, so that the SUSY transformation rule 
in \cite{WessBagger} can be used without modification.
The covariant derivatives are of the form 
\begin{equation}
 \partial_\mu + i A^a_\mu t^a,
\end{equation}
and hence field strength is 
\begin{equation}
F_{\mu\nu} = \partial_\mu A_\nu - \partial_\nu A_\mu + i [A_\mu, A_\nu]. 
\end{equation}

Holomorphic coordinates $u^m$ ($m=1,2$) are introduced in the 
internal 4-dimensions. Four real local coordinates $y_{4,5,6,7}$ are 
combined into $u^1 = y_4 + i y_5$ and $u^2 = y_6 + i y_7$.
Derivatives and gauge fields in the holomorphic coordinate system 
are defined by 
\begin{equation}
 \partial_1 \equiv \frac{1}{2}\left(\partial_4 - i \partial_5 \right),
   \quad 
 \bar{\partial}_{\bar{1}} \equiv 
           \frac{1}{2}\left(\partial_4 + i \partial_5 \right), \qquad 
 A_1 = \frac{1}{2}(A_4 - i A_5), \quad 
 A_{\bar{1}} = \frac{1}{2}(A_4 + i A_5), \label{eq:hol-coord-1}
\end{equation}
\begin{equation}
 \partial_2 \equiv \frac{1}{2}\left(\partial_6 - i \partial_7 \right),
    \quad 
 \bar{\partial}_{\bar{2}} \equiv 
           \frac{1}{2}\left(\partial_6 + i \partial_7 \right), \qquad 
 A_2 = \frac{1}{2}(A_6 - i A_7), \quad 
 A_{\bar{2}} = \frac{1}{2}(A_6 + i A_7),
\label{eq:hol-coord-2}
\end{equation}
Covariant derivatives are of the form $\partial_m + i A_m$ and 
$\bar{\partial}_{\bar{m}} + i A_{\bar{m}}$ ($m=1,2$). 
The lowest (complex scalar in $\R^{3,1}$) components of 
chiral multiplets $A_{\bar{m}}(x,y)$ are identified with 
the gauge fields $A_{\bar{m}}$ defined as above, and hence 
they have unambiguous normalization now. 

Using the local holomorphic coordinates $u^m$ ($m=1,2$),  
K\"{a}hler metric is given by 
\begin{equation}
 ds^2 = \frac{1}{2}g_{m\bar{n}}du^m \otimes d\bar{u}^{\bar{n}}
   + \frac{1}{2} g_{n\bar{m}} d\bar{u}^{\bar{m}} \otimes du^n,
\end{equation}
where $g_{m\bar{n}}$ is Hermitian, and the K\"{a}hler form by 
\begin{equation}
 \omega = \frac{i}{2} g_{m\bar{n}} du^m \wedge d\bar{u}^{\bar{n}},
\end{equation}
so that 
\begin{equation}
 \frac{1}{2} \omega \wedge \omega = 
  \left(\frac{i}{2} du^1 \wedge d\bar{u}^{\bar{1}}\right) \wedge 
  \left(\frac{i}{2} du^2 \wedge d\bar{u}^{\bar{2}}\right) \; 
  ({\rm det} g_{m\bar{n}}) = {\rm vol}
\end{equation}
is the volume form. Euclidean metric 
$ds^2 = \sum_{k=4}^7 d y_k \otimes dy_k$ corresponds to taking 
$g_{m\bar{n}} = \delta_{mn}$.

Differential forms can be used in writing down the Lagrangian.
Following the convention of \cite{BHV-1}, 
\begin{equation}
 \bar{A} = A_{\bar{m}} d\bar{u}^{\bar{m}}, \qquad
 \psi_{\alpha} = \psi_{\alpha, \bar{m}} d\bar{u}^{\bar{m}}, \qquad 
 {\cal G} = {\cal G}_{\bar{m}} d\bar{u}^{\bar{m}}, 
\end{equation}
\begin{equation}
 \varphi = \varphi_{mn} du^m \wedge du^n, \quad 
 \chi_\alpha = \chi_{\alpha, mn} du^m \wedge du^n, \quad  
 {\cal H} = {\cal H}_{mn} du^m \wedge du^n. 
\end{equation}
Note that there is no $1/2$ inserted in the definition of these 
(2,0)-forms. On the other hand, we use a notation 
\begin{equation}
 F^{(2,0)} = \frac{1}{2} F_{mn} du^m \wedge du^n, \quad 
 F^{(0,2)} = \frac{1}{2} F_{\bar{m}\bar{n}} 
   d \bar{u}^{\bar{m}} \wedge d \bar{u}^{\bar{n}}, \quad 
 F^{(1,1)} 
           = F_{m\bar{n}} du^m \wedge d\bar{u}^{\bar{n}}. 
\end{equation}
\begin{equation}
 F^{(1,0)}_\mu = F_{\mu m} du^m, \qquad 
 F^{(0,1)}_\mu = F_{\mu \bar{m}} d \bar{u}^{\bar{m}}.
\end{equation}
Here, $F_{MN} = \partial_M A_N - \partial_N A_M + i [A_M , A_N]$, as 
before, where $M, N$ can be any one of 
$m=1,2$, $\bar{m}=\bar{1},\bar{2}$ in 
(\ref{eq:hol-coord-1}, \ref{eq:hol-coord-2}) and 
$\mu=0,\cdots,3$.

{\bf Lagrangian}

All the conventions are fixed, and now one can follow the logic 
of the appendix C of \cite{BHV-1} to obtain a local field theory 
Lagrangian of 8-dimensions. Purely bosonic part is 
\begin{eqnarray}
 {\cal L}_{8D}^{\rm bos} & \propto & 
  \frac{1}{2} \tr {}' \left[ 
    \omega \wedge \omega \left( 
       \frac{1}{2} D^2 - \frac{1}{4} F_{\mu \nu} F^{\mu\nu}
       - \frac{\theta}{8} F_{\mu\nu} F_{\kappa\lambda} 
         \epsilon^{\mu\nu\kappa\lambda}
                         \right)
      \right. \nonumber \\
  & & \quad \left. + 2 \omega \left(
        i {\cal G}^* \wedge {\cal G} - D F^{(1,1)} 
      - i F_{\mu}^{(1,0)} \wedge F^{(0,1) \mu}  
                        \right)  \right. \nonumber \\
  & & \quad \left. 
      - 2 \alpha^* \left( F^{(2,0)} \wedge \overline{\cal H} 
               + {\cal G}^* \wedge \partial \overline{\varphi} \right)
      - 2 \alpha \left( {\cal H} \wedge F^{(0,2)} 
               + {\cal G} \wedge \bar{\partial} \varphi \right)
             \right. \nonumber \\
  & & \quad \left. 
       + |\alpha|^2 \left( {\cal H} \wedge \overline{\cal H} 
		       + [\varphi, \overline{\varphi} ] D 
                       - D_\mu \varphi D^\mu \overline{\varphi}
                 \right)          
             \right],
\label{eq:bosonic}
\end{eqnarray}
and a part bilinear in fermions is given by 
\begin{eqnarray}
 {\cal L}_{8D}^{\rm f-bilinear} & \propto & \frac{1}{2} \tr {}' \left[
   \omega \wedge \omega i (D_\mu \eta) \sigma^\mu \bar{\eta}
  - 2 \omega \wedge (D_\mu \bar{\psi}) \wedge \bar{\sigma}^\mu \psi
   + |\alpha|^2 i (D_\mu \chi) \wedge \sigma^\mu \bar{\chi} 
 \right. \nonumber \\
  & & \quad +
   \left(2 \sqrt{2} i \omega \wedge \partial \eta \wedge \psi
    + 2 \sqrt{2} i \omega \wedge \bar{\psi} \wedge \bar{\partial}\bar{\eta}
   \right) \nonumber \\
  & & \quad + |\alpha|^2 
   \left( \sqrt{2} i [ \overline{\varphi}, \eta ] \wedge \chi 
        + \sqrt{2} i \bar{\chi} \wedge [ \varphi, \bar{\eta}]
   \right) \nonumber \\
  & & \left. \quad + 
   2 \alpha^* \left(\bar{\chi} \wedge \partial \bar{\psi} 
         - \frac{i}{2} \bar{\psi}[ \overline{\varphi}, \bar{\psi}]
     \right) 
   + 2 \alpha \left(\chi \wedge \bar{\partial} \psi 
         - \frac{i}{2} \psi [ \varphi, \psi ] \right) \right].
\label{eq:fermi-bilin}
\end{eqnarray}
The chiral multiplet $\Phi = (\varphi, \chi, {\cal H})$ appears 
in this Lagrangian only in a form $\alpha \Phi$, where $\alpha$ 
is a complex-valued parameter. This reflects the fact that 
we have not fixed a normalization of $\Phi$ yet. Different 
value of $\alpha$ corresponds to different normalization of 
$\Phi$, and does not have any physical meanings. 
$\tr'$ stands for $T_R^{-1} \tr_R$ for a representation $R$ 
of $\mathfrak{g}$, and $T_R$ its Dynkin index.

Among the various terms in the Lagrangian 
(\ref{eq:bosonic}--\ref{eq:fermi-bilin}), those involving 
$\omega \wedge \omega$ correspond to $I_3$ in (C.13) and 
those proportional to $2 \omega$ to $I_4$ in (C.17) of \cite{BHV-1}.
Terms involving only $\alpha \Phi$ or $\alpha^* \Phi^\dagger$ 
correspond to $W_S$ in (C.18) and $I_2$ in (C.10) of \cite{BHV-1}, 
respectively, and those involving $|\alpha|^2 \Phi^\dagger \Phi$
to $I_1$ in (C.9) of \cite{BHV-1}. The relative normalization 
among those four groups of terms are determined by requiring 
that the gauge-field kinetic terms becomes $\SO(7,1)$ invariant. 
By integrating out auxiliary fields ${\cal H}$ and $D$ from 
(\ref{eq:bosonic}), one finds indeed that the gauge-field 
kinetic terms are 
\begin{eqnarray}
& & - \frac{1}{4} \frac{\omega \wedge \omega}{2} \tr {}' \left[ 
  F_{\mu\nu} F^{\mu \nu}
+ 2 \times \left(F_{\mu m} F^{\mu}_{\; \bar{n}} (2g^{\bar{n}m})
+ F_{\mu \bar{m}} F^\mu_{\; n} (2 g^{\bar{m}n}) \right)\right] 
 \nonumber \\
& & - \frac{1}{4} \frac{\omega \wedge \omega}{2} \tr {}' \left[
 2 \times F_{\bar{m}n} F_{m\bar{n}}(2g^{\bar{m}m})(2g^{\bar{n}n})
- 2 \times F_{\bar{m}\bar{n}} F_{mn} (2g^{\bar{m}m})(2g^{\bar{n}n})  
\right] \\
& & - \frac{1}{4} \frac{\omega \wedge \omega}{2} \tr {}' \left[
 2 \times F_{\bar{m}\bar{n}} F_{mn} (2g^{\bar{m}m})(2g^{\bar{n}n})  
\right] \times 2, \nonumber 
\end{eqnarray}
where the second line comes from integrating out $D$, 
and the third line from ${\cal H}$.
All these terms are nothing but $\SO(7,1)$ invariant 
${\rm vol} \times (-1/4) \tr' [ F_{MN} F^{MN} ]$, 
where $M, N = 0, \cdots, 7$. 

{\bf BPS Conditions}

In the process of completing square for $D$, ${\cal H}$ and ${\cal G}$, 
one also finds that 
\begin{eqnarray}
D & = & \frac{4}{{\rm det} g_{m\bar{m}}} \left(
  \omega \wedge F^{(1,1)}
 - \frac{|\alpha|^2}{2} [\varphi,\overline{\varphi}]
                                         \right)_{12\bar{1}{\bar{2}}}
 = - i 2g^{\bar{n}n} F_{n\bar{n}} 
  - \frac{8 |\alpha|^2}{{\rm det} g_{m\bar{m}}} 
      [\varphi_{12},\overline{\varphi}_{\bar{1}\bar{2}}], \\
 \alpha {\cal H} & = & 2 F^{(2,0)},  \qquad 
  \alpha {\cal H}_{mn} = F_{mn},  \\
 i \omega \wedge {\cal G} & = & 
 \alpha^* \partial \overline{\varphi}.
\end{eqnarray}
Here, $(\psi)_{12\bar{1}\bar{2}}$ of a (2,2)-form $\psi$ is defined by 
$\psi = (\psi)_{12\bar{1}\bar{2}} 
 du^1\wedge du^2 \wedge d\bar{u}^1 \wedge d\bar{u}^2$.
Background field value of $\varphi$ and $A_m$ has to be chosen so that 
all of $\vev{D}$, $\vev{{\cal H}}$ and $\vev{{\cal G}}$ vanish. 
Vanishing $\vev{\overline{\cal H}}$ means that the gauge bundle 
can be chosen as holomorphic vector bundles (vanishing $F^{(0,2)}$).
Vanishing $\vev{\overline{\cal G}}$ also says that 
$\vev{\bar{\partial} \varphi} = 0$, that is $\vev{\varphi}$ depends 
only holomorphically on complex coordinates $u^m$ ($m=1,2$). 

\section{Zero-Mode Solution in the Doublet Background}
\label{sec:0-mode}

This appendix is dedicated to study of zero-mode wavefunctions.
From the zero-mode equations (\ref{eq:EOM3}, \ref{eq:EOM4}), 
\begin{equation}
 \tilde{\psi}_{\bar{m}}(u_1,u_2) = \frac{1}{\tau(u_1,u_2)} \; 
  \bar{\partial}_{\bar{m}} \tilde{\chi}(u_1,u_2) \qquad (m=1,2);
\label{eq:psi--chi}
\end{equation} 
here we simply write $\lambda_i \tau_i$ as $\tau$. 
$\vev{\varphi}_{12}$ is assumed to be diagonal here, 
and $\tau$'s are diagonal entries of 
$\rho_{U_i}(2\alpha \vev{\varphi_{12}})$. Substituting 
$\tilde{\psi}_{\bar{m}}$ above into another zero-mode 
equation (\ref{eq:EOM1}), we obtain 
\begin{equation}
 \left[ 
   \frac{\partial}{\partial v} \frac{1}{\tau(u,v)} 
   \frac{\partial}{\partial \bar{v}}
 + \frac{\partial}{\partial u} \frac{1}{\tau(u,v)} 
   \frac{\partial}{\partial \bar{u}}
 - \bar{\tau}(\bar{u},\bar{v}) 
 \right] \; \tilde{\chi}(u,\bar{u},v,\bar{v}) = 0, 
\label{eq:chi-eq}
\end{equation}
where a new set of coordinates $(u,v) = (u_1, -u_2)$ are 
introduced just for better readability.

In the main text, we encountered cases with $\tau(u,v) \propto u$, 
and 
\begin{equation}
 \tau(u,v)_\pm = - u \pm \sqrt{u^2 + v}, 
\label{eq:tau-pm}
\end{equation}
where the coordinate $v$ here corresponds to $-u_2$ 
in (\ref{eq:bg-AB}). 
The doublet background $(\tau_+ , \tau_-)$ also determines 
the zero-mode wavefunctions of ${\bf 10}+\overline{\bf 10}$
multiplets of $\SU(5)$ GUT models around type (a) codimension-3 
singularities, those of ${\bf 5}+\bar{\bf 5}$ multiplets 
of $\SU(5)$ GUT modes around type (c1) codimension-3 
singularities, and those of ${\bf 16}+\overline{\bf 16}$ 
multiplets of $\SO(10)$ GUT models around type (a) codimension-3 
singularities. The zero-mode wavefunction is known to 
decay in a Gaussian profile $e^{-|u|^2}$ in the case 
$\tau(u,v)$ is linear in a single coordinate, but it is not 
easy to find a solution for cases with general $\tau(u,v)$.
In this appendix, we will find a zero mode solution under the 
doublet background that is valid in a region of sufficiently 
small $|u^2 / v|$.

The background field vev $\tau_\pm$ in (\ref{eq:tau-pm}) has 
an expansion 
\begin{equation}
 \tau_{\pm} = \pm \sqrt{v} - u \pm \frac{1}{2}\frac{u^2}{\sqrt{v}} 
   + \cdots. 
\end{equation}
This is a $|u^2/v|$-expansion, and leading order terms should 
give a good approximation to $\tau_{\pm}$ for sufficiently small 
$|u^2/v|$. The leading order solution $\chi_0$ of (\ref{eq:chi-eq})
is determined by the condition
\begin{equation}
 \left[
   \frac{\partial}{\partial v} \frac{1}{v^\nu}
   \frac{\partial}{\partial \bar{v}}
 + \frac{\partial}{\partial u} \frac{1}{v^\nu}
   \frac{\partial}{\partial \bar{u}}
 - \bar{v}^\nu
 \right] \chi_0 = 0, 
\label{eq:chi-eq-nu}
\end{equation}
where $\nu = 1/2$ for the leading order of the doublet background.
($\nu = 1$ for the $\tau(u,v) = v$ case.) 
We drop $(u,\bar{u})$ dependence from $\chi_0$ for now, because
$\tau_\pm$ does not depend on the coordinate $u$ at the leading 
order. Now the solution $\chi_0$ is obtained by series expansion in 
$v,\bar{v}$; one can see explicitly that the following series 
\begin{eqnarray}
 \chi_{0,A,(\beta)}(v,\bar{v}) & = & v^{\nu+\beta} 
   \sum_{n=0}^{\infty} \frac{1}{n! \Gamma(n + \mu_A)}
      \left(\frac{(v\bar{v})^{\nu+1}}{(\nu+1)^2}\right)^n 
  \qquad \quad \left(\mu_A = 1 + \frac{\beta}{\nu+1} \right), \\
 \chi_{0,B,(\gamma)}(v,\bar{v}) & = & v^\nu \bar{v}^\gamma 
   \sum_{n=0}^{\infty} \frac{1}{n! \Gamma(n + \mu_B)}
      \left(\frac{(v\bar{v})^{\nu+1}}{(\nu+1)^2}\right)^n 
  \qquad \quad \left(\mu_B = 1 + \frac{\gamma}{\nu+1} \right), 
\end{eqnarray}
satisfy (\ref{eq:chi-eq-nu}). Thus, the leading-order solution 
$\chi_0$ should be given in a linear combination 
\begin{equation}
 \chi_0 = c_{\beta} \chi_{0,A,(\beta)}(v,\bar{v}) + 
          \tilde{c}_\gamma \chi_{0,B,(\gamma)}(v,\bar{v}).
\end{equation}

We are interested in a solution localized along the matter curve,
which is now $v=0$. To see the constraint imposed on $\beta$, 
$\gamma$, $c_\beta$ and $\tilde{c}_\gamma$ for a localized 
solution, let us study the asymptotic behavior of the 
solutions above. To this end, it is important to note that 
the series expansion form of $\chi_{0,A,(\beta)}$ 
and $\chi_{0,B,(\gamma)}$ is written by generalized hypergeometric 
functions, 
\begin{equation}
 {}_p F_q (\alpha_1, \cdots, \alpha_p, \beta_1, \cdots, \beta_p; 
           \rho_1, \cdots, \rho_q, \mu_1, \cdots, \mu_q; z)
  = \sum_{n=0}^\infty 
   \frac{\prod_i \Gamma(\alpha_i n + \beta_i)}
        {n! \; \prod_i \Gamma(\rho_i n + \mu_i)} z^n.
\end{equation}
For the case of our interest, ${}_0 F_1$ is relevant, and 
$z = (v\bar{v})^{\nu+1}/(\nu+1)^2$, $\rho_1 = 1$ and 
$\mu_1 = \mu_{A,B}$. Asymptotic behavior of the generalized 
hypergeometric functions for large $|z|$ was studied 
in \cite{Wright}:
\begin{eqnarray}
 {}_0 F_1(\rho_1 = 1, \mu; |Z|^2) & \sim &
   (2|Z|)^{\frac{1}{2}-\mu} e^{+2|Z|} 
  \left(A_0 + \frac{A_1}{2|Z|} + \frac{A_2}{(2|Z|)^2} + \cdots \right)
\label{eq:GHG-asymptotic}  \\
 & & + (e^{\pi i} 2|Z| )^{\frac{1}{2} - \mu} e^{-2|Z|} 
  \left(A_0 - \frac{A_1}{2|Z|} + \frac{A_2}{(2|Z|)^2} + \cdots \right) .
   \nonumber 
\end{eqnarray}
The coefficients of the asymptotic expansion $A_m$ ($m = 0, 1 , \cdots$) 
do not depend on $|Z|$, the first two of which are 
\begin{equation}
 A_0(\mu)  =  \frac{2^\mu}{\sqrt{2\pi}}, \qquad 
 A_1(\mu)  =  \frac{2^{\mu-1}}{\sqrt{2\pi}}\left(\frac{1}{4}- (\delta\mu)^2
					\right),
\end{equation}
where $\delta\mu \equiv \mu - 1$. Looking at the leading order 
term in the asymptotic expansion, we see that all of 
$\chi_{0,A,(\beta)}(v,\bar{v})$ and $\chi_{0,B,(\gamma)}$ grow 
exponentially for large $|v|$:
\begin{eqnarray}
 \chi_{0,A,(\beta)} & \sim & (v/\bar{v})^{\frac{\beta}{2}} \times 
 (v \bar{v}^{-3})^{1/8} \; \exp \left(+ \frac{4}{3}|v|^{\frac{3}{2}}
				\right), \\
 \chi_{0,B,(\gamma)} & \sim & (v/\bar{v})^{-\frac{\gamma}{2}} \times
 (v \bar{v}^{-3})^{1/8} \; \exp \left(+ \frac{4}{3} |v|^{\frac{3}{2}}
                                \right),
\end{eqnarray}
where we used $\nu = 1/2$, and overall coefficients are ignored. 
In order to cancel this exponential growth 
to obtain a localized zero-mode solution, $\chi_{0,A,(\beta)}$ and 
$\chi_{0,B,(\gamma)}$ with $\beta = - \gamma \neq 0$ have to be 
employed and their coefficients have to be properly adjusted. 
Requiring that the zero-mode solution $\chi_0$ does not have a pole 
at $v = 0$, and it is either single valued\footnote{Remember that 
there is an $\mathfrak{S}_2$ monodromy
exchanging $\tau_+ \sim +\sqrt{v}$ and $\tau_- \sim - \sqrt{v}$. 
$x$ is a coordinate introduced in (\ref{eq:coord-x}).} 
in $x \equiv \sqrt{v}$, 
\begin{equation}
 \beta = - \frac{1}{2}, \quad \gamma = + \frac{1}{2}, \quad 
 \mu_A = \frac{2}{3}, \quad \mu_B = \frac{4}{3}
\end{equation}
is the only possibility. Therefore, the zero-mode solution is 
given by 
\begin{equation}
 \chi_0(v,\bar{v}) = c_{\beta = -1/2} \left(
  \chi_{0,A,(\beta = -1/2)}(v,\bar{v}) - 
  (2/3)^{2/3} \chi_{0,B,(\gamma = 1/2)}(v,\bar{v})\right).
\end{equation}
To see that this solution is really localized around small $|v|$, 
we need to note the following. The coefficients $A_m(\mu)$ 
in the asymptotic expansion (\ref{eq:GHG-asymptotic}) are those 
appearing in the asymptotic expansion \cite{Wright}\footnote{
For the case $\tau(u,v) \propto v$, $\nu = 1$, $- \beta = \gamma = 1$  
and $\delta \mu = \mp 1/2$. In this case, $A_m = 0$ for all 
$m=1,\cdots, \infty$, and the $A_0$ terms become the Gaussian-form 
wavefunction.}
\begin{equation}
 \frac{1}{2^{2t} \Gamma(t+1) \Gamma(t + \mu)} 
= \frac{1}{2^{2t} \Gamma(t+1) \Gamma(t+1 + \delta\mu)} \sim 
\sum_{m=0}^{\infty} \frac{A_m(\mu)}
  {\Gamma\left(2t + \frac{3}{2} + \delta\mu + m \right)}.
\end{equation}
With a change of variable $t' = t + \delta\mu$, one can show that 
\begin{equation}
 A_m (1 - \delta\mu) = A_m (1 + \delta\mu) 2^{- 2 \; \delta\mu}.
\end{equation}
Therefore, $A_m(\mu_B = 4/3) = A_m(\mu_A = 2/3) \times 2^{2/3}$, 
and the ratio $A_m(\mu_B)/A_m(\mu_A)$ are all the same 
for any $m = 0,\cdots, \infty$. As we set the coefficients 
$c_{\beta = -1/2}$ and $\tilde{c}_{\gamma = 1/2}$ so that the 
$A_0$-term of the growing exponentials cancel, all the 
higher order $A_m/(4|v|^{3/2}/3)^m$-terms multiplying the growing 
exponential also cancel between $\chi_{0,A}$ and $\chi_{0,B}$. 
Only the decaying exponential part survives because of the 
$e^{\pi i}$ phase factor in the second line of
(\ref{eq:GHG-asymptotic}), and the zero-mode solution $\chi_0$ 
behaves asymptotically as 
\begin{equation}
 \chi_0 \sim c_{\beta = -1/2} \; \sqrt{\frac{3}{2\pi}} \; (12)^{1/6} \;
(v \bar{v})^{- \frac{1}{8}} \exp\left(- \frac{4}{3} |v|^{\frac{3}{2}} \right)
\label{eq:chi-asymptotic}
\end{equation}
at large $|v|$ at the leading order. Compared with the case 
$\tau \propto v$, when the zero-mode wavefunction decays as 
the standard Gaussian profile, $\chi \sim e^{-|v|^2}$, the wavefunction 
decays more slowly, $e^{-|v|^{3/2}}$, when $\tau \propto \sqrt{v}$. 
The zero-mode wavefunction is still localized along the matter curve 
characterized by $\tau=0$ even in cases where $\tau \sim \sqrt{v}$.

For small $|v|$, this solution behaves as 
\begin{eqnarray}
 \chi_0 & = & c \left(
   \frac{1}{\Gamma(2/3)} + \frac{(2/3)^2 |v|^3}{\Gamma(5/3)}
 + \frac{(2/3)^4 |v|^6}{\Gamma(8/3)} + \cdots 
 - \frac{(2/3)^{2/3} |v|}{\Gamma(4/3)} 
 - \frac{(2/3)^{8/3} |v|^4}{\Gamma(7/3)} - \cdots \right), \\
  & = & c \left(\frac{1}{\Gamma(2/3)} 
     - \frac{(2/3)^{2/3} }{\Gamma(4/3)} x \bar{x} + \cdots \right).
\label{eq:chi-on-C}
\end{eqnarray}
The $\tilde{\psi}_0 \equiv i \psi_{\bar{m}} d u_{\bar{m}}$
part\footnote{
The subscript ${}_0$ means that this is a solution at the leading
order.} 
of the zero-mode wavefunction is obtained by using (\ref{eq:psi--chi}).
\begin{eqnarray}
 i \psi_0 & = & c \; \frac{1}{2\sqrt{\bar{v}}} \; d\bar{v} \;  \left( 
  - \frac{(2/3)^{2/3}}{\Gamma(4/3)}
  + \frac{3 (2/3)^2}{\Gamma(5/3)} v\bar{v}  - \cdots \right), \\
  & = & c \; d \bar{x} \;  \left( 
  - \frac{(2/3)^{2/3}}{\Gamma(4/3)}
  + \frac{3 (2/3)^2}{\Gamma(5/3)} x^2 \bar{x}^2  - \cdots \right).
\label{eq:psi-on-C}
\end{eqnarray}
Note that the zero mode solution $(\psi_0, \chi_0)$ behaves much 
better when it is expressed on the 2-fold covering space $C$ (with 
coordinate $x$). No half-integral power of $x$ or $\bar{x}$ appears in 
$\chi_0$ in (\ref{eq:chi-on-C}), and an apparent $1/\sqrt{\bar{v}}$ 
singularity in $\psi_0$ is gone in (\ref{eq:psi-on-C}). 
This is another indication\footnote{For the other indication associated
with the $\mathfrak{S}_2$ monodromy of the fields around codimension-3 
singularity points, see the main text.} 
that the covering space $C$, rather than the base space $S$, 
is where the zero-mode wavefunctions sit in the doublet background.

\end{document}